# COMBINING SHAMIR & ADDITIVE SECRET SHARING TO IMPROVE EFFICIENCY OF SMC PRIMITIVES AGAINST MALICIOUS ADVERSARIES

A Thesis presented to

the Faculty of the Graduate School

at the University of Missouri

In Partial Fulfillment

of the Requirements for the Degree

Doctor of Philosophy

by

Ken Goss

Dr. Wei Jiang, Thesis Supervisor

May, 2020

The undersigned, appointed by the Dean of the Graduate School, have examined the dissertation entitled:

COMBINING SHAMIR & ADDITIVE SECRET SHARING
TO IMPROVE EFFICIENCY OF SMC PRIMITIVES
AGAINST MALICIOUS ADVERSARIES

presented by Ken Goss,

a candidate for the degree of Doctor of Philosophy and hereby certify that, in their opinion, it is worthy of acceptance.

_________________________________

Dr. Wei Jiang

_________________________________

Dr. Dan Lin

_________________________________

Dr. Rohit Chadha

_________________________________

Dr. Stephen Montgomery-Smith

# ACKNOWLEDGMENTS

I would like to express my sincere gratitude to the National Science Foundation, for their provision of the SFS program and the support I received through this program. The country is receiving an invaluable contribution through the support of research and development of young professionals in this program. Further, I would like to thank my advisor, Dr. Wei Jiang, for the thoughtful comments and recommendations on this dissertation. He has been a constant source of counsel, encouragement, and support. I can think of no better advisor to have and am beyond appreciative for all he has done for me. Additionally I am thankful for the time, expertise, and comments of all the members of the committee who provided useful feedback and recommendations, Dr. Lin, Dr. Chadha, and Dr. Montgomery-Smith. I am also thankful for the University of Missouri, the College of Engineering, and the EECS Department in particular, and all the faculty and staff for all their considerate guidance and help navigating the entire process. To conclude, I cannot forget to thank my family and friends for all the unconditional support in this very intense academic year.

# Contents

# List of Tables

# List of Figures

# ABSTRACT

Secure multi-party computation provides a wide array of protocols for mutually distrustful parties be able to securely evaluate functions of private inputs. Within recent years, many such protocols have been proposed representing a plethora of strategies to securely and efficiently handle such computation. These protocols have become increasingly efficient, but their performance still is impractical in many settings. We propose new approaches to some of these problems which are either more efficient than previous works within the same security models or offer better security guarantees with comparable efficiency. The goals of this research are to improve efficiency and security of secure multi-party protocols and explore the application of such approaches to novel threat scenarios. Some of the novel optimizations employed are dynamically switching domains of shared secrets, asymmetric computations, and advantageous functional transformations, among others. Specifically, this work presents a novel combination of Shamir and Additive secret sharing to be used in parallel which allows for the transformation of efficient protocols secure against passive adversaries to be secure against active adversaries. From this set of primitives we propose the construction of a comparison protocol which can be implemented under that approach with a complexity which is more efficient than other recent works for common domains of interest. Finally, we present a system which addresses a critical security threat for the protection and obfuscation of information which may be of high consequence.

# Chapter 1

# Introduction

A need for the ability to perform secure computation in a practical manner is increasing proportionally to the need of our society to securely analyze the data we are amassing, which is itself increasing at a constantly accelerating rate.

Society is increasingly interested in a pair of mutually exclusive pursuits. There is strong and approaching universal interest in a wide range of statistics such as mean standard deviation and the like with respect to a similarly wide range of attributes such as salary. Simultaneously, many people are becoming more interested in maintaining control over their private information, as well as digital privacy in general, and distrustful of institutions which were trusted implicitly in the past. If the interest for these analytics is there, but no trusted central authority to handle their computation, how can the interest be satisfied? This is a perfect example of a situation in which SMC protocols may provide the requisite functionality, while maintaining the privacy of all individual user data, only revealing the aggregate statistics of interest. Secure multi-party computation (SMC) in general has been a very active area for recent research and there have been numerous improvements in the efficiency of these protocols. In our research, inquiries are planned concerning a number of interesting and important areas, such as the following:

- Provide a secure protocol compiler for transforming protocols secure against semi-honest adversaries into protocols secure against malicious adversaries such that the transformation is efficient, information-theoretically secure, and only requires an honest majority

- Further improve the efficiency of secure multi-party comparisons

- Efficient secure comparisons will additionally allow for other interesting applications such as secure and oblivious sorting

- Explore the novel threat scenarios where such techniques would be advantageous and provide a solution based on our foundational research

## 1.1 Organization

The following will address some foundational information on which much of the research is based in Section 1.2. Then we will enter the main contributions of the work.

### 1.1.1 General Protocol Compiler

In Chapter 2 we propose a general purpose compiler which allows for the transformation of any protocol secure against passive adversaries into a protocol secure against active adversaries.

The area of research related to such compilers has been very active recently and a wide array of options exist with an equally wide array of attributes. The focus of our research is on a novel combination of techniques which allows for the cost of the transformation to remain efficient, the threshold of resilience against malicious adversaries to remain less than half ($t < n/2$ rather than $t < n/3$), and information-theoretic security to be maintained.

Many of the most efficient protocols are secure against and threshold $t < n/3$ malicious adversaries. Other approaches, though at a greater cost, can tolerate up to $t < n/2$ such adversaries. The approach given here allows this method to maintain a threshold of $t < n/2$ malicious adversaries with a cost on par with those existing works with a threshold of $t < n/3$. This represents a significant improvement in complexity for simple honest majority compilers, or a security increase for the same efficiency in comparison with the other works.

### 1.1.2 Comparison Operator

In Chapter 3 we consider an operation of greater complexity than the basic requirements for SMC, namely secure comparison. We propose a protocol which can be implemented using the compiler proposed in Chapter 2 which is very efficient compared against other recent results.

Within recent years, secure comparison protocols have been proposed using binary decomposition and properties of algebraic fields. These protocols have become increasingly efficient, but their performance has seemingly reached a plateau. We propose a new approach to this problem that transforms the comparison function into comparing specialized summations and takes advantage of dynamically switching domains of secret shares and asymmetric computations for intermediate calculations among the participating parties. As a consequence, according to our analysis, communication and computation costs have been brought to a very low and efficient level. Particularly, the communication costs have been considerably reduced both in order as well as the dominating term's order of magnitude. In addition, we propose a secure protocol under the malicious setting which maintains our transformation and is more efficient than the existing work for common domain sizes and is able to be implemented efficiently under our approach presented in Chapter 2.

### 1.1.3 Secure and Oblivious Firewall

In Chapter 4 we propose a risk to security due to insider threats and propose a system to mitigate the threat which can be implemented based on the preceding work introduced in Chapters 2 and 3.

Firewalls have long been in use to protect local networks from threats of the larger Internet. Although firewalls are effective in preventing attacks initiated from outside, they are vulnerable to insider threats, e.g., malicious insiders may access and alter firewall configurations, and disable firewall services. In this paper, we develop an innovative distributed architecture to obliviously manage and evaluate firewalls to prevent both insider and external attacks oriented to the firewalls. Our proposed structure alleviates these issues by obfuscating the firewall rules or policies themselves, then distributing the function of evaluating these rules across multiple servers. Thus, both accessing and altering the rules are considerably more difficult thereby providing better protection to the local network as well as greater security for the firewall itself. We achieve this by integrating multiple areas of research such as secret sharing schemes and multi-party computation, as well as Bloom filters and Byzantine agreement protocols. Our resulting solution is an efficient and secure means by which a firewall may be distributed, and obfuscated while maintaining the ability for multiple servers to obliviously evaluate its functionality.

## 1.2 Preliminaries

In all the following work, we make use of secret sharing schemes. Research into secret sharing schemes, and means by which they may be improved or used is the core of this work. Very often we make use of additive secret sharing. The approach under additive secret sharing makes use of a different set of mathematical principles to achieve secure multi-party computation than the more widely referenced Shamir scheme [1], though modular arithmetic still lies at the core of its security. We also will make use of circular shifts to hide some information in the context of our scheme, and finally, we will introduce the adversary models and our approach to security concerns and proof arguments.

### 1.2.1 Shamir Secret Sharing

**Foundations**

Secret sharing in general seeks to divide data among many parties such that the data can be efficiently reconstructed given only the shares, and, possession of a share, or in some cases even a subset of shares, yields no increase in knowledge concerning the secret. In these schemes, unlike homomorphic encryption, there are no encryption keys, only the shares of the data itself. In this type of scheme, it is the data, in the form of its divided shares, that serves as both the encryption and the key. Specifically, secret sharing requires an "all-or-nothing" approach where having less than the sufficient number of shares provides no information about the original value [2].

The foundational and eminent work in Secret Sharing comes from the paper of Shamir [3], also of immense recognition for his contribution to the RSA public key encryption algorithm [4]. In his proposed approach, shares are generated as points along a polynomial with the y-intercept, or constant term, of the polynomial being assigned the value to be kept secret with the rest of the coefficients being generated from a uniformly random distribution of values in the same domain as the secret. For the polynomial coefficients, and the shares, this domain is a field defined by a prime, $p$, greater than the domain required by the magnitude of the secrets. All operations are done according to the restrictions and methods of the field defined by this prime. Then indices and evaluated values are distributed together as a tuple constituting the shares. In order to rebuild the secret a number of shares one greater than the degree of the originating polynomial is required. The most common and intuitive method to achieve this is that of Lagrange polynomial interpolation.

This ability to select the degree of the originating polynomial in a manner largely independent from the number of parties in the scheme, gives rise to the possibility of a threshold scheme for the shares of the secret. Given any desired number of shares, $k$, required to rebuild the secret, the polynomial used to construct those shares should be of degree $k-1$. It is then possible to generate as many shares as required for any number of parties, $n$, while maintaining the ability to rebuild the secret by any subset of the $n$ parties which has a magnitude of at least $k$. This is formally referred to as a $(k, n)$ threshold scheme.

It is clear from general mathematical principles that, given any $k-1$ shares from the set of shares, not only is it impossible to reconstruct the polynomial in this field, but it is also impossible to gain any knowledge at all about the secret. This is true unconditionally, and forms the greatest strength of all methods falling in this class of secure computation methods. Much of modern cryptography relies heavily, and at present, reasonably, on the presumption that $p \neq np$. This reliance on the current intractability of problem classes is the source of the strength for many protocols. If a proof were to be produced which negated that widely held belief, a vast swath of modern cryptographic protocols would be obsolete. Secret sharing, however, would remain due to this lack of reliance on intractability replaced by the fact that reconstructing the secret is not difficult, it is mathematically impossible without a sufficient number of shares, and with anything less than that sufficient number of shares, no information, at all, is gained by the would be attacker.

**Operations**

**Constructing shares**  Formally, a secret value, s, is placed within a $k-1$ degree polynomial function $f$ of the form:

$$f(x) = (c_{k-1}x^{k-1} + c_{k-2}x^{k-2} + \cdots + c_2x^2 + c_1x + s) \mod p$$

in which all constants $c_{k-1}$ through $c_1$ are chosen from a uniform random distribution on the domain of the secret. For all indices, $i$, in the range $[1, n]$ the share, $s_i$ for each party is constructed:

$$s_i = (i, f(i))$$

**Rebuilding Secrets**  Lagrange Polynomial Interpolation is the method of choice for these operations. Due to our interest being the y intercept the general formula,

$$f(x) = \sum_{j=1}^{n} y_j \prod_{k=1:k\neq j}^{n} \frac{x - x_k}{x_j - x_k}$$

may be somewhat simplified because every variable $x$ in the preceding equation, for our case, is 0.

$$s = f(0) = \sum_{j=1}^{n} y_j \prod_{k=1:k\neq j}^{n} \frac{-x_k}{x_j - x_k}$$

**Addition**  In the Shamir Secret Sharing Scheme, addition is "free" in that no communication is required between parties, and no lengthy local computations are necessary. For two values, $a$ and $b$, which are shared among the parties, the shared sum, $c$ is computed by each party $p_i$ locally computing their new share by adding their shares of the secrets $a$ and $b$:

$$\forall i \in [i, n], c_i = a_i + b_i$$

This is the same for the addition of shares with some publicly known constant. The constant is treated as a constant "polynomial" and each party adds the same constant into their locally held share.

**Multiplication**   Due to the fact that this scheme is inherently a linear scheme, multiplication is a more complex process. If we were to simply multiply the shares as we added them in the previous case, we would get a set of shares representing a polynomial of an order twice that of the what it was previously. This would be disastrous for the system as a whole. For some data, twice the desired number of shares would be required for the reconstruction of the secret. The system would be made up of inconsistent share representations based on polynomials of differing degree. Additionally, the resulting polynomial would no longer be as random thus providing opportunity to leak information.

In order to achieve multiplication for two shared values $a$ and $b$, and get the product, shared, $d$, we must compute the product of all the shares. However, it is necessary to also handle some additional communications and calculations by which the final polynomial maintains the same degree as was previously desired, as well as more uniform randomness. It is important to note that if multiplication is a desired operation, the number of parties holding shares, $n$, must be greater than or equal to $2k - 1$. This is required in order to be able to reduce the degree of the polynomial. The introduction and proof of this method is given by Gennaro et al [5].

Each party initially computes the product of their shares $q_i$ which is an intermediate value for each party in this protocol. They each independently, and uniformly randomly construct polynomials of the same degree desired for the system and set the constant, or y-intercept to their intermediate calculated value, $q_i$. Each of the local equations will be of the form:

$$h_i(x) = (c_{k-1}x^{k-1} + c_{k-2}x^{k-2} + \cdots + c_2x^2 + c_1x + q_i) \mod p$$

Each party then generates, exactly as the dealer would with a new value to be shared, a set of shares from this polynomial for each of the parties in the scheme. This is done by randomly and uniformly creating shares of the product of the data points by each party and for each player. These must be shared so that a new and similarly uniform random $k-1$ order polynomial can be calculated by each peer.

$$\forall i, j \in [1, n], R_{ij} = (i, h_j(i))$$

Parties then exchange these shares $\{R_{ij}\}$ of their intermediately calculated values, party $i$ sending its shares to each other player, $j$, in the range $[1, n]$ keeping the value applicable for its own index, where $i = j$.

Thus each party gains a set of values from each other party in the scheme. These values must be combined with weights that reflect the reduction of order desired. This is achieved through the construction and use of a Vandermonde matrix which neatly handles the geometric properties required. The shares of the lower order polynomial are calculated by exploiting a special property of Vandermonde Matrices, which represent a geometric progression in each row. Given a desired degree $t$ we should construct a $2t + 1 \times 2t + 1$ Vandermonde Matrix. The necessary matrix may be calculated independently by each player. This means that no communication needs to take place between the parties for each one to be able to independently build this matrix locally.

$$\begin{bmatrix} 1 & 1 & \ldots & 1 \\ 1 & 2 & \ldots & 2^{2t} \\ \vdots & & \ddots & \\ 1 & 2t+1 & \ldots & (2t+1)^{2t} \end{bmatrix}$$

This matrix must be inverted and the resulting first row, related to the 0 power of the geometric form, is what is used as a series of weights to combine the results from all the parties locally. The inner product of the top row of the inverted Vandermonde, and the vector of shares from each of the other parties is computed to result in the each party receiving the correct, and securely computed, share $d_i$ of the product $a \cdot b$. All the calculations, in every step of the protocol, must be carried out in accordance with the modulus by the prime $p$.

Following this approach ensures that the resulting constructed polynomial is of the desired degree, and that it is random because the matrix inversion process and resulting vector dot product will not remove the random attributes placed there in the uniform random construction of the individual polynomials built by the parties to share their locally computed intermediate values.

$$\forall i \in [1, n], d_i = \sum_{j=1}^{n} V_{1j}^{-1} \cdot R_{ij}$$

A series of numerical examples are provided in the appendices illustrating each of these operations.

### 1.2.2 Additive Secret Sharing

**Foundations**

This scheme makes use of a different set of mathematical principles to achieve secure multiparty computation, though modular arithmetic still lies at the core of its security.

The underlying security is dependent on the fact that adding any value to a uniformly randomly selected value, modulus a value delimiting the field, $p$, is still uniformly random. It is therefore impossible to say what the non-random component of the sum was, when considering only the resulting sum. In this context the sum is therefore unconditionally secure since any adversary, unbounded by limits on computational power, can do no better than also simply randomly guess at what the two original values may have been. This security is similar in principle to the unconditional security guarantees of the One Time Pad protocol.

In the context of this scheme many of the operations are much more computationally efficient than was the case in the Shamir Secret Sharing Scheme, but, as always, there is a tradeoff. With this secret sharing scheme, there is no ability to have less than all the shares that were originally generated involved in the reconstruction of the secrets. So, while it is fairly trivial to distribute the shares and reconstruct the secret, all shares are required. This is a special case of the idea previously discussed in Section 1.2.1, of a threshold scheme in which $k = n$.

**Operations**

**Constructing Shares**  To create any desired number of shares, $n$, in this scheme, the secret, $S$, to be shared is placed in an equation as follows in which all $s_1$ through $s_{n-1}$ are uniformly randomly selected integers and $s_n$ satisfies the equation in the domain of the secret defined by an integer, $p$ greater than that of the desired size of the secrets:

$$s_n = S - s_1 - s_2 - \cdots - s_{n-1} \mod p$$

**Rebuilding Secrets**  Rebuilding secrets in this case is also trivial since it can clearly be seen from the preceding all one need do is sum all the $n$, shares held by

the parties modulo $p$:

$$S = \sum_{i=1}^{n} s_i \mod p$$

$$S = s_1 + s_2 + \cdots + s_n \mod p$$

**Addition**   Since this scheme simply makes use of addition modulus some prime, it is possible for each party in the scheme to acquire the share for the sum of two secrets by simply adding, modulo p, their local shares of the two secrets. This can be seen for the two secrets $A$, and $B$, here via some arithmetic manipulation:

$$A + B = (a_1 + a_2 + \cdots + a_n) + (b_1 + b_2 + \cdots + b_n) \mod p$$

$$A + B = (a_1 + b_1) + (a_2 + b_2) + \cdots + (a_n + b_n) \mod p$$

**Multiplication**   Multiplication is affected among three parties through the generation, distribution, and use of multiple additional uniform random values from the field. They are combined and shared among the parties in a multistep process which creates a complex, but secure, uniform random sharing of the desired product among the parties. The protocol given in the work of Bogdanov [2, 6], is as follows in Algorithm 1 for two secrets $U$, and $V$, with shares $u_1, u_2, u_3$ and a similar set for V. The proof of correctness is verbose, but straightforward, we would direct the reader to the thorough treatment it is given in [2, 6].

A series of numerical examples are provided in the appendices illustrating each of these operations as well.

**Algorithm 1:** Additive Secret Sharing Multiplication

1. Party 1 generates $u_{13}, v_{13} \in \mathbb{Z}_N$
2. Party 2 generates $u_{21}, v_{21} \in \mathbb{Z}_N$
3. Party 3 generates $u_{32}, v_{32} \in \mathbb{Z}_N$
4. $P_1 \to P_2 : [u]_N^{P_1} - u_{13}, [v]_N^{P_1} - v_{13};\ P_2 \to P_1 : u_{21}, v_{21}$
5. $P_2 \to P_3 : [u]_N^{P_2} - u_{21}, [v]_N^{P_2} - v_{21};\ P_3 \to P_2 : u_{32}, v_{32}$
6. $P_3 \to P_1 : [u]_N^{P_3} - u_{32}, [v]_N^{P_3} - v_{32};\ P_2 \to P_1 : u_{13}, v_{13}$
7. Party 1 and 2 compute: $u'_1 = [u]_N^{P_1} - u_{13} + u_{21}$ and $v'_1 = [v]_N^{P_1} - v_{13} + v_{21}$
8. Party 2 and 3 compute: $u'_2 = [u]_N^{P_2} - u_{21} + u_{32}$ and $v'_2 = [v]_N^{P_2} - v_{21} + v_{32}$
9. Party 3 and 1 compute: $u'_3 = [u]_N^{P_3} - u_{32} + u_{13}$ and $v'_3 = [v]_N^{P_3} - v_{32} + v_{13}$
10. Party 1 compute: $[uv]_N^{P_1} = u'_1 v'_1 + u'_1 v'_3 + u'_3 v'_1$
11. Party 2 compute: $[uv]_N^{P_2} = u'_2 v'_2 + u'_2 v'_1 + u'_1 v'_2$
12. Party 3 compute: $[uv]_N^{P_3} = u'_3 v'_3 + u'_3 v'_2 + u'_2 v'_3$

### 1.2.3 Operations & Notations in Linear Secret Sharing Schemes

We require that any secret sharing scheme to be used have the ability to perform the following operations, and will adhere to the following notational conventions in the much of the work:

- Share: given a particular value $x$, generate shares of $x$, denoted by $[x]_N^{P_j}$, in a group defined by a modulus $N$ for each party $P_j$. This must be done in a way that they can be uniquely recombined in a method applicable to the scheme to reconstruct the original value.

- Reveal: a sufficient number of shares $[x]_N^{P_j}$ can be recombined to reveal the original value $x$.

- Add with a public constant: given shares $[x]_N^{P_j}$, and a public constant $c$, execute the necessary operations to calculate $c + [x]_N^{P_j} = [c + x]_N^{P_j}$. In a linear secret sharing scheme, this can be done locally.

- Add shares: given two shared values $[x]_N^{P_j}$ and $[y]_N^{P_j}$, calculate the sum of their values in a shared form. This can be executed without communications by using the addition operation implemented in a linear secret sharing scheme, i.e., $[x]_N^{P_j} + [y]_N^{P_j} = [x+y]_N^{P_j}$.

- Multiplication by a public constant: given shares $[x]_N^{P_j}$, and a public constant $c$, execute the necessary operations to calculate $c[x]_N^{P_j} = [cx]_N^{P_j}$. In a linear secret sharing scheme, this can be done locally without communication among the parties.

### 1.2.4 Random Shift

In our protocol, we will make use of a random shift permutation. This permutation $\pi$ is encoded as an integer in $\mathbb{Z}_\ell$ requiring $\log_2 \ell$ bits. We use a specific kind of cyclic permutation called a circular shift. It is important to note that this is not a fully random permutation, in the sense that the placement of all values in the vector after permutation are totally independent from their previous location. In our permutation, we are only concerned about shifting one element in an array by a uniformly random amount. In a circular shift, every element that was previously adjacent to another element will maintain that relation in the permuted vector. Given a number encoding the permutation, the values are circularly shifted right that number of index locations. This is denoted as applied to a vector/array $v$ by $\text{shift}_\pi(v)$, and similarly for an inverse $\text{shift}_\pi^{-1}(v)$. We only need the location, indexing the most significant bit difference, being able to move to any other locations with a uniform probability. In our use, as in Protocol 4, every other element in the array is meaningless and uniformly random by design, and therefore, does not leak any information. This is similar to the shift permutation in [7].

## 1.2.5 Security and Adversary Model

Aside from the correctness and efficiency of any existing protocols, an important consideration is the security guarantee that can be derived from their execution. Toward the end of proving security, we have generally followed the proof style and conventions of simulation [8, 9] under the universal composability paradigm [10, 11] as reduced and simplified in [12]. The goal in this setting is to demonstrate for an arbitrary functionality $f$, there exists an equivalence of information disclosure between the ideal execution of a protocol and a real execution of a protocol implementing the same ultimate functionality. This is constrained for cases in which adversaries and their ability to corrupt the execution of the protocol are identical. In the ideal setting a functionality is executed by a trusted third party which is not present in the real model. In the semi-honest model, or passive adversary setting, adversaries may invest an arbitrary amount of computational power to attempt to extract additional undesirable information from the record of the communications they may locally maintain. This should be provably unsuccessful in this model, but the adversaries are assumed to follow the prescribed steps of the protocol exactly. In the malicious model, or active adversary setting, not only should the preceding condition hold, but the protocol should also be provably secure against arbitrary deviations from the prescribed protocol up to the level of deviance tolerated in the ideal setting, such as refusal to participate.

When we move our protocol from the semi-honest model into the malicious model, we will require some additional tools. As it stands, the protocol to be presented in the semi-honest setting has many optimizations which are infeasible or inefficient to make secure in the malicious setting. While preserving our comparison transformation to be discussed in Claim 1, we use methods proposed by [5] in order to secure our protocol in the malicious setting. Other frameworks for transforming between semi-honest and malicious settings exist, such as [13], but as with these two in particular, for arbitrary sets of parties, there is no considerable difference in complexity, and any difference would shift all the protocols together but maintain their comparative ordering in terms of efficiency. Additionally, the scheme weakens security since some of its primitives are only computationally, not information theoretically secure, and it does not guarantee fairness [13]. Therefore, any framework for affecting this transformation may be used though we have elected to base our results on the scheme in [5].

# Chapter 2

# Share with Shamir or Adopt Additive? Why not both?

## 2.1 Introduction

Secure multiparty computation (SMC) has been a very active and useful area of research in recent years. As such many schemes are constantly being proposed for various security settings and efficiency concerns from protocols for various primitives, suites of protocols tuned for a particular applications, to general compilers which present the ability to transform any general computation into a protocol computable in the SMC setting. A number of such compilers have been proposed in the literature, again, each having their emphasis or specialty. Those seeking the strongest security guarantees are particularly interesting as they must prevent an an adversary from being able to disrupt the output of the SMC protocol up to equivalence with an idealized setting. Many are able to do this, though there are legitimate concerns with respect to their efficiency.

In general, SMC protocols seek to define a means by which a set of parties can cooperate in evaluating a functionality of interest on some privately held information.

The goal is that this should be done in a manner which is functionally equivalent to there being a trusted third party present.

This goal is achievable, with varying costs in a variety of scenarios. The two primary types of settings are those which include passive or semi-honest adversaries, and those which include active or malicious adversaries. Passive adversaries are trusted to follow the prescribed steps of the protocol, though one must guard against attempts of any participating party to be able to extract additional information from the transcript of the protocol's execution. In addition, if malicious adversaries are to be considered, the protocol designers can no longer trust the steps they specify to be followed. They must then include mechanisms to force potential adversaries into compliance, or at least detect their deviation.

Transforming a protocol from being secure against passive adversaries to also being able to handle active ones is certainly non-trivial. This transformation forms the basis of a considerable amount of recent research seeking to mitigate the costs of this transition. It is with respect to this goal that we present our research.

### 2.1.1 Organization and Our Contributions

In the sections to follow we will present and summarize some of the most pressing details about the proposed schemes of recent works in this area, as well as salient security definitions (Section 2.2). We then follow with Section 2.3, in which we present our novel scheme which is very efficient and maintains information theoretic security. Finally we summarize with our conclusions in Section 2.5.

## 2.2 Related Work

In all the discussion and analysis to follow we will adopt some uniform notation as follows:

- $n$: the number of parties involved in the scheme

- $t$: the threshold of the scheme

- $\kappa$: the bit-width of the field for the shares, and security parameter

- $C$: the size of the arithmetic circuit in question, in terms of a count of its gates. $C_M$ is used to refer to the count of multiplication gates as many schemes incur communications only during that operation

- $D$: depth of the circuit in question, i.e., the length of the longest path from input to output

- $[s]_N^{P_j}$: a share of a secret $s$ for party $P_j$ in the field of characteristic $N$ where $N$ is a $\kappa$ bit prime.

### 2.2.1 Secret Sharing

Shamir's scheme [14], in general allows for a coalition of $n$ parties to share a secret among themselves with information theoretic security in terms of privacy against a threshold subset of at most $t \leq n$ adversaries. However, guarding against malicious adversaries attempting to falsify or manipulate the shared secrets is challenging. A straightforward use of Shamir's scheme does not provide guards against such activities as noted by [15]. Their scheme, while efficient, and providing the ability to detect manipulation, does not preserve the homomorphic properties necessary for general SMC. While their remediation to the scheme does not allow it to be used for SMC it is important to recognize the threat they describe which is exposed by using Shamir's scheme naively. It is possible, with non-negligible probability, for a malicious adversary or coalition of such adversaries to deceive honest parties *undetectably*. This implies that in such a situation there would exist no means by which an honest adversary could identify whether or not malicious activity has occurred, such as incorrect polynomial degree or other criteria. This is a point we will return to and emphasize in our proposed construction.

The additive secret sharing approach is a special case of threshold secret sharing in which $t = n$ such that all parties are required to participate honestly in order for the secrets to be recovered. As will be discussed later, any malicious party can trivially manipulate the shared secrets in a manner which is undetectable by honest parties in the basic scheme.

### 2.2.2 Gennaro et al

The scheme of Gennaro et al [5] is dependent on the verifiable secret sharing scheme of Pedersen et al and uses the same type of commitments. In this method, each party generates the shares of their input along with commitments to each share and a commitment to the input itself. These commitments are broadcast to all the parties, and each party is given their share of the input. The functionality to be calculated is then carried out. Being a linear scheme, additions are handled locally and the commitments are updated due to their homomorphic properties. Multiplications require considerably more communications since each party generates an intermediate sharing (along with commitments) of the product of their shares and must prove, in zero-knowledge, that they have behaved honestly. This has a high cost since not only does it involve an all-to-all exchange of shares, it also involves an all-to-all exchange of Pedersen commitments which are comparatively large since they are based on Diffie-Hellman style primitives.

This complexity has lead many researchers to explore other techniques rather than attempt to improve the efficiency of this scheme. Specifically, the complexity is $\mathcal{O}(n^2\kappa)$ for each multiplication and the constants hidden in that asymptotic analysis are quite large, again, due to the Pedersen commitments built on Diffie-Hellman style structures. Overall then the complexity is $\mathcal{O}(C_M n^2 \kappa)$.

### 2.2.3 Beerliová-Trubíniová and Hirt

Hyperinvertible matrices used in conjunction with Shamir's secret sharing scheme are used in this method to allow for the amortization of the cost of multiplications [16]. The hyper invertible matrices allow for many multiplications to be handled in parallel. Thus if there are $n$ parties, and there are $n$ multiplications which can be parallelized, the asymptotic cost of the multiplication can be decreased from $\mathcal{O}(Cn^2\kappa)$ to $\mathcal{O}(Cn\kappa)$. While this is very efficient, and a significant improvement over previous results, the maximal threshold tolerable under the scheme is $t < n/3$. We seek to maintain the linear complexity relative to the number of parties, at least for the online phase, and allow for the threshold to rise to a simple honest majority, or $t < n/2$ without the use of amortization, which may or may not always be possible.

### 2.2.4 Goyal et al

The scheme of Goyal et al [17] is very similar in many respects to that of Beerliová-Trubíniová and Hirt. The key innovation is that they avoid the dependency on the multiplicative depth of the circuit through the use of a different management of the redundancy in the scheme. In this scheme the shared secrets are distributed such that partitions of the parties can recover their partition's information in the presence of faults and, once this is achieved, the original complete matrix can be recovered. These matrices represent sets of shared secrets of degree $t$ and $n - 1$ respectively.

The lower degree shares being inconsistent with the higher degree shares is detectable up to the specified threshold. Since there are two sharings differing only in degree, the need to distribute and exchange large commitments is avoided. Malicious parties are discovered only by one party disputing with them in which case both parties are suspected of potential malicious behavior. This is due to there being no immediate means to tell which of the two parties is lying. In terms of efficiency, this is the most efficient work to date of which we are aware using Shamir's scheme as a basis. The complexity is the same asymptotic complexity as would be the case for a protocol against semi-honest adversaries up to a multiplicative constant which is ignored by the asymptotic analysis anyway. Crucially, every operation in the online phase of the scheme is linear in the number of parties, the size of the arithmetic circuit, and the size of the field being used i.e., $\mathcal{O}(Cn\kappa)$. If malicious activity is detected, a re-evaluation and dispute resolution phase is required. This expensive operation may cost as much as $\mathcal{O}(n^3\kappa)$. Also, their threshold is not up to a simple honest majority. They require $t < n/3$.

### 2.2.5 SPDZ/Overdrive and MASCOT

We combine these methods due to their very closely related nature [18, 19, 20]. Their online phases are essentially identical, and the functional requirements of the precomputation are the same. Though the means of implementing the desired precomputation functionality is different, this effects primarily the complexity of the schemes rather than the functionality itself or its security. There are also some similarities between these schemes and that of Gennaro et al. For every shared secret, additional shared information is distributed among the parties. As a basis these schemes use additive secret sharing which means that their threshold $t = n$. While this is good in terms of protecting privacy, it also means that the scheme can be brittle since any party can perform a DoS attack trivially at any time. Other honest parties have no recourse.

## 2.3 Efficient SMC through some redundancy

In each of the preceding systems a linear secret sharing scheme (either additive or Shamir) was used and additional information is generated or used along the way to provide some means of detection for malicious activity. This is achieved by a number of methods from commitments to MACs and hyperinvertible matrices. The focus of our proposition is, rather than using only Shamir or Additive, use them in conjunction to provide an efficient means to carry on general SMC as well as be able to detect malicious activity. Briefly, we wish to pair additive secret sharing with Shamir's scheme so that the best of each scheme is preserved and the overall composite scheme is strong and efficient. Most notably, we point out that protocols which allow for a simple honest majority have a higher complexity than ours, and the protocols which require an honest 2/3 majority, meaning $t < n/3$, though they have a similar complexity, require more parties to be honest. Thus we maintain the lower complexity of the protocols tolerant of $t < n/3$ malicious parties, but in our approach can tolerate $t < n/2$.

### 2.3.1 Protocol

The novelty of our approach is in using the resilience of Shamir's scheme in conjunction with the brittleness of the additive approach. Combining them makes the overall scheme resilient to data loss since any subset of honest users still forming a majority will not experience loss of the data, but malicious activity will be detectable due to the manner in which we propose their composition. The overall flow of the protocol, similar to many such compilers, is given in Algorithm 2.

### 2.3.2 Output

The key check here is that everybody effectively commits to their locally held information by altering their Shamir share according to their additive share's value. For all honest parties, this should result in a Shamir shared zero. No honest party will continue beyond this step if zero is not reconstructed. It effectively binds every party

**Algorithm 2:** SMC Compiler Protocol

**Input:** Circuit to be computed
**Output:** Result of computation

(a) Input: If gate $C_i$ is an input gate for party $P_j$, then party $P_j$ shares their input by transmitting shares of their input to every party

(b) Addition: If gate $C_i$ is an addition gate, all parties perform the local computations required using the additive homomorphic property of the shares in both schemes in parallel

(c) Multiplication: If gate $C_i$ is a multiplication gate, the multiplication protocols for both the additive scheme and Shamir scheme are invoked and executed.

(d) Output:

  i Every party computes the amount they would have to subtract from their Shamir share to subtract the value of their additive share from the Shamir shared secret

  ii Every party subtracts the appropriate value from their Shamir share and broadcasts it

  iii Every party locally reconstructs the Shamir secret which, if all parties were honest, should yield valid and consistent reconstruction of 0.

  iv Every party then broadcasts their additive shares

  v Rebuild the secret and check its correctness/consistency to their following inputs, and serves as a zero knowledge proof of knowledge for every party that they know what the correct sharings should be. If a zero is not revealed at this point, there exists two possible reasons why. One, the adversary in question is terribly inept, or the computations have become corrupted to the point that they are inconsistent. In either case, the parties should "rewind" to the last known stable state and re-evaluate.

**Altering Shamir shared secrets**

Altering additive shares is trivial since adding or subtracting from a party's share directly affects the same change in the shared secret. Altering the underlying secret by manipulating one's Shamir shares is straightforward, though not as obvious as the additive case. It is also not often addressed in the literature. We give a brief outline to the steps here. Crucially, it is the difference in these manipulative techniques which we will later use to make changes detectable. It is important to note that such manipulation is normally considered malicious activity. Here we use it in the context of our zero knowledge proof of knowledge. The value each party should subtract from their Shamir share can be calculated from their additive share as follows.

We wish to apply a desired change $\delta$ to a Shamir shared secret $s$, in our specific case this is the the additive inverse of their additive share of the secret. For a particular party $P_j$, the party has to compute the value to be added in to their share taking in to account the Lagrange interpolation coefficient which will be applied to their share by the other parties during reconstruction. Thus they need to calculate a $\delta'$ relative to their desired $\delta$ which they will add in to their share to shift the secret $s$ by the desired amount $\delta$. This is done by the following construction given as an example for three parties, in which $P_2$ is the "malicious" actor, and $\mathcal{L}_j$ denotes the Lagrange coefficient from the basis polynomial relative to the point $j$. This is a specific example for one case, but is generalizable directly to an arbitrary number of parties.

$$\begin{aligned}
\mathcal{L}_1[s]_N^{P_1} + \mathcal{L}_2([s]_N^{P_2} + \delta') + \mathcal{L}_3[s]_N^{P_3} &= s + \delta \\
\mathcal{L}_1[s]_N^{P_1} + \mathcal{L}_2[s]_N^{P_2} + \mathcal{L}_2\delta' + \mathcal{L}_3[s]_N^{P_3} &= s + \delta \\
\mathcal{L}_2\delta' &= s + \delta - \mathcal{L}_1[s]_N^{P_1} + \mathcal{L}_2[s]_N^{P_2} + \mathcal{L}_3[s]_N^{P_3} \\
\mathcal{L}_2\delta' &= \delta + \left[s - \cancel{\mathcal{L}_1[s]_N^{P_1} + \mathcal{L}_2[s]_N^{P_2} + \mathcal{L}_3[s]_N^{P_3}}\right] \\
\mathcal{L}_2\delta' &= \delta \\
\delta' &= \delta\mathcal{L}_2^{-1}
\end{aligned} \quad (2.1)$$

In general, any change to a local share $\delta'$ to achieve a desired $\delta$ to be applied to the secret $s$ for a party $P_j$ can be calculated in this way, i.e., $\delta' = \delta\mathcal{L}_j^{-1}$. Finally add $[s]_N^{P_j} + \delta'$.

**Proceeding with the output**

Once all parties have performed the required alteration to their shares and zero is reconstructed by all parties, everyone knows that everyone else knows valid Shamir and additive secrets that are consistent. What the parties must do next is broadcast the values of their additive shares. This permits two operations. One, the secret can be reconstructed by summing together all the received additive shares, and the correctness of the secret can be checked. The check takes the form of reversing the previous operation. Recall that each party holds a Shamir share of zero which was built from a Shamir share of the *same secret* by each party locally manipulating their shares. Each party is now in possession of the information used as the basis of that manipulation. Therefore that alteration can now be reversed. When the appropriate values are added to each received share from the previous step, which just amounts to performing some of the preceding calculations again and subtracting in the final step instead of adding, the result should be a set of Shamir shares with the same value as the constant term as was revealed in the summation of the additive shares, and overall degree $t$. If this is not the case, not only is there an active adversary in the group, assuming an honest majority, the identity of the adversary can be revealed. Thus we make parallel arguments to recent and very efficient protocols which transform passively secure protocols into protocols secure against active adversaries by the addition of fault detection and localization functionalities [21, 16]. Well known techniques such as the Berlekamp-Welch algorithm and others exist which are useful for identifying erroneous or potentially malicious data in this setting [22].

### 2.3.3 Proof Sketch

Consider the standard interpolation theorem which will form the basis of our argumentation:

> **Theorem 1** For any field F, for any k distinct elements $x_1, x_2, \ldots, x_k \in F$, and for any k elements $y_1, y_2, \ldots, y_k \in F$, there exists a unique polynomial $q(x) \in F[x]$ with degree less than k such that $q(x_i) = y_i \forall 1 \leq i \leq k$.

In our construction, we have multiple polynomials, of degree $t$ as well as $2t$, and often more points involved than the minimal set necessary in order to perform a correct interpolation.

In the context of our scheme, we have Shamir sharing of a secret $s$, in a degree $t$ polynomial and an additive sharing of the same secret. We construct a polynomial from the shares of the additive secret such that it has degree $2t$. This is done by each party locally multiplying their additive share by the multiplicative inverse of the Lagrange coefficient associated with their index. The result of this operation yields valid Shamir shares for a random polynomial containing the same secret as the additive shares originally did. Furthermore, it is with probability arbitrarily close to 1 that the resulting polynomial will be of degree $2t$.

**Lemma 1** A polynomial interpolated from $n = 2t+1$ random field elements is degree $2t$ with overwhelming probability based on the bitwidth of the modulus for the field $\kappa$.

This statement is true from a statistical inspection of the nature of interpolation in finite field. For a $\kappa$ bit prime modulus for the field, there are $2^{n\kappa}$ possible polynomials up to degree $n - 1 = 2t$. Being interested in polynomials of degree $2t$, we wish to find the probability that the degree of the polynomial is exactly that degree. This is possible to calculate directly from applying the same logic to the subset of cases we are interested to avoid. There are $2^{(n-1)\kappa}$ possible polynomials up to degree $n - 2$. Thus the probability of getting one of those as a result of our process is:

$$\frac{2^{(n-1)\kappa}}{2^{n\kappa}} = \frac{1}{2^{\kappa}}$$

This probability can therefore be made negligibly small in the security parameter $\kappa$, and we can expect to produce a polynomial of degree $2t$ as we desired.

We then calculate the difference between these polynomials, and reveal the result. We have defined the polynomials, we will call them $f_t(x)$ and $g_{2t}(x)$ given subscripts relative to their degree, such that they have uniform random coefficients for all terms of nontrivial degree, and the constants are equal implying $f_t(0) = s = g_{2t}(0)$. When the difference is calculated, the result is such that the constant term must be zero due to the means by which we have defined the polynomials. Other terms will be uniform random elements, except the highest degree term which is expected to be non-zero by Lagrangian interpolation, Theorem 1, and Lemma 1. Alternatively, since the degree $2t$ polynomial will have higher degree coefficients that are uniform random in $\mathbb{Z}_F^*$ and the difference being calculated is with respect to a degree $t$ polynomial, it must be the case that terms of a higher degree than $t$ remain in the polynomial represented by the shares.

If some party or parties were to attempt to manipulate their shares in a manner inconsistent with the steps we have specified, it is possible that a value other than zero will be revealed. It is possible, as we will later show, to manipulate shares in manner which will still reveal zero at this stage, but this causes other problems. In any event, at this stage, all we seek is a polynomial of degree $2t$ with zero as its constant term i.e., $f_t(0) - g_{2t}(0) = 0$. This requires participation by all the parties, and once completed will yield a unique polynomial in the field for all honest parties according to Theorem 1. This serves two purposes, it is a zero-knowledge proof of knowledge that all parties know of a set of shares between the Shamir and additive schemes that are consistent. Additionally, this effectively binds together each party's Shamir and additive shares. The binding occurs through the summation of a degree $t$ and degree $2t$ polynomial as we have said previously.

In the case that parties have behaved honestly, in revealing their additive shares, parties effectively share sufficient information for everyone to reconstruct the degree $2t$ polynomial related to their additive shares by the same means previously defined. Adding this back into the previous polynomial amounts to adding back in the additive inverse of what was previously calculated i.e., $f_t(x) + g_{2t}(x) - g_{2t}(x) = f_t(x)$.

Therefore, all that should be left is the terms of the original degree $t$ sharing, if and only if all parties have behaved honestly. Any attempt to behave maliciously during any of the steps will result in terms of a higher degree than $t$ remaining in the interpolated polynomials because there will exist no means for them to be removed due to the inconsistent/malicious activities of active adversaries. The only way to end with a degree $t$ polynomial in this setting is for the parties to behave honestly, and any malicious activity is detectable by the uniqueness property of Theorem 1.

## 2.3.4 Security

Due to the preceding, we claim that it will not be possible for a malicious adversary to hide the fact that they have attempted to manipulate the outcome of the calculation. We achieve this by linking the independent additive and Shamir shares of the secrets during the process of revealing results. This has the effect of requiring an adversary to either behave honestly, or act in a manner which will reveal their activity, so their behavior is at least detectable. This is due to the manner in which an adversary must manipulate Shamir and additive shares differently in order to alter the respective shared secrets. The difference in these manipulations causes one to allow for a reduction of polynomial degree following some calculation when no malicious activity is present, while if there has been malicious manipulation of the shares, the degree of the polynomial will not be decreased through the same computations being applied. Thus the degree reduction of the polynomial is the key flag to serve as the indicator of the presence of malicious machinations. Because we never share the Shamir shares of the result directly, this assertion will hold. We require the Shamir shares of the result to be constructed from altered degree $2t$ Shamir shares of zero and additive shares of the result. That whole process will yield consistent degree $t$ Shamir shares of the result if and only if no malicious activity has occurred, otherwise, as has been stated, the degree of the interpolating polynomial will be off. Malicious parties have an incentive to enter into this situation due to the fact that computation will not proceed as long as the initial check does not reveal a zero that all parties can agree on. As we have explained, it is also that step which enables our next step to be robust. In any situation in which malicious activity is detected, the treatment of the shares as a Reed-Solomon code [23] and the use of the Berlekamp-Welch algorithm [22] will allow the secret to still be reconstructed assuming $t < n/3$. At least, and most importantly, malicious activity is detectable and privacy is maintained, for all $t < n/2$. That will allow for the honest parties to know there are bad actors and force reevaluation for $n = 2t + 1$. We thus assert that our protocol has information theoretic security against $t < n/2$ malicious adversaries.

### 2.3.5 Complexity

Since additive and Shamir secret sharing is employed, the complexity of these methods is the sum of the individual complexity of either one operating by itself. This is true even in our setting where secrets are revealed under our protocol since this amounts to 2 secrets being reconstructed. There is some choice available here since the multiplications evaluated under either scheme can be implemented in a few different ways. For the online phase, the most efficient means of multiplying shared secrets is the method involving Beaver triples. In the context of additive secret sharing, the "conventional" approach, without relying on on offline or precomputation phase requires parties to perform somewhat more communication than Shamir, but the overall increase, even in that situation, would be only slight above double. However, given that Beaver triples are used, then the complexity can be the same as is the case under Shamir's secret sharing scheme. Thus, to transform a protocol secure against passive adversaries into a version secure against active adversaries, the complexity increase can be approximately a multiplicative constant of 2 for the online phase, for small $n$. It is important to note that our complexity analysis makes no use of insignificance assumptions, or amortization arguments, which are themselves built on other assumptions regarding the ability to parallelize operations which may not hold in all situations.

## 2.4 Example

Suppose we have the result of some computation which the parties wish to reveal. They also want to have assurance that no party has maliciously altered the compu-tation. According to our proposed methods, the following is an example of the steps to be executed. Let us assume the result of the computation is 10, shared in both Shamir and additive schemes among 5 parties in a field of characteristic 101. For the Shamir shares, the degree $t = 2$. Thus we have the following:

Table 2.1: Shamir and additive shares of 10

|  | Shamir ||||| Additive |||||
| --- | --- | --- | --- | --- | --- | --- | --- | --- | --- | --- |
| Parties | $P_1$ | $P_2$ | $P_3$ | $P_4$ | $P_5$ | $P_1$ | $P_2$ | $P_3$ | $P_4$ | $P_5$ |
| Shares of 10 | 62 | 10 | 56 | 99 | 38 | 100 | 51 | 65 | 82 | 15 |

Table 2.2: Lagrange coefficients and their modular multiplicative inverses with respect to each index

|  | 1 | 2 | 3 | 4 | 5 |
| --- | --- | --- | --- | --- | --- |
| Lagrange Coefficients in $\mathbb{Z}_{101}$ | 5 | 91 | 10 | 96 | 1 |
| Multiplicative Inverses in $\mathbb{Z}_{101}$ | 81 | 10 | 91 | 20 | 1 |

Table 2.3: Calculating and applying $\delta'$

| $P_1$ Share of 0 | $62 - 100 \cdot 81$ | $\equiv 42$ |
| --- | --- | --- |
| $P_2$ Share of 0 | $10 - 51 \cdot 10$ | $\equiv 5$ |
| $P_3$ Share of 0 | $56 - 65 \cdot 91$ | $\equiv 100$ |
| $P_4$ Share of 0 | $99 - 82 \cdot 20$ | $\equiv 75$ |
| $P_5$ Share of 0 | $38 - 15 \cdot 1$ | $\equiv 23$ |

As discussed previously in our proof sketch, the result of this process transforms the additive shares into Shamir shares. In this case, the resulting polynomial from applying the transformation (multiplying each by the multiplicative inverse of the respective Lagrange interpolation coefficient) is as follows:

$$f(x) = 74x^4 + 9x^3 + 99x^2 + 30x + 10$$

When this is subtracted share-wise from each party's locally held Shamir share of the same secret the following is the polynomial result. Note that it is degree $2t$ as expected:

$$f(x) = 27x^4 + 92x^3 + 51x^2 + 74x$$

The only reason the two highest order terms are different than they were previously is that in this polynomial they are the additive inverses in the field with respect to the coefficients of the polynomial due to the subtraction. You can easily see that $74 + 27 \equiv 0 \mod 101$ and $9 + 92 \equiv 0 \mod 101$.

### 2.4.1 Honest Conduct

When parties receive the set of shares given in Table 2.3, i.e., $\{42, 5, 100, 75, 23\}$, it is straight forward to reconstruct the secret contained by applying the Lagrange coefficients for evaluating the polynomial at 0 we've already calculated in Table 2.2 as a dot product of that vector:

$$[42\ 5\ 100\ 75\ 23][5\ 91\ 10\ 96\ 1]^T \mod 101 = 0$$

Thus proving that every party knows a Shamir share and an additive share which are consistent. It is important to note that this is a degree $2t$ polynomial, specifically, for this case, the resulting polynomial is:

$$f(x) = 27x^4 + 92x^3 + 51x^2 + 74x$$

In this case every party has behaved honestly as will be borne out by the following steps. Later will we demonstrate that this does not hold for an example in which some party has behaved maliciously, attempting to alter the result of the computation.

Once every party is in possession of these shares of 0, and has verified that the secret contained in the shares is 0, the next step is for each party to broadcast their additive shares of the result. These values are given in Table 2.1. Once all the parties have the additive shares of the result, they can perform two sets of calculations. First they can reveal the secret contained in the additive shares:

$$100 + 51 + 65 + 82 + 15 \mod 101 = 10$$

They can also now check if that result is correct. This is achieved by 'undoing' the results of the previous calculation. This means each party adds to their Shamir shares of 0 the $\delta'$ each party previously subtracted. This is possible only now since each party now knows every other party's additive share and $\delta'$ is directly derived from each party's additive share and publicly available/computable knowledge.

The result of that process, in this case, because no one has acted maliciously, is the Shamir shares of 10 that we started with in Table 2.1. The polynomial underlying these shares is

$$f(x) = 49x^2 + 3x + 10$$

Notice that by reversing the calculations performed previously, we have simultaneously reduced the degree of the polynomial back to $t$ instead of $2t$. The higher degree terms all cancel out in the process when carried out honestly. This is crucial for detecting malicious behavior as we will now address.

### 2.4.2 Malicious activity

If a party tries to influence the outcome of the computation, for example, by party 3 incrementing the result, the following is what occurs:

In order to "pass" the initial test of providing a Shamir share of 0, Party 3 has two choices, they can either manipulate both shares in the computation, or neither. Any other approach will lead to a value other than zero being revealed in this step in which case the malicious activity is immediately discovered. We will consider both cases.

**Manipulating both shares in the initial test**

Incrementing an additively shared value is trivial since it is enough to simply increment one's share. For the Shamir shares, since it is known that all the shares will be used in reconstruction, party 3 must take action according to that in manipulating their Shamir share. Thus to increment their shamir share, it is necessary to add into their locally held share the Lagrange interpolation coefficient relative to their index, in this case 10.

Thus the state of the shares is now as shown in Table 2.4. Compare with the shares given in Table 2.1 to see the effect of the described manipulation.

Table 2.4: Shamir and additive shares altered to increment the secret

|  | Shamir | | | | | Additive | | | | |
| --- | --- | --- | --- | --- | --- | --- | --- | --- | --- | --- |
| Shares of 10 | 62 | 10 | 46 | 99 | 38 | 100 | 51 | 66 | 82 | 15 |

We now continue with the same steps described previously. Everyone subtracts from their Shamir secret the $\delta'$ which will lower the shared secret by the amount of their additive share. This step is shown in Table 2.5.

Table 2.5: Calculating and applying $\delta'$

| $P_1$ Share of 0 | $62 - 100 \cdot 81$ | $\equiv 42$ |
| --- | --- | --- |
| $P_2$ Share of 0 | $10 - 51 \cdot 10$ | $\equiv 5$ |
| $P_3$ Share of 0 | $46 - 66 \cdot 91$ | $\equiv 100$ |
| $P_4$ Share of 0 | $99 - 82 \cdot 20$ | $\equiv 75$ |
| $P_5$ Share of 0 | $38 - 15 \cdot 1$ | $\equiv 23$ |

Due to this crafty manipulation, a zero is still revealed here, thus computation can proceed, but party 3 has baked in their manipulation and it will soon be revealed. Now the parties are all obliged to broadcast their additive shares. In order to reveal an incorrect answer, party 3 must broadcast 66. This results in every party calculating Shamir shares identical to those in Table 2.4. These shares also do indeed reconstruct to yield a secret value 11 which is consistent with the value revealed in summing the additive shares. There is an issue that serves as a red flag though: the polynomial is now not of the correct degree:

$$f(x) = 48x^4 + 30x^3 + 78x^2 + 97x + 11$$

The degree is $2t$ and this serves as a flag for malicious influence in accordance with the arguments we have made in Section 2.3.3. We can see that if we run the same interpolating algorithm on the points, but exclude the share of party 3, this is the result:

$$f(x) = 49x^2 + 3x + 10$$

as we would expect. Confirmation that the malicious party is party 3 since the degree of the polynomial must increase to "fit" their input because they have shifted it to influence the result.

**Manipulating neither share in the initial test**

If party 3 chooses not to manipulate both their shares in the initial test, since they have maintained their honest behavior up to this point, a 0 is revealed as illustrated in Section 2.4.1. We pick up with this example immediately following that step. The next step for the parties to execute is for each party to broadcast their additive share. Party 3 wants the result to be incremented, so must now broadcast an additive share of 66 instead of 65, even though the previous step's calculations were done with 65 as the value being used.

When the parties reveal their additive shares, 11 is reconstructed, as party 3 desires. However, during the check for malicious activity, problems arise. Recall that in this scenario, each party is in possession of a set of Shamir shares like those in Table 2.3. Now they are attempting to reverse those calculations. We show this process in

Table 2.6. We have bolded the values which have changed from when the shares of zero were calculated vs the additive share that was broadcasted by party 3.

Table 2.6: Calculating and applying $\delta'$

|  | Shares of 0 |  | Shares of the secret |  |
|---|---|---|---|---|
| $P_1$ | $62 - 100 \cdot 81$ | $\equiv 42$ | $42 + 100 \cdot 81$ | $\equiv 62$ |
| $P_2$ | $10 - 51 \cdot 10$ | $\equiv 5$ | $5 + 51 \cdot 10$ | $\equiv 10$ |
| $P_3$ | $56 - \mathbf{65} \cdot 91$ | $\equiv 100$ | $100 + \mathbf{66} \cdot 91$ | $\equiv 46$ |
| $P_4$ | $99 - 82 \cdot 20$ | $\equiv 75$ | $75 + 82 \cdot 20$ | $\equiv 99$ |
| $P_5$ | $38 - 15 \cdot 1$ | $\equiv 23$ | $23 + 15 \cdot 1$ | $\equiv 38$ |

This set of shares, i.e., $\{62, 10, 46, 99, 38\}$, which is the same set of shares produced at this step in the other method, and it yields the same polynomial on interpolating it:

$$f(x) = 48x^4 + 30x^3 + 78x^2 + 97x + 11$$

While the constant term of the polynomial agrees with the secret reconstructed from the additive shares, once again the degree of the polynomial resulting from an attempt to reverse the calculation is incorrect. This is again in accordance with the arguments we have made in Section 2.3.3. Finally, excluding the share from party 3 again allows the parties to reconstruct the correct shared result.

## 2.5 Conclusion

In this work we have presented a scheme which allows for the transformation of an SMC protocol secure against semi-honest adversaries to be transformed into a protocol secure against malicious adversaries as long as a majority of the parties remain honest. This transformation is information theoretically secure, not relying on any cryptographic intractability or hardness assumptions, and comes at a cost of only a multiplicative constant for the online phase.

Future work in this area could entail means of making the pre-computation more efficient, or using differing properties, or composing different schemes which may have further advantageous results for either security or efficiency. In any case, it is our thought that the present work is of sufficient novelty and benefit to warrant independent interest as a contribution to the field.

# Chapter 3

# Secure Multi-Party Comparison

## 3.1 Introduction

At the heart of many analysis problems, and the current progress of our research, is the simple comparison operator which has been identified, repeatedly, as a computational bottleneck for further performance improvements [24]. Considering a variety of problems, the comparison operator will be heavily used in many cases. These problems range from the traditional problem of the Millionaires associated with Yao [25], secure auctioning [26], as well as queries on a secure database [27]. Thus, finding an efficient approach solving this problem - in any given situation - will have a positive performance effect on the rest of the overall system's performance. The boost in performance from improving comparisons will naturally be proportional to the use of this operation in the overall system. This result is well known from the work of Amdahl and his eponymous Law [28].

In the last few decades, several strategies have been proposed to allow for secure computations which would mitigate or eliminate the concerns and threats to data confidentiality. They represent multiple general approaches to solving the problem with various benefits and costs. Within the realm of secret sharing, two main classes of these approaches exist. First, there are methods based on binary decomposition. Another class of protocols exploits properties of finite field arithmetic. This is done through indirect comparisons with transformed values. These intermediate results are logically combined to form the results of the desired comparison.

In general, the approaches currently known have been steadily improving. However, it seems that a plateau has been reached. These approaches represent constant rounds solutions, but the costs of their communications are still high with respect to the values on which they operate. This is at least partially due to the trade-off made in each of their work to achieve constant rounds. In order to achieve constant rounds, greater local computations and communications costs were sacrificed. This is related to the results for unbounded fan-in multiplication from the work of Bar-Ilan and Beaver [29]. Given that these approaches require this result to be able to reduce their round complexity to a constant, and by self-admission [7], it seems unlikely to drop considerably further within the currently explored veins of inquiry. It is our desire to present a different approach which avoids the use of such products as much as possible.

### 3.1.1 Our Contributions

The goal of this work is to develop highly efficient and secure multi-party protocols to implement the comparison functionality:

$$f(a,b) = \begin{cases} 1 & a \geq b \\ 0 & \text{otherwise} \end{cases}$$

An important link in the existing work is that most of the current secret sharing based protocols are symmetric; that is, all parties perform identical computations on their locally held shares of the data and communicate in the same way. For our proposed protocols in the semi-honest setting, we propose an asymmetric secure comparison, based on secret sharing, that allows for greater efficiency than previous protocols with respect to required execution rounds, local computational requirements and communications. Another optimization we have adopted in the semi-honest setting is to use shares from different domains to perform the secure computation which can minimize both local computation and communication costs. The uniqueness of our design that leads to the efficiency gain in this setting is summarized below:

- Without using Shamir's secret sharing scheme, no two parties need to execute the exact same instructions. This compatibility with asymmetry allows for the control of individual pieces of knowledge to further minimize computation and communication complexities. However, it is important to note, that this introduction of asymmetry makes the resulting protocol more difficult to prove secure in general, as well as more difficult to secure against malicious adversaries.

- Where possible, we take advantage of differing representations of values and field magnitudes for randomization and intermediate secure computations. This further reduces computation and communication costs. Most notably, by making use of the group $\mathbb{Z}_2$, we are able to compute `xor` locally, without secure multiplications.

The main limitation of our use of these optimizations is that the proposed protocols are only secure in the semi-honest model and cannot be easily modified to be secure against malicious adversaries. While maintaining the key transformation on which our protocol is based, though removing these additional efficiency boosting attributes, we also later present a version of our protocol secure against malicious adversaries assuming an honest majority, which is still more efficient than the existing protocols for practical integer applications.

### 3.1.2 Performance Overview

Our approach is very efficient with respect to other contributions in the field, as can be seen by consulting Tables 3.1 and 3.2 later in our analysis, representing online and total complexities respectively. Please note that all the complexities are with respect to security of each protocol under the semi-honest adversary model (performance under the malicious model is discussed in Section 3.4.3). Overall, our proposed protocol requires 5 rounds, and the communication costs are on the order of $7\ell \log_2 \ell + 26\ell + 11 \log_2 \ell + 32$ bits for private inputs consisting of at most $\ell$ bits such that $\ell = \log_2 N$ for some $N$ defining the group used for the underlying secret sharing scheme and sufficiently large to represent the values being compared. No secure multi-party multiplication invocations are required. Furthermore, our local computations are guaranteed to be less than those of the other protocols currently existing in the literature for the semi-honest case, and on par with others in the malicious model.

For the semi-honest case, this is due to our local operations being almost exclusively shifts, additions and multiplications, in a simple sense. We do not rely heavily on a large number of group elements being multiplicatively inverted, nor do we require solutions to a large set of systems of equations to be solved. Many of these operations are required by other works. It is important to note one difference between some of the protocols. Most of the protocols operate on previously shared values where our proposed protocol operates on locally held private values. This is a difference we address in Section 3.4.2, but does not change the complexities considerably, and does not invalidate their comparisons for the overall differences in complexities.

### 3.1.3 Organization

The rest of the paper is organized as follows: Section 3.2 provides an overview of the existing secure comparison protocols. In Section 1.2, we describe the necessary properties of the secret sharing scheme and the adversary models adopted in our protocol. Then, we present our proposed protocol in Section 3.3, describing its functionality and procedure along with analysis concerning its complexity, security, and correctness. Section 3.4 discusses variations of the proposed protocol. Specifically, we show how our protocol can be generalized to larger groups of parties than merely three, and their input values already exist in a bitwise shared format. The final version of our protocol, presented in Section 3.4.3, demonstrates via construction a means by which the key feature of our design can be made secure against malicious adversaries in the presence of an honest majority. We conclude by summarizing our contribution in Section 2.5, and indicate areas for future research directions.

## 3.2 Related Work

Yao's well known construction, garbled circuits and his associated motivator, the Millionaire's Problem, is an early approach to the possibility of a secure multi-party protocol to affect the comparison of two private values [25, 30]. There have been a number of concerns regarding the efficiency of this scheme, and there are many dramatically optimized solutions [31, 32, 33, 34]. Yet, for large private data sets, they are still not feasible for a practical situation dealing heavily with arithmetic operations. Secret sharing approaches are generally preferable in this setting as noted in [35] and there are desires for stronger security guarantees.

A number of methods also exist for the implementation of this functionality based on homomorphic encryption [36, 37, 38, 39, 40, 41]. These methods operate on a very different set of principles and security guarantees than our present work. In general, they are prone to be less efficient than those based on secret sharing schemes [42, 43], particularly with respect to the time required for large exponentiations as empirically verified in [24]. Thus, the vast majority of methods based on homomorphic encryption remain confined to the realm of paper rather than practical implementations [44]. While homomorphic encryption based secure comparison protocols are necessary in the two-party setting, in this paper, we focus on secret sharing based approaches, requiring at least three parties.

The two main approaches based on secret sharing, as introduced earlier, are methods based on changing the representation of the shared values via bit decomposition [45], and those employing a series of tests related to the properties of finite field arithmetic [46]. Later improvements and optimizations to each of these methods have been published, subsequent to the original proposition of the first of each approach. We have chosen to present and discuss the work of Damgård et al [45] due to its foundational place in the literature, clear representation of the underlying principles, and relation to our general approach. Some of the optimizations on this method were proposed in [46, 7, 47].

The latter strategy, that of exploiting properties of finite field arithmetic, seeks to affect a comparison through intermediate comparisons and some logic to bring the meaning of these intermediate comparisons together to form the desired solution. This method was introduced in the work of Nishide and Ohta [46]. Other optimizations have since come which cut back on the complexity and the number of intermediate calculations necessary based on some restrictions to the domain of values which are shared and compared [7, 48]. We do not give these results as thorough a treatment due to the drastic difference in the approach, unrelated to our methods, though their complexities are important for consideration relative to the results of others as well as comparison with the complexity of our proposed solution, all reported in Tables 3.1 and 3.2.

### Damgård et al

The first known constant rounds result in this area is due to the work of Damgård et al [45]. In this setting, secretly shared values must first be bit decomposed among the parties involved in the computation. Alternatively, the values may exist as bitwise shares initially. This means the protocol takes as input bit decomposed shares of the private values to be compared. If this procedure is necessary, though expensive, it is potentially beneficial when other bit-wise operations may be seen as advantageous. The cost incurred in this scheme for bit decomposition may be amortized somewhat across all those sub-protocols which require it. Though there is a fairly high computational complexity and communication cost, this important result demonstrates constant rounds secure comparison is indeed possible and well within feasibility. The bulk of the complexity is contained in the PRE∨ or prefix `or` operation. It is also important to note that, given improvements in some of the primitives on which this is based, the complexity of this approach can be made much more efficient. Specifically, making use of the prefix operations as described in [46] has a considerable positive effect on the complexity.

**Algorithm 3:** BIT-LT from Damgård et al [45]

**Input:** bitwise shares of integers $a, b$ denoted $[a]_B, [b]_B$ with individual bits denoted $[a_0]_p \ldots [a_{\ell-1}]_p$ where $\ell = \lceil \log_2 p \rceil$

**Output:** shares of the single bit result $[c]_p$

1 **begin**
2     **for** $i = 0, \ldots, \ell - 1$ **do**
3        $[e_i]_p \leftarrow \text{XOR}([a_i]_p, [b_i]_p)$
4     **end**
5     $([f_{\ell-1}]_p, \ldots, [f_0]_p) = \text{PRE}_\vee([e_{\ell-1}]_p, \ldots, [e_0]_p)$
6     $[g_{\ell-1}]_p = [f_{\ell-1}]_p$
7     **for** $i = 0, \ldots, \ell - 2$ **do**
8        $[g_i]_p \leftarrow [f_i]_p - [f_{i+1}]_p$
9     **end**
10    **for** $i = 0, \ldots, \ell - 1$ **do**
11       $[h_i]_p \leftarrow \text{MULT}([g_i]_p, [b_i]_p)$
12    **end**
13    $[h]_p \leftarrow \sum_{i=0}^{\ell-1} [h_i]_p$
14    Output $[h]_p$
15 **end**

## 3.3 The Proposed Protocols

In this paper, we propose multi-party secure comparison protocols for two settings, the semi-honest and malicious models. In the semi-honest model, unlike the existing secret sharing based secure comparison protocols, the inputs and intermediate computations are not symmetric among the three parties. Expansions to larger numbers of parties are fairly trivial, and can be done based directly on the three-party examples given in the following section. For malicious security, a number of our optimizations which are possible in the semi-honest case are no longer possible, and many alterations are necessary. However, we preserve the core transformation of our protocol and demonstrate its efficiency. Each protocol considers a different scenario.

- The first protocol Section 3.3.1 allows for a group of three parties to securely compare two private inputs $a$ and $b$ held by two of the three parties. The purpose of the third party is to facilitate secret sharing based secure computations. The protocol returns shares of 1 if $a \geq b$.

- The second protocol, presented in Section 3.4.1, presents an option concerning the protocol complexity, with a lower number of rounds and still returns shares of 1 if $a \geq b$.

- For both of the preceding cases, we present an extension of our protocols in Section 3.4.2, and it assumes that the three parties do not know the values they compare, and that they are already shared in a bit decomposed format. This is similar to [45], listed in Tables 3.1 and 3.2.

- Finally, we present a version of our approach which is secure against malicious adversaries in Section 3.4.3.

Each proposed protocol targets a unique class of applications. However, they together cover most scenarios related to secure comparison based applications. The design of our protocol is based on the following key claim:

**Claim 1 (Comparison Reduction)** Given two non-negative integers $a$ and $b$, let $\alpha = 2a + 1$, $\beta = 2b$ and $i$ denote the most significant bit location where the $i^{th}$ bit of $\alpha$ is not equal to the $i^{th}$ bit of $\beta$, i.e., $\alpha_i \neq \beta_i$. Suppose $\ell$ is the bit length of the domain of $\alpha$ and $\beta$. Define an $l+1$-bit binary vector $h$ by setting $h_i = 1$ and $h_j = 0$ for $0 \leq j \leq \ell - 1 \wedge j \neq i$. Additionally, we define:

- $s_\alpha = \sum_{j=0}^{l-1} \alpha_j$

- $s'_\alpha = \sum_{j=0}^{l-1} [(h_j - \alpha_j) \mod 2]$

- $f = (s_\alpha - s'_\alpha + 1)/2$

Then we have $f = 1 \implies a \geq b$ and $f = 0 \implies a < b$.

The above claim can be verified based on the property of $h$. Again let $i$ be the most significant bit location where $\alpha_i \neq \beta_i$. According to the definition of $h$ and the modulo 2 operation, we have:

- $(h_i - \alpha_i) \mod 2 = 1 - \alpha_i$

- $(h_j - \alpha_j) \mod 2 = \alpha_j$, for $j \neq i$

As a consequence, we can conclude:

- $s_\alpha - s'_\alpha = 1$, when $\alpha_i = 1$ (implying $a \geq b$)

- $s_\alpha - s'_\alpha = -1$, when $\alpha_i = 0$ (implying $a < b$)

The rest of the claim follows directly from the above analysis. The proposed secure comparison protocol adopts the logic or function reduction/transformation presented in the claim and implements the process securely. However, it is still challenging to ensure both security and efficiency in designing the protocol. This claim is the core aspect of our comparison.

### 3.3.1 Key Steps and Correctness Analysis for Protocol 1

The main steps of our first proposed protocol are presented in Algorithm 4. Without loss of generality, $P_1$ has a private input $a$ and $P_2$ has a private input $b$. At the end, the protocol returns the secret shares of the comparison result to only $P_1$ and $P_2$. The comparison result 1 indicates $a \geq b$, and 0 otherwise. At the beginning of the algorithm, an indexing variable $i$ is defined for arrays of shared values. Unless otherwise noted, $i$ is used to span the entire array. There are two instances in Step 5 when this is not the case. Overall, a few key design principles are important to recall:

1. `xor` can be done locally in $\mathbb{Z}_2$, without secure multiplication.

2. We employ a summation of bits to avoid other costly multiplications and required communication.

3. Mapping between groups in which the shares are constructed allows us to further reduce communication costs.

Aside from groups $\mathbb{Z}_N$ and $\mathbb{Z}_2$, we also make use of a group $\mathbb{Z}_{N_2}$. We define $N_2$ to be a prime such that $\lceil \log_2 \ell \rceil + 1 < \log_2 N_2 < \lceil \log_2 \ell \rceil + 2$. This is due to the use of the values which will be shared in this group. By inspection in the protocols to follow, specifically Step 5, the maximum possible value which may be shared in this group is $2(\ell + 1) - 1$ related to $\gamma_0'$. Thus, the modulus should define a group of sufficient order to represent these values, and this gives us the bounds we have defined for its modulus. It is known that a prime must exist in this range from the proof of Bertrand's Postulate which states that, for all $n > 3$ there exists at least one prime $p$ such that $n < p < 2n$ [49, 50, 51].

The following enumeration gives a high level intuitive view of the operation of our proposed protocol.

- *Input transformation* – Steps 1-2 either double or double and increment the input values to assure that there is at least one location with a differences

- *Compute bitwise differences* – Steps 3-4 focus on producing the bitwise xor of the inputs and handling their mapping between sharing domains.

- *Locate the most significant difference* – Steps 5-6 identify the location of the most significant difference in the bitwise `xor`.

- *Deriving $s_a'$:* Steps 7-9 use the constructed vector of bits $h$ with the private input $a$ to construct the vector of bits $h'$ which will be different from $a$ in exactly one bit index, in accordance with Claim 1. Then the protocol sums the bits of $h'$ calculates the difference between this result and the count of set bits in $a$ calculated earlier and maps this to the desired comparison result, again as justified in Claim 1.

- *Share transformation* – Steps 10-11 in which the final goal is to map the result of the transformed functionality back into the group in use for the secret sharing scheme in general.

The protocol unfolds as follows along with correctness justifications. In the following discussions, note that we refer to values without the level of specificity that is given in the algorithm. For example, shares of $a$ belonging to $P_1$ and $P_2$ in $\mathbb{Z}_2$ may simply be referred to as $[a]$. Full details are given below and a concrete numerical example is given in the appendix.

- *Step 1*: In the first step, $P_1$ generates the necessary values for later permutation and randomization. The private input of $P_1$ is also doubled and incremented before shares are generated (i.e., $a = 2a + 1$). The purpose for this alteration of the private input is to make sure there exists one bit position in which the parties have a difference. Otherwise in the case of equality the protocol would fail and information would be leaked. With this alteration, in the case of equality, the least significant bit which is now different, will allow the protocol to correctly return 1, indicating that $a \geq b$. The value $s_a$ is merely a count of the number of set bits in $a$. This will be used later to determine the result of the comparison in accordance with Claim 1. Finally, the shares for $P_2$ along with other randomly generated values for later use are transmitted to $P_2$.

- *Step 2*: This step consists of $P_2$ generating shares of $2b$ and sending one share to $P_1$. Once again, the two private values are modified to avoid what would have otherwise been a risk for protocol failure leading to an information leak in the case of equality.

**Algorithm 4:** $\text{SC}(\langle P_1, a \rangle, \langle P_2, b \rangle, \langle P_3, \bot \rangle) \to (\langle P_1, [f]_N^{P_1} \rangle, \langle P_2, [f]_N^{P_2} \rangle)$

**Input:** Public info: $N$ and $N_2$, $N > N_2$, $N$ is an integer and the modulus of the secret sharing scheme, $\ell$ is the required bitwidth for the domain of $a$ and $b$, $N > 2^\ell$. $N_2$ is a prime such that $\lceil \log_2 \ell \rceil + 1 < \log_2 N_2 < \lceil \log_2 \ell \rceil + 2$, $0 \le i \le \ell$, and $j \in \{1, 2\}$

**Output:** $f$ is secretly shared between $P_1$ and $P_2$. $f = 1$ if $a \ge b$; otherwise, $f = 0$

1. $P_1$
   - (a) $a = 2a + 1$
   - (b) $s_a \leftarrow \sum_i a_i$
   - (c) Generate $r_i, r_i' \in_R \mathbb{Z}_2$, $r'' \in_R \mathbb{Z}_2$, $\tau_i \in_R \mathbb{Z}_{N_2}^*$, and random shift $\pi$
   - (d) Generate $[a_i]_2^{P_j}$, and $[s_a]_{N_2}^{P_j}$
   - (e) Send $[a_i]_2^{P_2}$, $[s_a]_{N_2}^{P_2}$, $r_i$, $r_i'$, $r''$, $\tau_i$ and $\pi$ to $P_2$

2. $P_2$
   - (a) $b = 2b$
   - (b) Generate $[b_i]_2^{P_j}$
   - (c) Send $[b_i]_2^{P_1}$ to $P_1$

3. $P_j$
   - (a) $[e_i]_2^{P_j} \leftarrow [a_i]_2^{P_j} + [b_i]_2^{P_j}$
   - (b) $[e_i]_2^{P_1} \leftarrow r_i - [e_i]_2^{P_1}$, if $r_i = 1$
   - (c) Send $[e_i]_2^{P_j}$ to $P_3$

4. $P_3$
   - (a) $e_i \leftarrow [e_i]_2^{P_1} + [e_i]_2^{P_2}$
   - (b) Generate $[e_i]_{N_2}^{P_j}$
   - (c) Send $[e_i]_{N_2}^{P_j}$ to $P_j$

5. $P_j$
   - (a) $[e_i]_{N_2}^{P_1} \leftarrow r_i - [e_i]_{N_2}^{P_1}$, if $r_i = 1$
   - (b) $[e_i]_{N_2}^{P_2} \leftarrow -[e_i]_{N_2}^{P_2}$, if $r_i = 1$
   - (c) $[\gamma_\ell']_{N_2}^{P_j} \leftarrow [e_\ell]_{N_2}^{P_j}$
   - (d) $[\gamma_i']_{N_2}^{P_j} \leftarrow [\gamma_{i+1}']_{N_2}^{P_j} + [e_i]_{N_2}^{P_j}$, for $i = \ell - 1, \ldots, 0$
   - (e) $[\gamma_\ell]_{N_2}^{P_j} \leftarrow [\gamma_\ell']_{N_2}^{P_j}$
   - (f) $[\gamma_i]_{N_2}^{P_j} \leftarrow [\gamma_{i+1}]_{N_2}^{P_j} + [\gamma_i']_{N_2}^{P_j}$, for $i = \ell - 1, \ldots, 0$
   - (g) $[\gamma_i]_{N_2}^{P_1} \leftarrow [\gamma_i]_{N_2}^{P_1} - 1$
   - (h) $[u_i]_{N_2}^{P_j} \leftarrow \tau_i [\gamma_i]_{N_2}^{P_j}$
   - (i) $[v_i]_{N_2}^{P_j} \leftarrow \text{Shift}_\pi \left( [u_i]_{N_2}^{P_j} \right)$
   - (j) Send $[v_i]_{N_2}^{P_j}$ to $P_3$

6. $P_3$

    (a) $v_i = [v_i]_{N_2}^{P_1} + [v_i]_{N_2}^{P_2} \mod N_2$

    (b) Find the unique index $k$, where $v_k = 0$

    (c) Generate $[h_i]_2^{P_j}$, where $h_k = 1$ and $h_i = 0$ for $i \neq k$

    (d) Send $[h_i]_2^{P_j}$ to $P_j$

7. $P_j$

    (a) $[h_i]_2^{P_j} \leftarrow \text{shift}_\pi^{-1}\left([h_i]_2^{P_j}\right)$

    (b) $[h'_i]_2^{P_j} \leftarrow [h_i]_2^{P_j} - [a_i]_2^{P_j}$

    (c) $[h'_i]_2^{P_1} \leftarrow r'_i - [h'_i]_2^{P_1}$, if $r'_i = 1$

    (d) Send $[h'_i]_2^{P_j}$ to $P_3$

8. $P_3$

    (a) $h'_i \leftarrow [h'_i]_2^{P_1} + [h'_i]_2^{P_2}$

    (b) Generate $[h'_i]_{N_2}^{P_j}$

    (c) Send $[h'_i]_{N_2}^{P_j}$ to $P_j$

9. $P_j$

    (a) $[h'_i]_{N_2}^{P_1} \leftarrow r'_i - [h'_i]_{N_2}^{P_1}$, if $r'_i = 1$

    (b) $[h'_i]_{N_2}^{P_2} \leftarrow -[h'_i]_{N_2}^{P_2}$, if $r'_i = 1$

    (c) $[s'_a]_{N_2}^{P_j} \leftarrow \sum_i [h'_i]_{N_2}^{P_j}$

    (d) $[f]_{N_2}^{P_j} \leftarrow [s_a]_{N_2}^{P_j} - [s'_a]_{N_2}^{P_j}$

    (e) $[f]_{N_2}^{P_1} \leftarrow [f]_{N_2}^{P_1} + 1$

    (f) $[f]_{N_2}^{P_j} \leftarrow 2^{-1}[f]_{N_2}^{P_j}$

    (g) $[f]_{N_2}^{P_1} \leftarrow r'' - [f]_{N_2}^{P_1}$, if $r'' = 1$

    (h) $[f]_{N_2}^{P_2} \leftarrow -[f]_{N_2}^{P_2}$, if $r'' = 1$

    (i) Send $[f]_{N_2}^{P_j}$ to $P_3$

10. $P_3$

    (a) $f \leftarrow [f]_{N_2}^{P_1} + [f]_{N_2}^{P_2}$

    (b) Generate $[f]_N^{P_j}$

    (c) Send $[f]_N^{P_j}$ to $P_j$

11. $P_j$

    (a) $[f]_N^{P_1} \leftarrow r'' - [f]_N^{P_1}$, if $r'' = 1$

    (b) $[f]_N^{P_2} \leftarrow -[f]_N^{P_2}$, if $r'' = 1$

- *Step 3*: $P_1$ and $P_2$ compute a bitwise `xor`, and $P_1$ additionally randomizes its shares by the uniform random bits $r_i$. Both $P_1$ and $P_2$ send the resulting shares to $P_3$.

- *Step 4*: $P_3$ uses the two shares received to rebuild the re-randomized secret and generate new shares, now in a group defined by $N_2$. This amounts to a mapping of shares between groups $\mathbb{Z}_2 \to \mathbb{Z}_{N_2}$. Note that, due to the randomization from Step 3, no information is leaked to $P_3$ in this process.

- *Step 5*: Recall that the shares from $P_1$ were randomized before sending to $P_3$. This randomization must be reversed so that the original values, now secretly shared in a new group, can be reclaimed. This is achieved through sharing the randomization vector $r$ between $P_1$ and $P_2$.

  - When $r_i = 1$, $P_1$ computes $r_i - [e_i]$ and $P_2$ simply computes the additive inverse of its share. This is accomplished in Steps 5(a) and 5(b) of Algorithm 4. These two parties now hold bitwise shares of the `xor` of their private inputs, each as elements in $\mathbb{Z}_{N_2}$.
  - A repeated prefix summation over the vector (5(c-f)) is computed to ensure that there is only one indexed location which is equal to 1.
  - In 5(g) $P_1$ subtracts one from every share of this vector, thereby ensuring that every value in the vector is non-zero with one exception, the location of the most significant bit difference.
  - The values contained in the vector are multiplied by the vector $\tau$ in step 5(h) consisting of random non-zero values which are known to $P_1$ and $P_2$ to hide the intermediate calculated values from $P_3$. Since we require $N_2$ to be prime, these values in $\tau_i$ will not yield a product equal to 0 under modulus $N_2$ when multiplied with any other group element, except 0, which is the flag of the location we wish to uniquely preserve.

- Now, every array value related to every bit position is a non-zero and uniformly random value with one exception, the position related to the most significant bit difference. Finally, to hide this information from $P_3$, the vector of values $[u_i]$ is shifted according to the random value $\pi$ agreed upon between $P_1$ and $P_2$ in step 5(i).

- *Step 6*: $P_3$ combines the shares received from $P_1$ and $P_2$ to reconstruct the vector $v$ and finds the unique index $k$ containing the value zero. $P_3$ then constructs a bitwise series of shares in $\mathbb{Z}_2$ such that all shares are shares of zero with the exception of the shares indexed by $k$. These shares are sent to the corresponding parties.

- *Step 7*: Upon receipt of these shares, the parties perform an inverse shift on the received vector to place the sole location containing a share of one back into the position related to the most significant bit difference. They then compute, for each location, the difference between the newly received vector of shares and the shares of $P_1$'s original input. This is directly related to our justification for the comparison reduction presented in Claim 1. $P_1$ additionally randomizes this result with the vector $r'$, similar to what was done in Step 3(b).

- *Step 8*: Similar again to what has preceded, $P_3$ reconstructs and performs another oblivious mapping in the generation of new shares between groups $\mathbb{Z}_2 \to \mathbb{Z}_{N_2}$.

- *Step 9*: The additional randomization is removed by $P_1$ and $P_2$ in steps 9(a) and 9(b). In step 9(c), all values in the vector $h'$ are summed by $P_1$ and $P_2$, and assigned to shares of $[s'_a]$. Similar to the calculation of $s_a$, this action merely counts the number of set bits in $h'$.

    - The difference $[s_a] - [s'_a]$ is calculated in step 9(d). Because there is exactly

one bit difference between $a$ and $h'$ in their binary representations, the sums of the set bits within these two values will have a difference of exactly magnitude 1. Again, in accordance with our logic presented in Claim 1, calculating the difference $[s_a] - [s'_a]$ results in either one or its additive inverse in $\mathbb{Z}_{N_2}$ being assigned to shares of $f$.

- Finally, in step 9(e-i), since the result of the computation so far is $f \in \{1, -1\}$ shared in $\mathbb{Z}_{N_2}$, it is necessary to map this set to $f \in \{0, 1\}$, and bring the shares into the field for the overall secret sharing scheme. The first part of these last steps is achieved through $P_1$ adding the value one to its share, thereby incrementing the set of possible values to get shares of either zero or two, and then both parties multiply their shares by the multiplicative inverse of two. In the case that $f$ is zero, it will remain unchanged, and in the case that $f$ is two, it will be reduced to one. Thus, the correct functionality has been achieved and $f = 1$ iff $a \geq b$. All that remains is to map the shares into the appropriate group. As we have done previously, this shared value $f \in \mathbb{Z}_{N_2}$ is re-randomized with another bit $r''$. This value is sent to $P_3$.

- *Step 10:* For the last time, $P_3$ reveals the randomized $f$ and builds shares of that revealed quantity in $\mathbb{Z}_N$ and returns these to the other parties.

- *Step 11:* As the last step of our proposed protocol, $P_1$ and $P_2$ remove the temporary additional randomization from the shared result $[f]_N$.

### 3.3.2 Security Analysis under the Semi-honest Model

For our security formulations and proofs to follow, we consider adversaries in the semi-honest or honest but curious model. We consider each player's role independently, due to the asymmetry used in our protocol. For each case, we will analyze the protocol focusing on communicated information and opportunities for leaks. Though it is lengthy, we discuss every step and every bit of communication in detail due to the asymmetry and unconventional nature of our protocol to demonstrate the equivalence between the idealized setting with a trusted third party available and the simulation of this behavior via our protocol in a real setting in terms of information theoretic security and lack of additional undesirable information disclosure. We follow the proof framework in [12].

**Semi-honest $P_1$**

The view of $P_1$ during the execution of the proposed protocol includes public domain information for inputs and outputs, and the value $a$ which is the private input for $P_1$. Additionally, $P_1$ generates random bit vectors $r$ and $r'$, a random bit $r''$, the vector of random values in $\mathbb{Z}_{N_2}^*$ $\tau$, and the random shift $\pi$. In the course of execution, more values are exchanged; these consist of shares of $b$, $e$, $h$, and $h'$.

- Step 1(c) and 1(d) are those which represent the generation of all the values $P_1$ will share with $P_2$. Specifically, these are: $r_i, r_i', r'', \tau_i, \pi, [a_i]_2^{P_2}$, and $[s_a]_{N_2}^{P_2}$.

- In Step 2(c), $[b_i]_2^{P_1}$, the $P_1$'s share of $P_2$'s bitwise shared input, is received.

- Step 4(c) consists of the receipt of the vector $[e_i]_{N_2}^{P_1}$.

- The vector $[h_i]_2^{P_1}$ is received in Step 6(d).

- In Step 8(c), the vector $[h_i']_{N_2}^{P_1}$ is received.

- Step 10(c) is the final communication in the scheme and consists of the receipt of $[f]_N^{P_1}$.

Since for each of the values listed above in the Steps 2-10, $P_1$ only receives one share, it is impossible for $P_1$ to rebuild the actual values. This is immediate from the security guarantee of the underlying secret sharing scheme.

The values generated in Step 1(c) and 1(d) are all uniformly random, and each consists of use of the randomness source provided in the real implementation of the protocol, which is permissible under the scheme used in the proof system. Since these consist of uniform random values or the result of calculations which are also uniform random, they do not pose any threat to the privacy of the protocol through their use in the manner we have described in the context of the protocol. The final value retained as the result of the protocol's completion is uniformly random and represents a share of the desired outcome from the evaluation of the functionality we implement. $f$ is an element in $\mathbb{Z}_2$, and its shares are in a larger group. Again, it is impossible to rebuild the underlying secret, or learn anything at all about it, from the possession of a single share. This is once more immediate from the underlying secret sharing scheme. Therefore, $P_1$ learns nothing beyond what is known at the beginning of the execution of the protocol, namely, the value of the private input already held $a$, the public domain information, and anything which may be deduced from these data.

**Semi-honest $P_2$**

The roles of $P_1$ and $P_2$ are fairly similar, though not identical. $P_2$'s view of the protocol differs from $P_1$ in only two ways, receipt from $P_1$ of some of the random values to be used to hide information from $P_3$, and the manner in which some of these values are used. The total execution view consists of the public domain information for inputs and outputs, and the value $b$ which is the private input for $P_2$. $P_2$ receives random bit vectors $r$ and $r'$, a random bit $r''$, the vector of random values $\tau_i$ in $\mathbb{Z}_{N_2}^*$, and the random shift amount $\pi$. In the course of execution, more values are exchanged. These consist of shares of $a$, $s_a$, $e$, $h$, and $h'$.

- Step 1(e), all values $P_1$ shared with $P_2$ are received. Specifically, these are: $r_i, r'_i, r'', \tau_i, \pi, [a_i]_2^{P_2}$, and $[s_a]_{N_2}^{P_2}$.

- Step 4(c) consists of the receipt of the vector $[e_i]_{N_2}^{P_2}$.

- The vector $[h_i]_2^{P_2}$ is received in Step 6(d).

- In Step 8(c), the vector $[h'_i]_{N_2}^{P_2}$ is received.

- Step 10(c) is the final communication in the protocol and consists of the receipt of $[f]_N^{P_2}$.

The understanding of the security of the protocol, with respect to an honest-but-curious $P_2$, is essentially the same as the case for $P_1$. Since $P_2$ is never in possession of both shares of any of the secret shared values, the values represented by the shares are impossible to reconstruct due to the properties of the secret sharing scheme. The possession of the true values of, for example, $r_i$ in no way compromises any security principles since the value is used only to hide information from $P_3$ and allows for the removal of this additional randomization after the shares have been mapped into their new domain. Specifically, where $P_1$ calculates, in Step 5(a), $[e_i]_{N_2}^{P_1} \leftarrow r_i - [e_i]_{N_2}^{P_1}$, if $r_i = 1$, $P_2$ calculates $[e_i]_{N_2}^{P_2} \leftarrow -[e_i]_{N_2}^{P_2}$, if $r_i = 1$.

Since $P_2$, parallel to many of the ideas presented with respect to $P_1$, cannot rebuild any of the secret shared values, learns nothing of additional information which should not be gained from the randomization values and the underlying values or result beyond what can be surmised from its privately held input and public information. Thus, the protocol is secure against a semi-honest $P_2$.

**Semi-honest $P_3$**

The role of $P_3$ in the overall protocol is very different from the other two. The view of $P_3$ during the execution of the proposed comparison protocol includes the public domain information for inputs and outputs, receipt of the shares of $v$, and bitwise shared values $e$, and $h$.

- In Step 3(c), the shares $[e_i]_2^{P_1}$ and $[e_i]_2^{P_2}$ are sent to $P_3$. In Step 7(d), the shares $[h'_i]_2^{P_1}$ and $[h'_i]_2^{P_2}$ are received. These are both bitwise shares of values related to $a$ and $b$, which are recombined to reconstruct the underlying values.

- Shares $[v_i]_{N_2}^{P_1}$ and $[v_i]_{N_2}^{P_2}$ are received in Step 5(j) to reconstruct the randomized values in $\mathbb{Z}_{N_2}$.

- Shares $[f]_{N_2}^{P_1}$ and $[f]_{N_2}^{P_2}$ received in Step 9(i) are two random integers in $\mathbb{Z}_{N_2}$.

$P_3$ receives no shares of the output of the protocol, and holds no shares of the inputs, so the above mentioned messages comprise the total view of the execution of the protocol from the perspective of $P_3$. The shares of these values have been built from the original values based on the algorithm standard for the secret sharing scheme in use. Since $P_3$ receives all necessary information to rebuild values $e$ and $h'$, and in fact does rebuild them, there would be a significant leaking of information for these cases if they were not additionally re-randomized by uniform values in $\mathbb{Z}_2$ before the shares are received by $P_3$. This additional layer of randomization allows for $P_3$ to rebuild the shares and generate new shares in a different group, namely $\mathbb{Z}_{N_2}$, without gaining any knowledge about the true values themselves. In the case of Step 5, and the reception of the shares of $v$, the only information of significance is the index of the bit position related to the most significant difference between the inputs. This position is flagged by being the only zero element in the array. This position is hidden by the permutation $\pi$, so even though $P_3$ once again rebuilds and maps the values to a different group, no meaningful information can be gleaned.

$P_3$ has no input and only processes results from others' shared inputs. Since the results have been uniformly re-randomized, they retain no information of the actual result beyond what can be ascertained in an ideal case. The best this party can do, independent on the availability of arbitrary computational power, is to randomly guess in the domain what the correct values could be. Thus, the protocol, with respect to a semi-honest $P_3$, is information theoretically secure.

For our protocol the final area of concern for the adversary is the ability of the adversary to control the routing of messages. As can be seen from our protocol's structure, it is roughly self-synchronizing since no honest party will proceed to calculations for which the necessary messages have not been received. This property allows us to conclude that any halting of message exchange in the intermediate calculations both yields no information to the adversary which has corrupted any party, and is equivalent to a corrupted party refusing participation in the protocol in the ideal model, which is acceptable.

For each of the preceding three cases, analyzing the algorithms for each of the parties and the messages passed through the adversary controlled router in the environment these were seen to be admissible in the sense that they neither yield nor leak information beyond that which is desired. Additionally, they constitute a correct implementation of the desired functionality as discussed in Section 3.3.1. Therefore the probabilistic difference in protocol result and the result of the parties interacting with an idealized functionality executed in an idealized setting is negligibly small. This immediately leads to their universally composable security in an information theoretic sense, the desired property we wished to demonstrate.

### 3.3.3 Complexity Analysis

The complexity of our protocol is analyzed from two perspectives: that of round complexity, and communications. We define a computational *round* as consisting of any arbitrary amount of local computation accompanied by at most one send and receive cycle, as established and explained in [52]. The round complexity of our protocol is simply a constant since the number of cooperative steps executed by the overall protocol is fixed and independent of the size of the input values. Specifically, the number of required rounds is 5. All the following step numbers refer to the algorithm of our protocol given in Algorithm 4.

The first round consists of Steps 1-3 since this it is in these steps that the parties are generating shares of values to be used, exchanging them, performing some computation and transmitting the response to $P_3$. This is analogous to one send and receive cycle with an arbitrary amount of local computation accompanying. During this phase, the most complex values to be communicated are generated and sent. Specifically, in Step 1(e), generating and sending $[a_i]_2^{P_2}$, $[s_a]_{N_2}^{P_2}$, $r_i$, $r_i'$, $r''$, $\tau_i$ and $\pi$ requires the transmission of $\ell \log_2 \ell + 5\ell + 3\log_2 \ell + 9$ bits. Step 2(c) requires only $\ell + 1$ bits in communication, and concludes round 1 since the next step is dependent on the values transmitted in this step. Now we get into the more strict send and receive relationship for the rest of the required rounds. Steps 3 and 4 together form a round since $P_1$ and $P_2$ calculate and send values to $P_3$, and receive back from $P_3$ a response based on those values. The same is true for pairs of Steps 5-6, 7-8, and 9-10.

With respect to Step 5, specifically beginning with Step 5(d), the protocol proceeds by computing the prefix sum of the bits from most to least significant shares of bits. This results in a series of shares such that the most significant bits are shares of zero until the first bit difference is encountered, which is one. Following this position, all bits are non-zero and non-decreasing in value with each successive bit. In Step 5(f), to ensure that exactly one value in the array is equal to one, we iterate through the bit positions again performing a prefix sum of the values from most to least significant bit positions. This is the justification for the bounds which we have given for $N_2$. If $a$ and $b$ have no bits in common, the vector $[e_i]$ will consist of shares of all ones. Thus, the prefix sum of the shares of these values will yield, at most, $\ell + 1$. Repeating this operation will yield a value of at most $2\ell + 1$; as a result, this serves as the defining factor for our bounds for $N_2$, thus having a direct effect on our complexity since this is the source of our logarithmic term in $\ell$.

Step 3(c) requires $2(\ell+1)$ bits in transmission, and 4(c) requires $2(\ell+1)(2+\log_2 \ell)$

bits. The next round requires $2(\ell+1)(2+\log_2 \ell)$ bits for Step 5(i) and $2(\ell+1)$ bits for Step 6(d). The penultimate round requires $2(\ell+1)$ bits to be sent in Step 7(d), and in step 8(c) $2(\ell+1)(2+\log_2 \ell)$ bits are communicated. The final round consists of Steps 9 and 10 in which $2(2+\log_2 \ell)$ and $2\ell$ bits are communicated respectively. The last step follows immediately from the values received in Step 10, and requires no further communication. Therefore, Step 11 is considered as a final part of round 5. For the total communication complexity, we consider the number of bits which need to be transmitted, according to the steps in which they occur:

- Step 1: $3(\ell+1)$ bits for $r_i, r'_i,$ and $[a_i]_2^{P_2}$, 1 for $r''$, $(\ell+1)(2+log_2\ell)$ for $\tau_i$, $1+\log_2 \ell$ for $\pi$, and $2+\log_2 \ell$ for $[s_a]_{N_2}^{P_2}$. The total: $\ell \log_2 \ell + 5\ell + 3\log_2 \ell + 9$

- Step 2: $\ell+1$ for $[b_i]_2^{P_1}$

- Step 3: $2(\ell+1)$ for $[e_i]_2^{P_1}$ and $[e_i]_2^{P_2}$

- Step 4: $2(\ell+1)(2+\log_2 \ell)$ for $[e_i]_{N_2}^{P_1}$ and $[e_i]_{N_2}^{P_2}$

- Step 5: $2(\ell+1)(2+\log_2 \ell)$ for $[v_i]_{N_2}^{P_1}$ and $[v_i]_{N_2}^{P_2}$

- Step 6: $2(\ell+1)$ for $[h_i]_2^{P_1}$ and $[h_i]_2^{P_2}$

- Step 7: $2(\ell+1)$ for $[h'_i]_2^{P_1}$ and $[h'_i]_2^{P_2}$

- Step 8: $2(\ell+1)(2+\log_2 \ell)$ for $[h'_i]_{N_2}^{P_1}$ and $[h'_i]_{N_2}^{P_2}$

- Step 9: $2(2+\log_2 \ell)$ for $[f]_{N_2}^{P_1}$ and $[f]_{N_2}^{P_2}$

- Step 10: $2\ell$ for $[f]_N^{P_1}$ and $[f]_N^{P_2}$

Summing these leads directly to the aforementioned figure for the overall communication requirement of our protocol: $7\ell \log_2 \ell + 26\ell + 11 \log_2 \ell + 32$ bits.

The communication complexities are described and compared in Tables 3.1 and 3.2 in terms of the required number of bits to be transmitted. This is a sharp contrast to the common convention of the currently optimized works in the area requiring a communication complexity represented by a count of secure multiplication invocations. The comparisons that follow are all calculated under the assumption that all the schemes are employing the secret sharing scheme chosen by their designers. In the semi-honest setting, a single secure multiplication under the Shamir secret sharing scheme requires, for $m$ parties, assuming appropriate choices for the degree and threshold, $m(m-1)\ell$ bits to be communicated between the parties. For 3 parties and a protocol requiring $24\ell + 26\log_2 \ell + 4$ multiplication invocations, this results in more than $144\ell^2 + 156\ell\log_2 \ell + 24\ell$ bits to be transmitted, a distinctly larger quantity than our proposed protocol. This is calculated directly from analyzing the number of bits required for a secure multiplication according to the method common for Shamir's secret sharing scheme[2]. The communication requirements and the number of required rounds would both grow even more, if these protocols were implemented under additive secret sharing because secure multiplication under additive secret sharing is more expensive (in rounds and bits transmitted) than that under Shamir's secret sharing. For all the following analysis, we assume 3 parties, the same as we propose for our own protocol. We then simply multiply the number of bits required for a single Shamir secure multiplication by the number of expected multiplication invocations to yield the figures we present. For our protocol consisting of an arbitrary number of parties, the complexity is dependent on both $\ell$ and $m$. We also generalize the analysis of the existing protocols when we discuss this extension in Section 3.4.2.

Table 3.1: Online round and communication complexity for secure comparison protocols in the semi-honest adversary model

| Presented in | Type | Online Rounds | Bits transmitted online |
|---|---|---|---|
| [45] | A | 37 | $126\ell^2 \log_2 \ell + 336\ell^2$ |
| [46] | A | 8 | $90\ell^2 + 30\ell$ |
| [7] | A | 4 | $108\ell^2 + 48\ell$ |
| [7] | R | 2 | $24\ell^2 + 6\ell$ |
| [47] | R | 2 | $12\ell^2$ |
| Sharemind[53] | R | $8 + \log_2 \ell$ | $2(\ell-1)^2 \log_2 \ell + 63\ell^2 + 6\ell$ |
| Sharemind[6] | R | $3 + \log_2 \ell$ | $5\ell^2 + 12\ell(\log_2(\ell) + 1)$ |
| Section 3.3.1 | A | 5 | $6\ell \log_2 \ell + 22\ell + 9 \log_2 \ell + 28$ |
| Section 3.4.1 | A | 4 | $2\ell^2 + 4\ell \log_2 \ell + 19\ell + 4 \log_2 \ell + 16$ |

Table 3.2: Total round and communication complexity for secure comparison protocols in the semi-honest adversary model

| Presented in | Type | Total Rounds | Total Bits transmitted |
|---|---|---|---|
| [45] | A | 44 | $1104\ell^2 \log_2 \ell + 1254\ell^2$ |
| [46] | A | 15 | $1674\ell^2 + 30\ell$ |
| [7] | A | 10 | $918\ell^2 + 2592\ell \log_2 \ell + 144\ell$ |
| [7] | R | 8 | $162\ell^2 + 216\ell \log_2 \ell + 30\ell$ |
| [47] | R | 9 | $45\ell^2 + 66\ell$ |
| Sharemind[53] | R | $8 + \log_2 \ell$ | $2(\ell-1)^2 \log_2 \ell + 63\ell^2 + 6\ell$ |
| Sharemind[6] | R | $3 + \log_2 \ell$ | $5\ell^2 + 12\ell(\log_2(\ell) + 1)$ |
| Section 3.3.1 | A | 5 | $7\ell \log_2 \ell + 26\ell + 11 \log_2 \ell + 32$ |
| Section 3.4.1 | A | 4 | $2\ell^2 + 5\ell \log_2 \ell + 23\ell + 6 \log_2 \ell + 22$ |

Additionally, we differ with respect to the round complexity analysis given by [47], in that we adopt a more strict definition of a computational round, with that adopted in Sharemind as discussed in Section 3.3.3. This stricter definition leads to a somewhat higher count of rounds than that given in the analysis published by the authors. To summarize the existing work in the field with respect to round and communication complexities, we offer Tables 3.1 and 3.2 based on the summary present in [7]. The following complexities for our table have been obtained directly from the table in the referenced work by the method mentioned previously. Similar to the work of Reistad and Toft [7], the "A" denotes arbitrary input in the group used for the secret sharing scheme, while "R" denotes a domain of input which is restricted to a proper subset of the group. Here, it is very clear that the communication costs of our protocol are significantly less than any other of which we are aware, by a considerable factor on the dominating term and a reduction of order overall.

Most of the recent research work has pursued constant rather than non-constant round secret sharing based secure comparison. Though there is a recent paper focusing on linear round secure bit decomposition [54], it is not directly related to our work. As the domain of the values being compared increases, communication cost generally dominates local computation cost. Thus, recent research has been primarily focused on reducing the communication complexity. The Sharemind system [53, 55] breaks with the norm in this respect since its round complexity is non-constant, and it has a communication complexity lower than the communication complexity of the existing constant rounds protocols, as well as non-constant round protocols.

In our performance analysis, the complexities of other protocols were directly taken from their papers under the semi-honest model, which is consistent with the adversary model we adopted in our main protocol. Also, we actually benefit the other protocols as much as possible by accepting the communication complexities given by their respective papers which, in some cases, explicitly neglects communications dealing with some operations, e.g., revealing shared values. Additionally, we benefit the other protocols by allowing them to remain in their chosen Shamir scheme. If we analyzed their protocols in the additive setting (which would make the comparison with our method more direct), both the communication complexity and the round complexity would increase considerably. The complexity for our protocol is precise, but it still represents a more than incremental increase in efficiency. The magnitude of the complexity as well as the order have both decreased with respect to other protocols with the same security guarantee in the semi-honest model.

## 3.4 Protocol Variations

In this section, we address a few variations on our approach which may be made to our protocol to alter its complexity or expand its generality. The first such proposed alteration allows for a trade-off in complexity which could be desirable in some settings, trading the elimination of a computational round for a slightly increased communication complexity. The other suggests a manner in which it is possible to generalize our three-party protocol to any number of parties greater than three with only slight increase in communication costs. Finally, we describe a version of our protocol which, though heavily altered, still implements the key transformation we identified in Claim 1 which is secure in the malicious model.

### 3.4.1 Variation for Alternative Complexity

If round complexity is of greater concern than overall communication costs, we offer a trade-off which can be implemented with respect to our general protocol. There is a one round complexity reduction which is bought at the expense of an increase in the communication complexity. In order to affect this change, the protocol as given previously stands for Steps 1-7. Beginning with Step 8, alterations are necessary. We present the Algorithm 5 which consists of the necessary steps to replace in Algorithm 4. The only other alterations occur in Step 1(d)-(e) where $s_a$ is shared in $\mathbb{Z}_N$ rather than $\mathbb{Z}_{N_2}$. In addition, $N$ is now required to be odd. Finally, Steps 10 and 11 are no longer necessary, and the protocol ceases with completion after the finish of Step 9. The communications required in this version of the protocol are as follows:

**Algorithm 5:** Replacement steps for alternative complexity for $\text{SC}(\langle P_1, a\rangle, \langle P_2, b\rangle, \langle P_3, \perp\rangle) \to (\langle P_1, [f]_N^{P_1}\rangle, \langle P_2, [f]_N^{P_2}\rangle)$

**8** $P_3$

    (a) $h'_i \leftarrow [h'_i]_2^{P_1} + [h'_i]_2^{P_2}$

    (b) Generate $[h'_i]_N^{P_j}$

    (c) Send $[h'_i]_N^{P_j}$ to $P_j$

**9** $P_j$

    (a) $[h'_i]_N^{P_1} \leftarrow r'_i - [h'_i]_N^{P_1}$, if $r'_i = 1$

    (b) $[h'_i]_N^{P_2} \leftarrow -[h'_i]_N^{P_2}$, if $r'_i = 1$

    (c) $[s'_a]_N^{P_j} \leftarrow \sum_i [h'_i]_N^{P_j}$

    (d) $[f]_N^{P_j} \leftarrow [s_a]_N^{P_j} - [s'_a]_N^{P_j}$

    (e) $[f]_N^{P_1} \leftarrow [f]_N^{P_1} + 1$

    (f) $[f]_N^{P_j} \leftarrow 2^{-1}[f]_N^{P_j}$

- Step 1: $3(\ell + 1)$ bits for $r_i, r'_i$, and $[a_i]_2^{P_2}$, 1 for $r''$, $(\ell + 1)(2 + \log_2 \ell)$ for $\tau_i$, $1 + \log_2 \ell$ for $\pi$, and $\ell$ for $[s_a]_N^{P_2}$ the total is $\ell \log_2 \ell + 6\ell + 2\log_2 \ell + 7$ bits

- Step 2: $\ell + 1$ for $[b_i]_2^{P_1}$

- Step 3: $2(\ell + 1)$ for $[e_i]_2^{P_1}$ and $[e_i]_2^{P_2}$

- Step 4: $2(\ell + 1)(2 + \log_2 \ell)$ for $[e_i]_{N_2}^{P_1}$ and $[e_i]_{N_2}^{P_2}$

- Step 5: $2(\ell + 1)(2 + \log_2 \ell)$ for $[v_i]_{N_2}^{P_1}$ and $[v_i]_{N_2}^{P_2}$

- Step 6: $2(\ell + 1)$ for $[h_i]_2^{P_1}$ and $[h_i]_2^{P_2}$

- Step 7: $2(\ell + 1)$ for $[h'_i]_2^{P_1}$ and $[h'_i]_2^{P_2}$

- Step 8: $2\ell(\ell + 1)$ for $[h'_i]_N^{P_1}$ and $[h'_i]_N^{P_2}$

The result of this alteration results in a complexity of $2\ell^2 + 5\ell \log_2 \ell + 23\ell + 6\log_2 \ell + 22$. There is some change in the groups used for shared values, but the functionality and overall procedure is changed in no substantive manner for the considerations of security and correctness. Therefore, we do not introduce detailed complexity and security discussion as we did for the primary version of our protocol. All the arguments are essentially the same.

## 3.4.2 Variation for Shared and Bit-decomposed Inputs among arbitrarily large numbers of parties

Table 3.3: Overall round and communication complexity for secure comparison protocols

| Presented in | Type | Total Rounds | Total Bits transmitted |
|---|---|---|---|
| [45] | A | 44 | $m(m-1)(184\ell^2 \log_2 \ell + 209\ell^2)$ |
| [46] | A | 15 | $m(m-1)(279\ell^2 + 5\ell)$ |
| [7] | A | 10 | $m(m-1)(153\ell^2 + 432\ell \log_2 \ell + 24\ell)$ |
| [7] | R | 8 | $m(m-1)(27\ell^2 + 36\ell \log_2 \ell + 5\ell)$ |
| [47] | R | 9 | $m(m-1)(7.5\ell^2 + 11\ell)$ |
| Section 3.4.2 | A | 5 | $9\ell \log_2 \ell + 4m\ell + 25\ell + 12\log_2 \ell + 35$ |

Similar to the previously referenced related protocol from [45], we can, with slight alteration, accept values which already exist in a shared and bitwise decomposed state as inputs to our protocol rather than only privately held values from two parties. This can allow arbitrarily many parties to be shareholders and still perform the comparison with minimal additional communication cost. Additive secret sharing is required due to its ability to recombine incomplete sets of shares into a share which is correct, and uniformly random, for a set of fewer parties. Recall that under additive secret sharing, shares of a value $x$ are constructed, for $m$ parties and modulus $N$, according to the following equation:

$$[x]_N^{P_m} = x - [x]_N^{P_1} - \cdots - [x]_N^{P_{m-1}}$$

where all $[x]_N^{P_1}$ through $[x]_N^{P_{m-1}}$ are randomly selected elements of $\mathbb{Z}_N$ and $[x]_N^{P_m}$ satisfies the above equation. Once these shares are distributed, all $m$ parties hold an equal share of the secret, and all $m$ are required to reconstruct the secret. If, however, $P_3$ through $P_m$ were to send their shares to $P_2$, then $P_2$ could sum them with its own share and hold a new share which would allow for $P_2$ and $P_1$ to compute functions without interaction from the rest of the parties. At the end of the desired functionality, shares can be redistributed to the remaining parties in a manner similar, though inverse, to the process by which they were combined. The share held by $P_2$ has gained no information since it is still uniformly random, and both security and correctness have been maintained. We exploit this property, which additive secret sharing possesses and Shamir's scheme does not, to maintain efficiency of our protocol. The steps 1-5 and 11, in need of alterations, are given in Algorithm 6. The messages exchanged in this version of the protocol are as follows:

- Step 1: $4(\ell + 1) + 1 + (\ell + 1)(2 + \log_2 \ell) + 1 + \log_2 \ell + 2\ell(m - 2)$ for $q_i$, $r_i$, $r'_i$, $r''$, $\tau_i$, $\pi$, and $[a'_i]_2^{P_1}$, as well as $[a_i]_2^{P_k}$ and $[b_i]_2^{P_k}$, where $k = 3 \ldots m$

- Step 2: $\ell + 1$ for $[a_i]_2^{P_2}$

- Step 3: $2(\ell + 1)$ for $[e_i]_2^{P_1}$ and $[e_i]_2^{P_2}$

- Step 4: $4(\ell + 1)(2 + \log_2 \ell)$ for $[e_i]_{N_2}^{P_1}$, $[e_i]_{N_2}^{P_2}$, $[a'_i]_{N_2}^{P_1}$, and $[a'_i]_{N_2}^{P_2}$

- Step 5: $2(\ell + 1)(2 + \log_2 \ell)$ for $[v_i]_{N_2}^{P_1}$ and $[v_i]_{N_2}^{P_2}$

- Step 6: $2(\ell + 1)$ for $[h_i]_2^{P_1}$ and $[h_i]_2^{P_2}$

- Step 7: $2(\ell + 1)$ for $[h'_i]_2^{P_1}$ and $[h'_i]_2^{P_2}$

- Step 8: $2(\ell + 1)(2 + \log_2 \ell)$ for $[h'_i]_{N_2}^{P_1}$ and $[h'_i]_{N_2}^{P_2}$

- Step 9: $2(2 + \log_2 \ell)$ for $[f]_{N_2}^{P_1}$ and $[f]_{N_2}^{P_2}$

- Step 10: $2\ell$ for $[f]_N^{P_1}$ and $[f]_N^{P_2}$

- Step 11: $2\ell(m - 2)$ for $[f]_N^{P_{j,k}}$, where $k = 3 \ldots m$ and $j = 1, 2$

This leads directly to a total communication cost of $9\ell \log_2 \ell + 4m\ell + 25\ell + 12\log_2 \ell + 35$ bits. Even though we are operating on already bit decomposed and shared values, our protocol is still very efficient in comparison to others due to the fact that we share our bit decomposed values in $\mathbb{Z}_2$ as opposed to $\mathbb{Z}_N$. This boosts efficiency through immediate savings in communication, but it is also the means by which we can compute `xor` locally without requiring secure multi-party multiplications. We present Table 3.3, which clearly demonstrates that not only is our proposed protocol of lower asymptotic complexity in both $m$ and $\ell$, and lower complexity in terms of required overall rounds, but it is also lower in the coefficients of the dominating terms, thus presenting a large reduction in required communication costs.

**Algorithm 6:** Replacement steps for shared and bit-decomposed inputs for $\text{SC}(\langle P_1, [a_i]_2^{P_1}, [b_i]_2^{P_1}\rangle \ldots \langle P_m, [a_i]_2^{P_m}, [b_i]_2^{P_m}\rangle) \rightarrow (\langle P_1, [f]_N^{P_1}\rangle \ldots \langle P_m, [f]_N^{P_m}\rangle)$

1. $P_1$
   (a) $[a_{i+1}]_2^{P_1} = [a_i]_2^{P_1}$ for $i = \ell - 1, \ldots, 0$
   (b) $[b_{i+1}]_2^{P_1} = [b_i]_2^{P_1}$ for $i = \ell - 1, \ldots, 0$
   (c) $[a_0]_2^{P_1} \leftarrow 0$
   (d) $[b_0]_2^{P_1} \leftarrow 0$
   (e) Generate $q_i, r_i, r'_i \in_R \mathbb{Z}_2$, $r'' \in_R \mathbb{Z}_2$, $\tau_i \in_R \mathbb{Z}_{N_2}^*$, & random shift $\pi$
   (f) $[a'_i]_2^{P_1} \leftarrow q_i - [a_i]_2^{P_1}$, if $q_i = 1$
   (g) Send $q_i$, $r_i$, $r'_i$, $r''$, $\tau_i$ and $\pi$ to $P_2$, and send $[a'_i]_2^{P_1}$ to $P_3$

2. $P_k$ ($k = 3 \ldots m$)
   (a) Send $[a_i]_2^{P_k}$ and $[b_i]_2^{P_k}$ to $P_2$

3. $P_2$
   (a) $[a_{i+1}]_2^{P_2} = \sum_{k=2}^m [a_i]_2^{P_k}$, for $i = \ell - 1, \ldots, 0$
   (b) $[b_{i+1}]_2^{P_2} = \sum_{k=2}^m [b_i]_2^{P_k}$, for $i = \ell - 1, \ldots, 0$
   (c) $[a_0]_2^{P_2} \leftarrow 1$
   (d) $[b_0]_2^{P_2} \leftarrow 0$
   (e) Send $[a_i]_2^{P_2}$ to $P_3$

4. $P_j$
   (a) $[e_i]_2^{P_j} \leftarrow [a_i]_2^{P_j} + [b_i]_2^{P_j}$
   (b) $[e_i]_2^{P_1} \leftarrow r_i - [e_i]_2^{P_1}$, if $r_i = 1$
   (c) Send $[e_i]_2^{P_j}$ to $P_3$

5. $P_3$
   (a) $e_i \leftarrow [e_i]_2^{P_1} + [e_i]_2^{P_2}$
   (b) Generate $[e_i]_{N_2}^{P_j}$
   (c) $a'_i \leftarrow [a'_i]_2^{P_1} + [a_i]_2^{P_2}$
   (d) Generate $[a'_i]_{N_2}^{P_j}$
   (e) Send $[e_i]_{N_2}^{P_j}$ and $[a'_i]_{N_2}^{P_j}$ to $P_j$

5. $P_j$
    (a) $[e_i]_{N_2}^{P_1} \leftarrow r_i - [e_i]_{N_2}^{P_1}$, if $r_i = 1$
    (b) $[e_i]_{N_2}^{P_2} \leftarrow -[e_i]_{N_2}^{P_2}$, if $r_i = 1$
    (c) $[a_i]_{N_2}^{P_1} \leftarrow q_i - [a_i]_{N_2}^{P_1}$, if $q_i = 1$
    (d) $[a_i]_{N_2}^{P_2} \leftarrow -[a_i]_{N_2}^{P_2}$, if $q_i = 1$
    (e) $[s_a]_{N_2}^{P_j} \leftarrow \sum_i [a_i]_{N_2}^{P_j}$
    (f) $[\gamma'_\ell]_{N_2}^{P_j} \leftarrow [e_\ell]_{N_2}^{P_j}$
    (g) $[\gamma'_i]_{N_2}^{P_j} \leftarrow [\gamma'_{i+1}]_{N_2}^{P_j} + [e_i]_{N_2}^{P_j}$, for $i = \ell - 1, \ldots, 0$
    (h) $[\gamma_\ell]_{N_2}^{P_j} \leftarrow [\gamma'_\ell]_{N_2}^{P_j}$
    (i) $[\gamma_i]_{N_2}^{P_j} \leftarrow [\gamma_{i+1}']_{N_2}^{P_j} + [\gamma'_i]_{N_2}^{P_j}$, for $i = \ell - 1, \ldots, 0$
    (j) $[\gamma_i]_{N_2}^{P_1} \leftarrow [\gamma_i]_{N_2}^{P_1} - 1$
    (k) $[u_i]_{N_2}^{P_j} \leftarrow \tau_i [\gamma_i]_{N_2}^{P_j}$
    (l) $[v_i]_{N_2}^{P_j} \leftarrow \text{Shift}_\pi \left( [u_i]_{N_2}^{P_j} \right)$
    (m) Send $[v_i]_{N_2}^{P_j}$ to $P_3$

11. $P_j$
    (a) $[f]_N^{P_1} \leftarrow r'' - [f]_N^{P_1}$, if $r'' = 1$
    (b) $[f]_N^{P_2} \leftarrow -[f]_N^{P_2}$, if $r'' = 1$
    (c) Generate $[f]_N^{P_{j,k}} \in_R \mathbb{Z}_N$, for $k = 3, \ldots, m$
    (d) $[f]_N^{P_j} \leftarrow [f]_N^{P_j} - \sum_{k=3}^m [f]_N^{P_{j,k}}$
    (e) Send $[f]_N^{P_{j,k}}$ to $P_k$, for $k = 3, \ldots, m$

    $P_k$ $(k = 3, \ldots, m)$
    (a) $[f]_N^{P_k} \leftarrow [f]_N^{P_{1,k}} + [f]_N^{P_{2,k}}$

### 3.4.3 Variation under the Malicious Model

Not all of the optimizations we have proposed thus far can be made secure in the malicious model in a straightforward manner, but important aspects of our protocol can be preserved and be made secure against malicious adversaries as this section will demonstrate. The key feature addressed in Claim 1, transforming the comparison into finding the difference in the number of set bits in an original argument and one which has been masked, is preserved. In this context, we will require commitments, secure multi-party multiplications, as well as a secure `xor` functionality constructed from secure multiplication in the usual way. Additionally, we only use symmetric operations in the context of Shamir's secret sharing scheme rather than the additive scheme we used in the semi-honest setting.

We present another comparison protocol under the semi-honest model. This version, presented in Algorithm 7, is securable in the malicious model according to the following, well known, techniques. The verifiable secret sharing we make use of is that presented by [5], including the ability to have constant rounds unbounded fan-in multiplication which we will denote with $mult^*$ [29]. Other frameworks for securing protocols into the malicious adversary setting exist, such as [13], but whatever framework would be used, they would shift all the protocols together and not change their ordering.

In the course of the protocol, we construct a vector which consists of the bitwise differences in the inputs. Based on this, we then construct matrix of values which is a triangular matrix. The values in the matrix are calculated by rows. The element at the index of the row is used to `xor` with every element of lower significance. For all leading 0's, the bits of lower significance will remain unchanged through the `xor`, when the first set bit is encountered, all the bits of lower significance will be inverted for that row. This ensures that, in the triangle of the matrix, every column consists of either all 0's, or a mix of 1's and 0's with a sole exception, the column associated with the most significant set bit. That column in the triangle will be the only one to consist of all 1's. Thus taking the product along the columns of the triangle's elements in the matrix will yield a vector of values which is all 0's except for the index associated with the most significant set bit. Thus our mask is constructed. All that remains is to apply the mask, sum across bits, calculate the difference with the sum of set bits in one of the original inputs, and finish the transformation of this value into the result of the comparison, as we did in the semi-honest approach.

Table 3.4: Total round and communication complexity for secure comparison protocols in the malicious adversary model

| Presented in | Type | Total Rounds | Total Bits transmitted |
|---|---|---|---|
| [45] | A | 132 | $13248\ell^2 \log_2 \ell + 15048\ell^2$ |
| [46] | A | 45 | $20088\ell^2 + 360\ell$ |
| [7] | A | 30 | $11016\ell^2 + 31104\ell \log_2 \ell + 1728\ell$ |
| Section 3.4.3 | A | 24 | $216\ell^3 - 36\ell^2 + 72\ell$ |

**Complexity Analysis**

Complexity analysis is given in Table 3.4. The protocols included are all those which operate on arbitrary input, not restricted to some proper subset of the group, and all of which are operating with three parties in the malicious adversary model. All are secure against arbitrary malicious behaviors assuming an honest majority is maintained, and based on the secure composable primitives from the verifiable secret sharing proposed in [5]. We have presented the rounds in Algorithm 7 as they would be in the semi-honest case. However, transforming each round in the semi-honest setting to the malicious setting requires 3 rounds.

As can be seen by inspection our complexity, while of a higher order in the malicious setting, due to the significantly smaller coefficients, it is still lower than the others in important cases in both rounds and overall required number of bits for communication. This is with respect to modulus of interest such as 32 and 64 bits used

**Algorithm 7:** $SC(\langle P_1, [a_i]_2^{P_1}, [b_i]_2^{P_1}\rangle \ldots \langle P_m, [a_i]_2^{P_m}, [b_i]_2^{P_m}\rangle) \rightarrow$
$(\langle P_1, [f]_N^{P_1}\rangle \ldots \langle P_m, [f]_N^{P_m}\rangle)$

**Input:** Public info: $N$ is an integer and the modulus of the secret sharing scheme, $\ell$ is the required bitwidth for the domain of $a$ and $b$ which are bitwise shared among the $m$ parties, $N > 2^\ell$. Additionally we use $0 \leq i \leq \ell$, and $j \in \{1 \ldots m\}$ for $m$ parties

**Output:** $f$ is secretly shared between all $P_1 \ldots P_m$. $f = 1$ if $a \geq b$; otherwise, $f = 0$

1. $P_j$ round 1
   (a) $[a_{i+1}]_N^{P_j} = [a_i]_N^{P_j}$ for $i = \ell - 1, \ldots, 0$
   (b) $[b_{i+1}]_N^{P_j} = [b_i]_N^{P_j}$ for $i = \ell - 1, \ldots, 0$
   (c) $[a_0]_N^{P_j} \leftarrow 1$
   (d) $[b_0]_N^{P_j} \leftarrow 0$
   (e) $[s_a]_N^{P_j} \leftarrow \sum_i [a_i]_N^{P_j}$
   (f) $[e_i]_N^{P_j} \leftarrow \text{xor}([a_i]_N^{P_j}, [b_i]_N^{P_j})$

2. $P_j$ round 2
   (a) For $i \in 0 \ldots \ell$ and $k \in 0 \ldots i-1$ in parallel
   $[E_{ik}]_N^{P_j} = \text{xor}([e_i]_N^{P_j}, [e_k]_N^{P_j})$

3. $P_j$ round $3 \ldots 7$
   (a) For $i \in 0 \ldots \ell - 1$ in parallel
   $[v_i]_N^{P_j} \leftarrow mult^*([E_{0i}]_N^{P_j}, \ldots [E_{ii}]_N^{P_j})$
   (b) $[v_\ell]_N^{P_j} \leftarrow [e_\ell]_N^{P_j}$

4. $P_j$ round 8
   (a) $[h_i]_N^{P_j} \leftarrow \text{xor}([a_i]_N^{P_j}, [v_i]_N^{P_j})$
   (b) $[s'_a]_N^{P_j} \leftarrow \sum_i [h_i]_N^{P_j}$
   (c) $[f]_N^{P_j} \leftarrow [s_a]_N^{P_j} - [s'_a]_N^{P_j}$
   (d) $[f]_N^{P_j} \leftarrow [f]_N^{P_j} + 1$
   (e) $[f]_N^{P_j} \leftarrow 2^{-1}[f]_N^{P_j}$

in many integer based applications.

The main point of complexity of our protocol is caused by large products involved in condensing the columns of the triangular matrix. This is due to the size of the products in terms of numbers of elements involved in the products and the necessity for the result which enables unbounded fan-in multiplications to be evaluated securely in constant rounds [29]. While we avoided this operation in the semi-honest case, it was not possible for us to avoid at present for the malicious setting. Each product operating on a list of $\ell$ shared values, for example, will require $5\ell$ multiplication invocations. In the triangular section of the matrix we will require such products for every column greater than length 3 up to $\ell - 1$ i.e., $\sum_{i=3}^{\ell-1} 5i \approx \frac{5\ell(\ell-1)}{2}$. We do save some complexity in this step by limiting our operations to the triangular subset of the matrix, and keeping the matrix as small as possible in dimensionality, but it is still costly. These efforts however are the source of the negative coefficient in the quadratic term of our complexity. Aside from this we have $\frac{\ell(\ell-1)}{2} + 2\ell + 1$ multiplications for the other steps. To get down to the level of bits in the malicious setting we have analyzed the methods proposed in [5], which require each round of multiplications in the semi-honest model to take 3 in the malicious case, and $12\ell m(m-1)$ bits in communication for $\ell$ serving as the bitwidth of the shares, and $m$ as the number of parties. In the analysis, we have fixed all $m$ at 3. To compare the analysis for the more general case of $m$ parties, one need only divide our analysis by 6 for each term and multiply each term by $m(m-1)$.

**Security Sketch**

As showed in Algorithm 7, the protocol only invokes previously proven unconditionally secure composable primitives, specifically multiplication with commitment as presented and proven secure against malicious adversaries [5]. Consequently, according to the universal composability theorem [10], our protocol is unconditionally secure under the malicious model assuming an honest majority.

# Chapter 4

# Distributing and Obfuscating Firewalls via Oblivious Bloom Filter Evaluation

## 4.1 Introduction

Firewalls are a very common security tool used to protect local networks from threats present in the larger Internet. This is achieved by various strategies, the most basic of which is a packet filter [56]. This type of firewall checks each incoming packet against some set of rules often represented by a "blacklist" or "whitelist". With the increasing sophistication of attacks on firewalls from both internal and external threat, there has been an interest to distribute the functionality and management of the firewall [57, 58, 59, 60]. While this does allow a large number of machines in the local network to potentially maintain protection without dependence on the functionality of a central server, it also presents the opportunity for individual users inside the firewall to disable or circumvent the rules of their now localized firewall. Furthermore, if the rules of these firewalls exist in plaintext form, as it is often the case in practice, the rules themselves may be considered sensitive information due to a desire to hide the identities and locations of whitelisted external users, or the identities and locations of those blacklisted sources the administration consider as threats.

If we consider a brief vignette, we may encounter Carl, a corporate employee, who has received some incentive to act in the professional interests of others besides exclusively his employer. Thus, depending on his actions whether they may be passive or active attacks, he can readily and more easily gain access to the corporate system infrastructure.

- In the passive case, he may copy the data contained in the firewall rules. If the firewall rules represent a blacklist, he may learn and share identity or location information of entities considered as threats to the corporation. Alternatively, if the firewall represents a whitelist, he may learn and leak information regarding allies or affiliates' identities or locations outside the corporate intranet.

- If instead his attacks are more active, he may be able to insert or delete entries from either blacklists or whitelists either of which could prove disastrous for the network, its assets, and desired functionalities.

To prevent these attacks, we may argue that firewall rules can be encrypted and access control policies can be used to restrict entities to access firewall configuration files. However, these techniques are not sufficient for the following reasons:

- *Encrypted firewall policies and rules*: When the firewall policies and rules are encrypted with a symmetric-key encryption scheme such as AES [61], only the person who has the secret key can decrypt these files to access the actual policies and rules. However, during firewall evaluation process for a new network package, the rules have to be decrypted first. This opens the door for attackers to access these policies and rules without knowing the secret key. In addition, when a security breach occurs, the secret key may get exposed, and consequently, so do the encrypted firewall files.

- *More restrictive access control policies*: Suppose that mandatory access controls restrict the access to firewall files to the root or a small group of privileged users. Nevertheless, security breaches, exploring unknown vulnerabilities of the network, can bypass access control mechanisms. In addition, human errors and mis-configurations of firewall policies can inadvertently expose sensitive firewall files to other internal users and outside attackers.

Insider threats to enterprise network security has been a topic of greatly increasing interest among researchers as they seek to address issues of identifying malicious insiders, understanding the limits of the threats they may pose, and the extent of damages which are possible from various thresholds of actions they may take [62, 63]. These researches have progressed into attempts to construct predictive models for identifying malicious inside entities and their attacks [64, 65]. The practical utility of such inquiries has been made clear due to the dramatic recent increases of specifically these types of malicious insiders exploiting the access and network privilege with which they have been entrusted.

Network security news and information security sites are rife with articles documenting cases of internal breaches and concern over future insider attacks of a similar nature [66, 67, 68, 69, 70]. For a concrete and specific example, some analysts have indicated that the recent Capital One incident was caused, at least in part, by issues pertaining to the security of firewalls and their configuration [71, 72]. Recent research has indicated that nearly 25% of insider threats are active and malicious insiders, and one of the chief threats that they present is the theft or exfiltration of sensitive data. Average annualized cost per threat of active malicious insider is $2.9M so there should be strong motivation to develop mitigation for sensitive data, such as firewall information if that is critical for a particular setting [73].

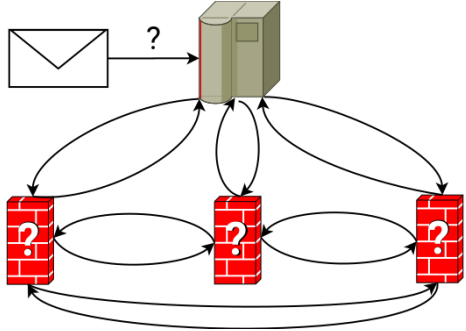

Figure 4.1: Oblivious firewalls

Based on the aforementioned observations, to maximize firewall protection from both internal and external attacks, we need a novel way to manage firewalls. Ideally, firewall rules should be encrypted and remain encrypted during firewall evaluation process. In addition, we should not rely on one server to manage the firewalls because when a security breach occurs, even encrypted data can get disclosed and firewalls services can be disabled. Therefore, to hide the firewall rules and policies and achieve certain degree of fault-tolerance, we need to utilize a distributed architecture and innovative ways to "encrypt" or "hide" the firewall rules and policies. Simultaneously, the architecture and firewall hiding method should allow secure and oblivious firewall policy evaluations without decrypting the hidden firewall information.

### 4.1.1 Our Contributions

We propose a means by which with relatively low additional overhead, a system is established to allow for the rules and functionality of the firewall system to be distributed in such a way that they are information theoretically secure against local inspection, and resistant to tampering due to the techniques we employ. Figure 4.1 shows a high level overview of the proposed distributed architecture which requires at least three servers. The firewall is secretly distributed among the servers with either additive or Shamir secret sharing schemes [14]; thus, each server alone will not be able to discover any information regarding the firewall. Unlike the previous distributed approaches, our proposed architecture and secure evaluation methods to distributing a firewall do not allow any one of the individual servers to circumvent the firewall.

Additionally, it could be made resilient to failures as cited as a motivating factor for many other distributed approaches. To achieve this we may allow the servers to continue functioning so long as, out of $m$ total servers, at least some threshold $t$ remain operational. This is possible when implemented, as we suggest, under the Shamir secret sharing scheme. This property of resilience in the face of component failure is a key feature of this scheme. We make use of some previous research, such as this scheme's resilience properties, as well as some novel strategies to achieve these results as will be discussed in Section 4.3.

Furthermore, we employ secure multi-party computation (SMC) techniques to securely and obliviously evaluate the firewall criteria checking function efficiently through the use of a distributed and secret shared Bloom filters. Following the sharing of the Bloom filter constructed from the information of a given desired firewall functionality, our proposed methods will proceed as follows:

1. During a brief and trusted initialization phase, the filter to represent the firewall is constructed and shared.

2. Following this initialization, any gateway receiving a packet invokes a set membership query with respect to the related information of interest in the filter.

3. The distributed firewall servers, acting as shareholders in the secret sharing scheme, cooperate to calculate shares of the result which is sent to the gateway.

4. The gateway reconstructs the result of the distributed computation and either rejects or forwards the packet.

The above steps can easily be adopted by many systems with little software modification in their existing routing and firewall protocols. There will be a non-negligible cost in latency for the machines within such a firewall, but increased security always comes with a cost. In our implementation, we empirically what that cost is for our experimental setup. In summary, the desirable properties of our system allows for a local network of arbitrary topology with respect to connection to the outside Internet, to securely, and obliviously, evaluate firewall rules in a distributed manner while maintaining the ability to tolerate some margin of component failure or malicious behavior. We provide analysis concerning the resilience of our scheme against the previously mentioned attacks to demonstrate its merit from the perspective of security, and we analyze the complexity of our approaches to demonstrate its efficiency. Finally, we present some options for various modifications depending on desires for security and the effects these alterations have on efficiency as an interesting trade-off. Next we formally define the threat models and adversary behaviors that can be prevented by using the proposed techniques.

### 4.1.2 Threat Model and Assumptions

Here we classify the participating entities into insiders and administrators and specify their allowable behaviors. In addition, we define our assumption regarding the initial state of the system and the trust model.

**Classification of Entities**

The following notations will be used throughout the rest of the paper:

- $P_1, \ldots, P_m$: a set of $m$ servers that obliviously store and manage our distributed firewalls where $m \geq 3$.

- $U_{P_i}$: a set of users who can access $P_i$, where $1 \leq i \leq m$.

- $G$: the gateway server controlling the network package to enter or exit the local area network.

- $U_A$: the system administrator.

Without loss of generality, we assume that there is one system administrator and one gateway server. Our threat model does not change even if there exist multiple system administrators and gateway servers.

### Normal System State and Trust Assumption

When the system or the organization's local area network is in its normal state, we make the following assumptions:

- No users in $U_{P_i}$ should be granted permissions to access other firewall hosting servers: $U_{P_i} \cap U_{P_j} = \emptyset$, where $1 \leq i, j \leq m$ and $i \neq j$.

- $U_{P_i}$ can observe and modify local area network traffics incoming to or outgoing from $P_i$.

- $U_A$ is trusted and has permissions to access $P_1, \ldots, P_m$.

- $G$ is only accessible by $U_A$.

The above assumptions are realistic and can be easily achieved. For instance, mandatory access control can implement the first assumption, and virtual local area network (VLAN) is a mechanism to achieve the second assumption. In addition, trust must exist somewhere in the network, and our protocol design assumes the system administrator is trustworthy.

### Modeling of Insider Threat

To model insider threat, we need to clarify who the adversary is. In our problem domain, an adversary is or possesses access control credential of one of these users $U_{P_1}, \ldots, U_{P_m}$. As a result, for the rest of the paper, we use $U_{P_1}, \ldots, U_{P_m}$ to represent the adversaries for insider threat. The behaviors of these adversaries are classified into two categories:

- Passive $U_{P_i}$:

  - Read access to firewall related files on $P_i$.

  - Observe LAN traffic passing through $P_i$.

- Active $U_{P_i}$:

  - Read and write access to firewall related files on $P_i$.

  - Modify LAN traffic passing through $P_i$.

  - Colluding with $U_{P_j}$ to gain read and write access to $P_j$'s data and its LAN traffic.

The above can also model external attacks or data breaches on these $P_1, \ldots, P_m$ servers.

Under the proposed system architecture, we develop two classes of secure protocols to obliviously manage firewalls. The first class offers security guarantees against passive adversaries, and the other class can detect malicious activities, identify the parties involved in these activities, and recover from their activities up to a threshold.

**Benefits of the Proposed Distributed Architecture**

Consider a comparison with a "standard" firewall, even if it is distributed across multiple machines and an insider has access to one of these machines, the insider can learning firewall information stored on that machine and corrupt its content. Because the way the firewall is distributed in our scheme, accessing the firewall data on one or more (up to a predefined threshold) of $P_1, \ldots, P_m$ servers, no information can be leaked regarding the firewall. Even when the firewall data is modified by a malicious insider, our firewall can still function properly.

### 4.1.3 Organization

The rest of the paper is organized as follows: Section 4.2 presents foundational information regarding firewalls and more details concerning recent concerns and developments. Then, we present our protocols in detail in Section 4.3, describing their functionality and procedure along with analysis concerning complexity, security, and correctness. Section 4.4 presents extensive empirical results to show the feasibility of the proposed protocols. Finally, Section 4.5 summarizes our contribution and future research work.

## 4.2 Related Work

Firewalls in general have been in existence nearly as long as computer networking. The earliest functions simply involved a router or gateway server maintaining a list of addresses to block or permit. These are so-called *blacklists* and *whitelists* respectively [56]. As sophistication increased, divergent interests produced multiple approaches to securing local subnets against threats from the Internet at large. These include rules to block entire domains or sub-domains, as well as a wide array of means by which these rules may be optimized for both security and efficiency. [74]. While very quick, this method potentially allows for a single point of failure in terms of both network functionality, as well as security. It is possible in this case that an adversary, particularly an internal one, may be able to access the list of permitted or prohibited addresses and use this information outside the desired context for nefarious purposes. Additionally, if access to the file containing the lists is obtained, it is possible for an adversary to grant access to other external undesirable and previously blocked parties, or hinder the ability of a legitimate party to connect. These are important and perhaps critical concerns depending on the setting and desire for security and control of critical information.

One means by which a firewall may be represented in terms of a whitelist or blacklist is through the structure of a Bloom Filter [75]. This type of data structure quickly and efficiently gives set membership in probabilistic terms. The efficiency of this scheme is particularly strong with respect to space. This important topic, on which our work is dependent, will be addressed more thoroughly in Appendix B.3. This data structure has been used in firewall constructions previously to good effect [76].

A considerable amount of research has been conducted in reliably distributing the functionality of a firewall throughout a network with varying rates of efficiency and security [77]. This is motivated by a variety of concerns for an array of settings from cellular networks to cryptocurrencies and enterprise systems [75, 78]. While advantageous from an resilience perspective, in a situation where restricted network topology is not possible or practical, this approach can have risks of its own [79]. If the filter for a particular device or user is local to the system in question, many users may take steps to circumvent this policy. This is much more readily achievable for users with direct access to the system implementing the firewall than if it were a separate dedicated and more secure system in the network infrastructure for the subnet. Our interest in distributing the firewall is to distribute it in an oblivious means, distinct from these approaches, such that the individual machines cannot simultaneously alter the result of the firewall and maintain anonymity. This is discussed in greater depth in the presentation of our proposed protocols in Section 4.3.

A combined approach allows for the distribution of the firewall, represented by a set of Bloom filters [80]. This is useful especially when network topology is not constrained to a single gateway making the bridge between the local network and the Internet, a common occurrence in mesh networks. In this setting, the filters are individually based on the acceptance of packets at each node [80, 81], and forwarding a packet from one neighbor to another is dependent on the filters associated with the forwarding table. A packet is forwarded if the destination node has whitelisted the source node. This approach still allows for potential manipulation of the filters without the ability to detect malicious activities, and it is less efficient than the centralized approaches.

Software defined networking has recently shown great promise to ameliorate the issues related to much of networking's inherent hardware dependence. It allows for both hardware abstraction, leading to greater flexibility, as well as a separation between data handling and forwarding and traffic controls [82]. This separation is certainly advantageous from the perspective of flexibility and efficiency, as witnessed by more than a decade ongoing effort to update the infrastructure from IPv4 to IPv6 so the connection is both feasible and reliable [83]. Even today, adoption is estimated at around only 23% [84]. While this has many advantageous properties, it does not inherently impart any greater security to the operations which is a primary motivator for our work. Firewalls implemented under this paradigm may still leak undesirable information, and they may still be maliciously altered.

For our proposed methods to follow, we require two additional pieces of research upon which our solution is built. These are Bloom Filters, as previously mentioned, as well as secret sharing. These two mathematical structures form the foundation of what we propose. It is important that these dependencies be recognized. However, the summary of the techniques and background information and the specifics of the schemes we employ are given in

## 4.3 The Proposed Protocols

In this section, we propose our solutions by which a firewall can be secured in an information-theoretic sense and distributed across multiple servers that evaluate the firewall functions quickly and efficiently while maintaining a higher level of security than previously proposed systems. The proposed solutions can also be expanded to include resilience in the face of some subset of servers failing and being additionally secure against attempts to manipulate the firewall via the primitives and nature of the underlying mathematical constructs. In the descriptions to follow, we have assumed a blacklist approach to the firewall construction due to the possible but unlikely event of a false positive test result, more tolerable in general for a firewall. The proposed protocols can be classified into three categories according to different aspects of firewall management: (1) firewall initialization, (2) firewall rule evaluation and (3) firewall rule or policy update.

- *Firewall initialization*: Given a false positive error rate, the size of the blacklist, the initialization protocols can decide the Bloom filter size and the number of hash functions. Then the Boom filter representation of the blacklist will be produced. Depending on the adversary model, either additive or Shamir secret sharing scheme, as discussed in Appendix B.3.2, will be used to generate secretly shared Bloom filter, and each share is provided to one of the designated servers for oblivious firewall management and evaluation. Thus, each server has shares of the Bloom filter representing the rules of the firewall, and they are collectively ready for firewall rule evaluations. This initialization of generating secret shares of the Boom filter is assumed to be a trusted process, and after that, we consider various threats to the security of the proposed methods to obliviously manipulate and evaluate the firewall policies and rules.

- *Firewall rule evaluation*: We developed several protocols for firewall rule evaluation with various security guarantees. In the proposed protocols, we assume that the firewall has been initialized, that is, the information of interest, such as IP addresses on the blacklist have already been hashed, and a Bloom filter has been constructed and exists as shares across the servers to be involved in the firewall evaluation.

- *Firewall rule or policy update*: When firewall rules and policies are updated, the Bloom filter representing this firewall also needs to be updated accordingly. In addition, secret shares of the Bloom filter need to be updated as well at each server. We developed a secure way to perform these actions to update our oblivious firewall.

For the rest of the paper, our proposed schemes are described with several key system parameters shown in Table 4.1. Before providing the details for implementing a secured and distributed firewall evaluation function among independent servers, we next clarify the network topologies we assumed.

| | |
|---|---|
| $\eta$ | The number of addresses to add to the Bloom filter |
| $\beta$ | Number of bit locations in the Bloom filter |
| $\kappa$ | Number of hash functions used in checking set membership via the Bloom filter |
| $t$ | Threshold from the Shamir secret sharing scheme |
| $m$ | Number of parties involved in the secret sharing scheme |
| $N$ | Prime modulus for the secret sharing scheme |

Table 4.1: Common notations

### 4.3.1 Initialization

The main challenge during the firewall initialization process is to determine the Bloom filter and the number of hash functions needed to construct the filter. Equations B.3 and B.4 given in Appendix B.3.1 can help the decision making in this regard. Since the size of the blacklist $\eta$ is known and the probability of a false positive is approximated by $0.6185^{\beta/\eta}$, fixing the false positive rate will lead to the size of the Bloom filter $\beta$. Consequently, the number of hash functions $\kappa$ can be calculated by Equation B.4. For example, suppose $\eta = 1,000,000$ (one million), and the false positive rate is 0.001. Then $\beta \approx 14.5$ million bits. As a result, the number of hash functions is about $14.5 \times \ln 2 \approx 10$. As for the hash functions, we can adopt SipHash with a set of keys of size $\kappa$ indexed by $t$. Our setting is exactly the type of setting for which SipHash was designed and it performs very well [85].

---

**Algorithm 8:** FWInit($\langle\text{Admin}, addr_s\rangle, \langle P_i, \perp\rangle) \to (\langle\text{Admin}, \perp\rangle, \langle P_i, [\phi_j]_N^{P_i}\rangle)$

**Input:** IP addresses ($addr_s$ from the blacklist), $1 \leq s \leq \eta$
**Output:** $\phi_j$ for $P_i$, $0 \leq j < \beta$ and $1 \leq i \leq m$

1 Admin

(a) $\phi_j \leftarrow 0$, for $0 \leq j < \beta$

(b) For $1 \leq s \leq \eta$ do
$\phi_j \leftarrow 1, \forall j \in \{hash_t(addr_s) \mod \beta, \text{ for } 1 \leq t \leq \kappa\}$

(c) Gen_Shares($\phi_j$), for $0 \leq j < \beta$

(d) Send $[\phi_j]_N^{P_i}$ to $P_i$, for $0 \leq j < \beta$ and $1 \leq i \leq m$

---

During a trusted initialization phase, the addresses forming the blacklist are hashed with the prescribed functions, and the Bloom filter is constructed. This vector of bits is secretly shared using the Gen_Shares function that can be implemented using methods introduced in Appendix B.3.2. The shares are sent to each of the shareholder parties to participate in the scheme. Once these two steps have been completed, the online operation may commence. The steps of the initialization process are presented in Algorithm 8.

## 4.3.2 Secure Distributed Firewall Evaluation under the Semi-honest Model

As mentioned before, we developed two secure protocols to evaluate firewall rules obliviously distributed among $m$ servers: $P_1$, ..., $P_m$. In this section, we discuss the first one secure under the semi-honest model where the participating parties follow the protocol. The second proposed protocol presented in Section 4.3.3 is secure against malicious attacks, and it can also mitigate errors introduced by the malfunction of the network components. The main evaluation condition behind the first protocol is based on the following condition determined by the construction of the Bloom filter:

- The IP address ($addr$) of the incoming message is on the blacklist if the sum of the values stored in the Bloom filter locations indexed by the $\kappa$ hash functions with $addr$ as their input is equal to $\kappa$.

If the above condition is true, the package will be blocked. The design of our first protocol is to ensure that the condition can be verified without disclosing the actual content of the Bloom filter. The steps of the protocol are presented in Algorithm 9.

**Key Steps and Correctness Analysis**

Note that the parameter $N$ denote the size of the secret shares. In this protocol, in order to represent the sum of the values from locations indexed by the $\kappa$ hash functions, $N$ needs to be bigger than $\kappa$. In step 1, the gateway receives a packet from which the IP address ($addr$) is extracted. The IP address is sent to the set of parties (i.e., the servers who manage the firewall) implementing the distributed secret sharing scheme. In step two, the parties hash the received address with all $\kappa$ hash functions modulo $\beta$, and this set of calculated values are the indices to the locations of the secretly shared Bloom filter, and the values (i.e., secret shares) stored in these locations should be summed. This summation is computed, and the result is sent back

**Algorithm 9:** firewallEval($\langle$Gateway, $addr\rangle, \langle P_i, [\phi_j]_N^{P_i}\rangle$) $\rightarrow$ ($\langle$Gateway, $\sigma\rangle, \langle P_i, \bot\rangle$)

**Input:** IP address (addr), and each party $P_i$ has shares $[\phi_j]_N^{P_i}$ of the Bloom filter, for $1 \leq i \leq m$ and $0 \leq j < \beta$
**Output:** $\sigma = \kappa$: the Gateway blocks $addr$

1 Gateway

   (a) $addr \leftarrow$ IP address extracted from a network packet

   (b) Send $addr$ to $P_i$, for $1 \leq i \leq m$

2 For each $P_i$:

   (a) $[\sigma]_N^{P_i} \leftarrow \sum_j [\phi_j]_N^{P_i} : j \in \{hash_t(addr) \mod \beta, \text{ for } 1 \leq t \leq \kappa\}$

   (b) Send $[\sigma]_N^{P_i}$ to Gateway

3 Gateway

   (a) $\sigma \leftarrow [\sigma]_N^{P_1} + \cdots + [\sigma]_N^{P_m} \mod N$

   (b) **If** $\sigma = \kappa$: block packet
       **else**: forward packet

to the Gateway. In step 3, the Gateway reconstructs the result of the computation and compares it against $\kappa$. Due to the nature of the Bloom filter, this result will be equal to $\kappa$ if and only if every location in the summation was 1, which is exactly the criteria indicating set membership according to the Bloom filter construction.

**Complexity Analysis**

In this setting, the Gateway sends the data of interest to the set of servers forming the shareholding parties in the secret sharing scheme. We assume that this is an IP address in the context of our presentation, and we will continue that common and useful assumption in the context of this analysis. As we note in Section 4.2, even though we are more than 10 years out from the acceptance of the IPv6 standards, IPv4 is still strongly in the majority for traffic, thus we assume IPv4 addresses of 32 bits. In the second step of the protocol, all $m$ parties send the result of their summations to the Gateway thus the required communications will be $m \log_2 N$. This send and receive cycle constitutes one computational round, and completes the communication requirements for this protocol. The total communication complexity is $m(32+\log_2 N)$ bits in a single round.

**Proofs of Security for Passive Attacks**

We define our notions of security in consonance with the proposed definitions and conventions of Goldreich [9]. As discussed at length there, a protocol is secure in the computation of a desired functionality given that the view of the execution of the protocol, or execution image, is indistinguishable from that generated by a simulator. Here we define our security in information theoretic terms for passive attacks as described previously. These attacks include attempts to extract information from what data may be visible on a compromised server, observing traffic on a compromised server, and observing visible network traffic.

In general, with respect to the last point, we assume that the servers are in one of two situations, they are all on different local subnets, or all the servers communications are encrypted in transit. Under these circumstances, the attackers can only observe encrypted internal network traffic. Thus, no information regarding the Bloom filter or the firewall rules and policies are leaked to the attackers. With respect to the two other concerns listed, we provide the following analysis and arguments for security.

Given a functionality $f(x,y)$ operating on the inputs $x$ and $y$ and a protocol $\Pi$ implementing this functionality, then an execution image for $\Pi$ is denoted by $\{\text{VIEW}_1^\Pi(x,y)\}_{x,y \in \{0,1\}^*}$ for party $P_1$. What is needed to prove security of $\Pi$ is an algorithm $S_1$ which, given the public information as well as the necessary private information from $P_1$, $P_1$ is able to produce a simulated execution image of $\Pi$, denoted by $\{S_1(x, f_1(x,y))\}_{x,y \in \{0,1\}^*}$ for some simulator algorithm $S_1$. If there is an equivalence between the simulated and real execution images in an information theoretic sense, then the protocol is secure with this guarantee for passive adversaries. Thus, the goal is to demonstrate the following equivalence which is given for a party $P_1$ though, in general, it would need to be demonstrated for all parties:

$$\{S_1(x, f_1(x,y))\}_{x,y \in \{0,1\}^*} \equiv \{\text{VIEW}_1^\Pi(x,y)\}_{x,y \in \{0,1\}^*}$$

Since our protocol is symmetric; that is, all parties $P_1, \ldots, P_m$ who are responsible for obliviously managing the firewalls perform the same operations, it is sufficient to show the equivalence from one party's perspective to prove the protocol is secure for all parties.

The view of the servers in this situation is comprised of two different components. First they receive shares of the Bloom filter, these shares disclose no information regarding the filter itself, or the firewall rules it represents. This is immediate from the underlying secret sharing scheme. They additionally receive IP addresses which are queried for membership in the set comprising the firewall rules. The only use of this information is in generating the indices for share summation. Summing the shares, or any other local operation performed, yields no information to a shareholder without bounds on available computational power. This too is immediate from the properties of the secret sharing scheme. Thus, a trivially implementable simulator for this view consists of sending a random string of 32 bits representing an IP address to query on the system. The servers are expecting arbitrary 32 bit strings and will proceed in their calculations revealing no information among themselves without malicious behaviors. This is the property we wished to demonstrate for passive adversaries.

**Protocol Implementation Issues**

The protocol given in Algorithm 9 can be implemented using either additive or Shamir secret sharing schemes. Under the additive scheme, the Gateway has to collect all the shares before constructing $\sigma$. As a result, when a share is lost, the protocol may not proceed properly. On the other hand, under Shamir, the Gateway only needs to collect $t$ out of $m$ shares to derive $\sigma$. If some degree of fault-tolerance (regarding packet loss) is desired, we can implement the protocol using Shamir. However, this will incur more computation cost. The empirical results presented in Section 4.4 will show the performance differences between the two schemes.

### 4.3.3 Secure Distributed Firewall Evaluation under the Malicious Adversary Model

As previously acknowledged, the first proposed protocol is secure if the servers follow the prescribed steps of the protocol. The protocol presented in this section deals with the situations where the servers may be controlled by an adversary, who could prevent a server from participating or modify the messages sent to the other servers. The key steps of the protocol are given in Algorithm 10.

**Key Steps and Correctness Analysis**

Comparing to the first protocol, the steps 2 and 3 are significantly different. Instead of a summation, a secure product is called for at step 2. This results in large changes for both the complexity and the amount of information to be hidden. At step 3, the revealed values will be either 0 or 1 exclusively. The revealed result will be one if and only if every indexed element involved in the product is 1. This is again directly analogous to the result of checking for set membership using a Bloom filter directly. This additional information hidden is the contents of the filter itself, which may be advantageous for some use-cases. For example, an adversary can predict the existence of an IP address on the blacklist with high probability by knowing the summation values. Since network faults may occur and malicious adversaries may disable one or more servers among the $m$ servers, the gateway could only receive $m'$ shares within

**Algorithm 10:** firewallEval($\langle$Gateway, $addr\rangle, \langle P_i, [\phi_j]_N^{P_i}\rangle$) $\rightarrow$ ($\langle$Gateway, $\pi\rangle, \langle P_i, \bot\rangle$)

**Input:** IP address (addr), and each party $P_i$ has shares $[\phi_j]_N^{P_i}$ of the Bloom filter, for $1 \leq i \leq m$ and $0 \leq j < \beta$

**Output:** $\pi = 1$: the Gateway blocks $addr$

1 Gateway

    (a) $addr \leftarrow$ IP address extracted from a network packet

    (b) Send $addr$ to $P_i$, for $1 \leq i \leq m$

2 For each $P_i$:

    (a) $[\pi]_N^{P_i} \leftarrow \prod_j [\phi_j]_N^{P_i} : j \in \{hash_t(addr) \mod \beta,$ for $1 \leq t \leq \kappa\}$

    (b) Send $[\pi]_N^{P_i}$ to Gateway

3 Gateway

    (a) Reveal $\pi$ for all valid combinations of $t$ out of $m$ shares

    (b) If all $\binom{m}{t}$ reveals do not agree: flag for malicious behavior

    (c) **If** $\pi = 1$: block packet
        **else**: forward packet

an expected time window. The additional combinations and criteria are directed at security concerns for the proposed protocol which is addressed in detail in Section 3.

**Complexity Analysis**

The initialization is the same as previously, the first transmitted message is the same, and the conclusion of the protocol is the same with all parties sending their share of the result to the Gateway. Thus the complexity is the same for these steps, requiring $m(32 + \log_2 N)$ bits. On top of this, we add additional communications and operations due to the fact that secure products cannot be evaluated without interaction among the parties. Assuming that Shamir's secret sharing (applicable for three or more parties) is the scheme in use and that $m = 2t - 1$. Step 2(a) can be executed in $\log_2 \kappa$ rounds, and each round requires $m(m-1)\log_2 N$ bits of communication. Thus, the total communication cost is $\log_2 \kappa \cdot m(m-1) \log_2 N + m(32 + \log_2 N)$ bits.

**Security Analysis**

With respect to security, the arguments of the previous protocol hold, and we also strengthen the ability to hide the contents of the filter itself. We do so by considerably lengthening the expected time for the filter's contents becoming known by changing from summation to a product of the Bloom filter indices mentioned previously.

Consider the previous protocol. Given a large number of queries, information may accrue which will allow the contents and structure of the Bloom filter to be reconstructed. Revealing the Bloom filter does not imply the firewall we are using it to represent has been revealed. Obviously some information will be revealed, but complete reconstruction is not a possibility. Nevertheless, our second protocol prevents even the filter's structure from being revealed quickly. This information may be extracted over time with the previous protocol version by saving a record of IP addresses and reconstructed results at the Gateway. Imagine a packet arrives with an IP address that hashes to 4, 5, and 6. If the rebuilt summation of these locations equals 2, then obviously 2 of the previously mentioned indices have set bits. Imagine a second packet arrives with an address which hashes to 1, 5, and 6. In this case too let us assume that the revealed value is 2. Finally a third packet arrives with an address which hashes to 1, 4, and 7. The revealed summation result here we shall take as 0 for this example. Given these three instances, we can deduce with certainty that the Bloom filter has these contents $[?_0, 0_1, ?_2, ?_3, 0_4, 1_5, 1_6, 0_7]$ for the respective subscript indices.

Again this may be tolerable in some instances and it may be preferable to hide this in others. In the previous example, if our second proposed protocol were followed, the result for every case would be 0. Therefore the vector would still consist of only uncertainties rather than having some of the values known. Information may still be gleaned from this second protocol by similar means, but the rate of its accrual is much slower since only combinations which return a 1 will yield any sure foundations for deduction.

**Malicious Adversaries**

Malicious adversaries may do anything a passive adversary can do and more. For example, a malicious adversary may deviate from the protocol or manipulate their intermediate calculations. These deviations we wish to at least detect, and possibly identify the malicious party as well. Specifically, the active attacks we consider are (1) disabling a participating server, (2) reading data which is normally expected to be protected, and (3) modifying data which is normally expected to be protected.

**A Combinational Approach**  With respect to the first concern listed, we would draw attention to the fact that this is "built in" to the Shamir's secret sharing scheme provided that $m > t$, since any set of $t$ shares can be used to reconstruct the secret. With respect to the other two concerns we provide the following analysis and arguments for security.

Shamir's secret sharing scheme affords us the opportunity to achieve detection and identification of malicious adversaries relatively efficiently due to the fact that it is a threshold secret sharing scheme. If a sufficiently large number of parties, greater than the threshold, participate in the scheme, it is straightforward to detect malicious manipulation of the shares, or, with more parties, identify a malicious adversary. As we have discuss in Appendix B.3.2, Shamir's scheme uses a threshold number of parties to reveal secret shared values, while an arbitrarily large number of parties may participate in the scheme. If the number of parties $m$ is set such that $m \geq t+1$, then a malicious adversary may manipulate their share of the computational result to change the revealed value, only when their share of the result is used in the revealing process. Thus, for a set of 4 parties, a threshold of three, a malicious $P_1$, and a shared value $s$, the following equivalence relations, meaning the output of the reveal function being incorrect, hold for the reveal functionality when $P_1$ has acted maliciously and altered $[s]_N^{P_1}$.

$$\text{reveal}\left([s]_N^{P_1}, [s]_N^{P_2}, [s]_N^{P_3}\right) \equiv \text{reveal}\left([s]_N^{P_1}, [s]_N^{P_3}, [s]_N^{P_4}\right)$$
$$\equiv \text{reveal}\left([s]_N^{P_1}, [s]_N^{P_2}, [s]_N^{P_4}\right)$$

However, the other available combination of shares is not equal to the other two, specifically

$$\text{reveal}\left([s]_N^{P_1}, [s]_N^{P_2}, [s]_N^{P_3}\right) \not\equiv \text{reveal}\left([s]_N^{P_2}, [s]_N^{P_3}, [s]_N^{P_4}\right)$$

by the simple and direct principles of polynomial interpolation on which this functionality depends. Therefore, in this case it is possible to detect malicious behavior due to the fact that all combinations of shares used for reconstruction of the secret do not agree. There is a problem still. The majority of the revealed values are all influenced by the malicious party's actions.

When the difference between $m$ and $t$ grows wider, e.g., $m \geq 2t + 1$, revealing shared values becomes more treacherous for a malicious adversary. In this case, the majority of revealed combinations will agree on the correct value, and identifying the malicious party becomes very easy. In the minimal case, where $m = 2t + 1$ there are $\binom{2t+1}{t}$ combinations possible for revealing the shared value, and only $\binom{2t}{t-1}$ of them will be able to be influenced by any malicious party. Therefore, the following expression gives the exact fraction of combinations of shares to be used in polynomial reconstruction which it will be possible for a malicious adversary to influence:

$$\frac{\binom{2t}{t-1}}{\binom{2t+1}{t}} = \frac{\frac{2t!}{(t-1)!(t+1)!}}{\frac{(2t+1)!}{t!(t+1)!}} = \frac{t}{2t+1}$$

Additionally, in general, a set of $x$ malicious parties, by the same reasoning, the maximum number of shares that can be influenced by the malicious parties is defined by:

$$\sum_{i=1}^{\min(x,t)} \binom{x}{i}\binom{m-x}{t-i}$$

Thus, in order for the honest adversaries to be able to agree on the correct value by a majority agreement scheme, such as that represented in the Byzantine General's problem, a majority of the share combinations must recombine to form the correct solution. Therefore, the following inequality should be preserved to guarantee the correctness of the protocol against actively malicious adversaries:

$$\binom{m}{t} > 2 \sum_{i=1}^{\min(x,t)} \binom{x}{i}\binom{m-x}{t-i}$$

In a specific case, where $t = 3$ and $m = 2t + 1 = 7$, there are 35 combinations valid for reconstructing the shared secret. It is evident that a malicious entity can influence 15 combinations for rebuilding the secret, while the majority 20 of the share combinations will remain uninfluenced. Thus not only will a malicious party's activity be detectable, the party is easily identifiable through the minority of reconstructed values consisting of those with exactly one member in common across all combinations of shares used in their reconstruction, specifically the malicious party.

Suppose $m'$ is the number of messages received by the gateway from the $m$ servers, where $t \leq m' \leq m$. The following conditions specify when malicious behaviors can be detected or prevented:

- Malicious behavior detection: When there are at least $t$ shares from honest servers, the detection of malicious behaviors is possible in the framework.

- Correctness guarantee (or malicious behavior prevention): When $\binom{m'}{t} > 2\sum_{i=1}^{\min(x,t)} \binom{x}{i}\binom{m'-x}{t-i}$ and $x < m'$, malicious behaviors, including disabling the servers and modifying the shares, will not influence the correctness of the protocol.

**Berlekamp-Welch Error Correcting Algorithm** The preceding approach is sufficient for small sets of parties with few to no adversaries and can be reasonably efficient for this setting. If large sets of parties are involved the means which we have proposed for checking for malicious behavior become cumbersome and prohibitive in complexity. When large sets of parties are involved another approach, will be more beneficial. We do not address the details of this alteration in any great depth, but do present the intuition of the additional steps for consideration of this potential case.

We base the additional steps on the well known, resilient, and efficient Berlekamp-Welch algorithm for error correction [22]. Other algorithms with additional beneficial properties exist such as the various algorithms useful in list decoding [86, 87]. Since these return a list of possible polynomials and we lack any means of distinguishing correct from incorrect possibilities, the gain by using such methods is negated for our application, and they simply induce additional computational cost. Since the parties already hold shares of a polynomial, all that is required is a different set of steps at the server buffering the packet in question, namely those required by the Berlekamp-Welch algorithm. Rather than rebuilding the shares to reclaim the secret as is normally done, the shares received are used to construct a linear system which is solved locally. This allows for the correct reconstruction of the polynomial underlying the shares representing the filter evaluation, which is equivalent to revealing the shared secret, but with some extra benefit due to special properties of the system. The identification of the error laden points in the interpolation is possible thereby allowing the honest parties to root out the adversary in their midst. This does not hold in all cases for any values of $m$ the number of parties which are shareholders in the scheme, and $t$ the threshold of the scheme, but only for at most $\frac{m-t+1}{2}$ malicious shareholders in the scheme. This is achieved by identifying the shares, really points on a polynomial in Shamir's scheme, which are causing the majority of other points to not lie on the interpolated polynomial. The local complexity of this operation is straightforward in $\mathcal{O}(m^3)$ which is obviously a large improvement over $\mathcal{O}\binom{m}{t} \approx \mathcal{O}(m!)$ for large groups of parties. Lower complexities than $\mathcal{O}(m^3)$ are possible through other methods but the details are beyond the scope of our presentation.

**Byzantine Agreement** In the previous presentation of our protocols the ultimate decision to block or forward a packet was made based on the computations of the Gateway alone. While this is a very efficient means, it may not meet the security requirements for all interested parties. We thus further extend the security of our protocol by requiring that the decision to accept or reject the packets be a multi-party computation as well. We achieve this by altering the final step of the protocols, as previously given, to send the shared computation result to not only the Gateway, but every other party involved in the scheme. This necessarily increases the complexity by an additional $m(m-1)\log_2 N$ bits, but allows for each of the servers along with the Gateway to rebuild the result of the computation. Once this has been achieved, they can each locally consider the rebuilt values, come to a decision, and then begin a Byzantine agreement process to make sure that they are all in agreement with the proper way to proceed. While this initially seems like a cumbersome step, recent research has lead to great improvements in Byzantine agreement and fault tolerance protocols [88, 89, 90]. Given the promising results of these research endeavors, Byzantine agreement has been made considerably more feasible and less costly of an operation than was previously the case. The additional complexity from the Byzantine agreement scheme itself is dependent on the choice of approach though we have proposed a few options an in depth analysis of the complexity of these protocols is somewhat outside the scope of our present work. However, similar to our use, this type of agreement mechanism is increasing in appearance due to some of the discoveries of performance improvements, most notably perhaps in cryptocurrencies [91, 92].

**Full Security Against Malicious Adversaries**   In our preceding discussion, our focus has been on presenting our general framework with some suggestions on what may be the most advantageous and efficient scenarios and settings. To that end, we have not been concerned with *preventing* malicious adversaries from deviating from the protocol or altering their shares, only the ability to *detect* that some such activity has occurred, and being resilient in the face of some threshold of these occurrences. At this point however, it is important to note that a number of general frameworks exist by which our straightforward semi-honest protocol may have some constant increase in the communication (though no asymptotically significant increase) and be transformed into a protocol fully secure against active or malicious adversaries. Extending our protocols through the use of the work of [93, 94, 5] can, in a straightforward manner affect this transformation, if desired, and again, do so in a reasonably efficient means.

### 4.3.4 Firewall Updates

In many situations, subsequent to the establishment of security measures, new threats may be identified and added to the blacklist. If the list structure were static, this could pose some problems since a new Bloom filter would have to be constructed and shared from scratch. Thus, we propose a simple means by which elements may be obliviously added to the Bloom filter representing the firewall rules without disturbing elements already encoded in that structure.

Assuming some kind of strict access control structure is in place, on each server participating in the secret sharing scheme, the following computation must be exe-cuted for the new address ($addr$) to be inserted in the filter.

$$[\phi_j]_N^{P_i} = [1]_N^{P_i} : j \in \{hash_t(addr) \mod \beta, 1 \leq t \leq \kappa\}$$

This operation is invoked by a trusted system administrator who can access the individual servers. Care must be taken in this operation since the shared values in the Bloom filter should be exclusively 0 or 1, and we cannot allow this manipulation to introduce shares of elements aside from those two. Thus, we cannot simply increment the shares for each location resulting from the modular hash of the address in question. In the above expression, each location, if it is previously 0, gets incremented to 1. Additionally, if it is previously 1, it remains unchanged, as desired.

## 4.4 Performance Evaluation

In this section, we empirically analyze the computational overhead of the proposed protocols. First, we clarify the storage requirement of each firewall server to store the secretly shared firewalls. As analyzed previously, the Bloom filter size is denoted by $\beta$, and each entry of the filter is secretly shared among three servers. Since the size of each share is bounded by $\log_2 N$, the storage requirement for each server is about $\beta \log_2 N$ bits. Following the example given in Section 4.3 where the size of the blacklist is $\eta = 1,000,000$ (one million) and the false positive rate is 0.001, then $\beta$ is about 14.5 million and the number of hash functions is around 10. As a result, we set $N$ to be 11, and each server needs $14.5 \times \log_2 11$ bits which is about 6M bytes. In general, one million for the blacklist size is quite reasonable. The rest of the performance was conducted on three computers each with the following specifications:

- Operation system: 64 bit Kubuntu 18.04

- Processor: Intel Xeon E-2186G, 6 Core HT, 12MB Cache, 3.8Ghz with 64Gb

RAM.

- Network: connection speed 1 GBit/s with 20 Mbps average download and upload, and latency 16 ms.

- The implementation was done using C++ and the GNU Multiple Precision Arithmetic Library [95]. We design the communication channel using socket over TCP/IP.

Note that the secure multiplication protocol we adopted under additive secret sharing only works for three parties, and there does not exist an efficient secure multiplication protocol for more than three parties. Therefore, to compare efficiency between Shamir and additive secret sharing schemes, we chose to use three servers to conduct our experiments, along with the baseline which only requires one machine.

### 4.4.1 Initialization

Here we analyze the runtime of the Bloom filter initialization process. Under three parties setup, the initialization process is affected by three parameters, the Bloom filter size $\beta$, the number of the hash functions $\kappa$ and the size of the blacklist $\eta$. In our experiments, we set these parameters as follows:

- The Bloom filter size $\beta$: we vary $\beta$ from $10^3$ to $10^8$. We fix the size of the blacklist to $1M$ while $\kappa = 10$. Figure 4.2 shows a very close performance between the three schemes up to $\beta = 1M$ with difference of less than 1.5s and 0.3s between Shamir and additive comparing to the base case respectively. It is apparent that the performance of additive scheme is closer to the base case through all different sizes of $\beta$. The maximum time required for this process was by Shamir scheme which is still less than 4s for $\beta = 1M$ and roughly 120s for $\beta = 10^8$. As shown in Section 4.3.1, $1 \times 10^8$ ($> 14.5M$) is large enough to provide 0.001 false positive rate.

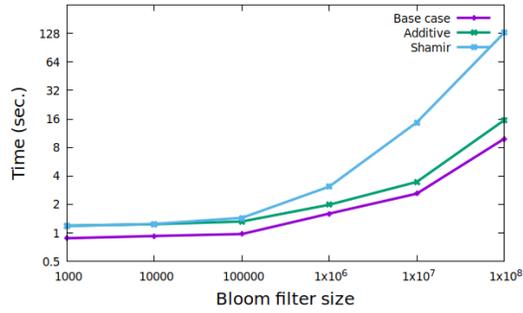

Figure 4.2: Initialization runtime varying Bloom filter sizes

- Number of hash functions: we range $\kappa$ from 5 to 20 while $\beta = 14.5M$ and the size of the blacklist $\eta = 1M$. Graph 4.3 demonstrates that the measured performance of the additive scheme is approaching the baseline with only 0.7s difference at the maximum $\kappa$. Even though Shamir shows the highest runtime, it only takes about 21s at the highest $\kappa$.

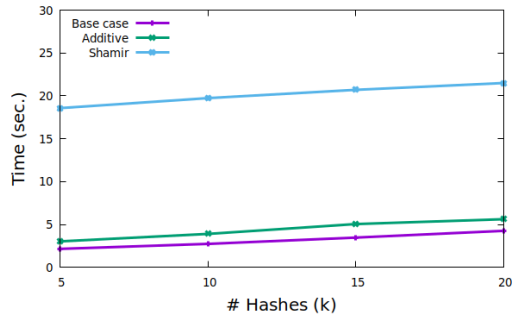

Figure 4.3: Initialization runtime varying number of hashes

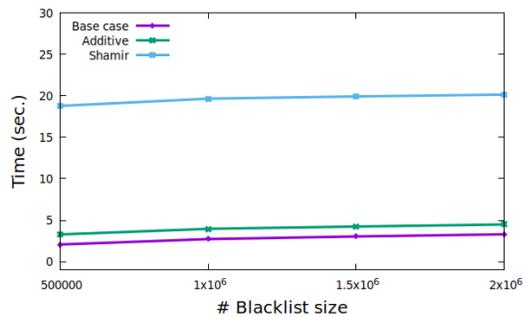

Figure 4.4: Initialization runtime for various blacklist sizes

- Blacklist size: we vary the size of the blacklist from $500K$ to $2M$. $\beta$ was set to $14.5M$ and $\kappa$ was fixed at 10. It is obvious from Figure 4.4 that the additive scheme has a similar performance to the base case along with different blacklist sizes. Shamir incurs the highest runtime. Yet, it takes around 20s for the maximum blacklist size.

For firewalls containing a very large number of entries, the initialization process can get somewhat lengthy, but it only needs to be performed once and its computations are highly parallelizable. In practice, taking advantage of such parallelization opportunities would yield dramatic performance improvements for large firewalls.

### 4.4.2 Firewall Evaluation

The runtime of firewall evaluation under three parties only depends on the number of the hash functions $\kappa$ and the number of the requests that are required to be filtered. As we offered two algorithms (Algorithm 9 and 10) for the evaluation process, we have fixed the size of the Bloom filter $\beta = 14.5M$. We emphasis that $\beta$ has no effect on this process. To show the performance of both algorithms under different $\kappa$, we fix the number of the requests at $50K$ and vary $\kappa$ from 5 to 20. On the other hand, to analyze the runtime for various number of requests we set $\kappa$ at 10 and range the requests from $10K$ to $50K$. The performance of the firewall evaluation process is summarized below.

**Algorithm 9**

In both scenarios mentioned above, the additive scheme outperforms the Shamir based implementation. With that being stressed, Shamir scheme still requires less than half a second to handle the highest $\kappa$ we set. and less than 0.4 sec. with the maximum number of requests as it can be seen in Figure 4.5 and 4.6.

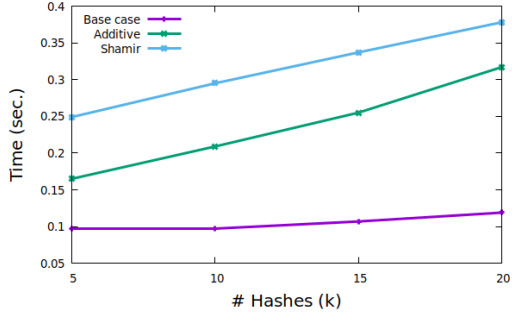

Figure 4.5: Algorithm 9 runtime varying number of hashes

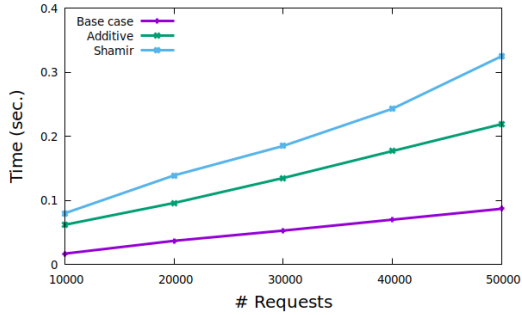

Figure 4.6: Algorithm 9 varying the number of requests

**Algorithm 10**

Recall that this approach invokes the secure multiplication protocol of either additive or Shamir scheme based on the trade off we presented previously. It requires $\kappa - 1$ rounds to perform the evaluation process as shown in the protocol. This is required for both schemes as they only offer a binary secure multiplication protocol $mul(u, v)$, where it can handle only two secretly shared values at once. While unbounded fan-in multiplication [29] is possible, for products of our size, the additional overhead and communication required negate much of the potential benefit. For a large number of hash functions $\kappa$, this affects the overall performance and increases the overhead dramatically. Therefore, we have designed the multiplication to handle pairs of shared values in parallel for all $\kappa$ values at once. We repeat the same process for the next rounds until we get the final result. This reduces the number of rounds from $\kappa - 1$ to $\log_2(\kappa)$. Therefore, in the case of $\kappa = 20$, it requires only 4 rounds rather than 19 rounds.

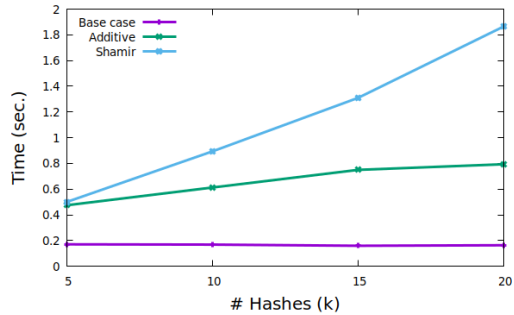

Figure 4.7: Algorithm 10 varying the number of hashes

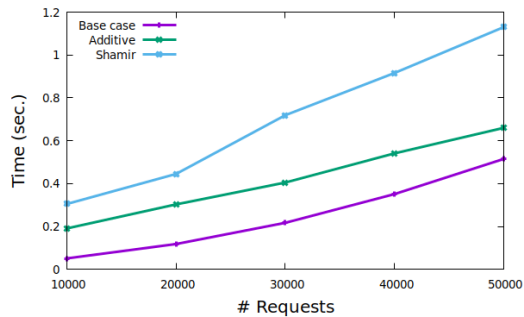

Figure 4.8: Algorithm 10 varying the number of requests

From Figure 4.7 and 4.8, again the additive scheme beats Shamir. As it requires roughly 0.6s at the maximum $\kappa$. Shamir shows that it can handle the $50K$ requests in almost 1s with $\kappa = 10$ which is enough to provide 0.001 false positive rate, with the respect of other parameters as explained previously. It is important to note though that while the additive implementation does outperform the Shamir based implementation, both are efficient and not very far removed from one another. Moreover, the additive scheme's performance is very close to the base case with at most 0.2s difference overall. For very high consequence systems and information, this cost in latency may be acceptable to gain the security the protocols offer.

## 4.5 Conclusion

Our contribution in this paper amounts to developing protocols for a secure, distributed and oblivious firewall evaluation. We have achieved this new and secure firewall management paradigm through the use of secret sharing schemes as well as Bloom filters to represent the firewall. In addition, we proposed multiple means for implementing this functionality with various options to increase efficiency or security, including fault tolerance and security against malicious attempts to disrupt the system. Furthermore, we have shown that our proposed approach need not be static, but can be dynamically updated if necessary to handle new threats or other security issues.

Our experimental results have shown that the overhead associated with the protocols are reasonable depending on design choices and security concerns. Securing firewalls against insider threat is only the first step, and we are planning to adopt the proposed methodology to develop a suite of secure and distributed protocols for managing network functionalities an resources, such as access control, user management and resource allocations.

# Chapter 5

# Summary and concluding remarks

Our contribution in this work amounts to a compilier allowing for the efficient and seucre transformatino of protocols from the semi-honest adversary model into the malicious adversary model, the proposed protocols for secure multi-partly comparison, which can be implemented very efficiently in a secret sharing scheme, and finally, an application for some of these SMC techniques to address a security concern.

### 5.0.1 Future Work

As research progresses we have a number of areas where we plan to focus further inquiry. Thus far we have presented some open problems, and one set of solutions to one of them. In the future, we plan to address the others. These goals include work to:

- Further improve the efficiency of secure multi-party comparisons

- Extend the methods employed in current research regarding comparison to secure bit decomposition

- Efficient secure comparisons will additionally allow for other interesting applications such as secure and oblivious sorting

- Explore the extent to which active security is truly necessary for every step of every protocol in order for overall active security to be achieved

- Improve other even more foundational primitives, such as secure multi-party multiplication, which would have further positive performance effects on SMC systems

- Explore if it is possible to use Shamir's scheme for degree $t$ and degree $2t$ sharings and if this has efficiency or security benefits for the work in Chapter 2.

It is our hope that completing these pursuits will have a substantive beneficial impact for the research community and provide another step of improvement toward the realization of SMC as a practical common system used to benefit society in daily life. While that goal may be lofty and somewhat distant, we hope to at least provide a beneficial contribution toward these goals of use to the academic community today.

# Appendix A

# Secret Sharing Numerical Examples

## A.1 Shamir Secret Sharing Illustrated

**Generating Shares** Suppose we have some individuals such that each has a secret value which they desire to secretly share, and it is further wished that any three shares may be able to reconstruct the secret, a set of quadratic polynomials would be constructed in accordance with the previously explained method. To fully demonstrate the system a set of five shares will be generated and utilized. For this example the following will suffice, consisting entirely of unsigned integers within the range representable by one byte. Specifically, we've chosen $p = 251$ which is the largest prime within that range. Our example opens with sharing a value of 18:

$$f(x) = 113x^2 + 88x + 18 \mod 251$$

Assuming we want to distribute shares among the five parties, thus requiring a simple majority to rebuild the secret, the set of shares contained in Table A.1. This would construct a (3,5) threshold scheme.

Table A.1: Example Shamir Shares

| Index | Shares of a |
|---|---|
| 1 | 219 |
| 2 | 144 |
| 3 | 44 |
| 4 | 170 |
| 5 | 20 |

**Rebuilding the Secret** To return to our example, we can see that selecting any three random shares will allow us to rebuild the original function and retrieve the secret stored within. We will demonstrate with shares 1, 2, and 3.

Following the Lagrange approach, we see the following generated from the same set of data:

$$L(0) = 219 \cdot \frac{-2}{1-2} \cdot \frac{-3}{1-3} + 144 \cdot \frac{-1}{2-1} \cdot \frac{-3}{2-3} + 44 \cdot \frac{-1}{3-1} \cdot \frac{-2}{3-2}$$

$$L(0) = 219 \cdot 3 - 144 \cdot 3 + 44 \mod 251 = 18$$

**Addition** As introduced previously, with respect to secure computation, this approach is additively homomorphic natively. If we now have a secondary secret value, 54, which we wish to secretly share, the same process can yield another quadratic polynomial, here, $167x^2 + 148x + 54$ with shares 1 through 5 being given as the indices and third column in Table A.2. If we need to securely calculate the sum of these two, each individual can simply add their two shares for the respective secrets. For each of the individuals, the result can be found in the table, in the final column. To verify that this does in fact represent shares of the correct secret, we can repeat the previous rebuilding process. Given the shares of this sum for parties 1, 2, and 3

Table A.2: Shamir Secure Addition Example Shares

| Index | Shares of a | Shares of b | Shares of a+b |
|---|---|---|---|
| 1 | 219 | 118 | 86 |
| 2 | 144 | 14 | 158 |
| 3 | 44 | 244 | 37 |
| 4 | 170 | 55 | 225 |
| 5 | 20 | 200 | 220 |

the following Lagrange Interpolation equation results:

$$L(0) = 86 \cdot 3 - 158 \cdot 3 + 37 \mod 251 = 72$$

Which is correct since here $x_0 = 72$ and $54 + 18 = 72$.

**Multiplication** Since our secret is in the constant term of a polynomial that is what we want in the product, but with simple multiplication of the polynomials themselves we get more and higher order terms than desirable represented in the shares. Consider the product of two polynomial functions below, $f_a(x) \cdot f_c(x) = f_ac(x)$ The bold portion below is our primary interest the rest must be securely handled For our second order examples so far:

$$\begin{aligned} f_{ac}(x) &= (a_2x^2 + a_1x + a_0)(c_2x^2 + c_1x + c_0) \\ &= a_2c_2x^4 + (a_2c_1 + a_1c_2)x^3 + (a_2c_0 + a_1c_1 + a_0c_2)x^2 + (a_1c_0 + a_0c_1)x + \mathbf{a_0c_0} \end{aligned}$$

(A.1)

For our case generating the product of 18 and 4 by using the shares spread among the parties, Table A.3 shows many of the necessary values. Given the shares of 18 and 4 each party computes the product of their shares. They each then construct a polynomial of the desired degree with random coefficients to share their product values according to the normal process. These polynomials are given in the right most column of the referenced table.

Table A.3: Shares for multiplication

| Index | Shares of a | Shares of c | Shares of a · c   mod p | h |
|---|---|---|---|---|
| 1 | 219 | 250 | 32 | $140x^2 + 75x + 32$ |
| 2 | 144 | 26 | 230 | $114x^2 + 185x + 230$ |
| 3 | 44 | 85 | 226 | $56x^2 + 96x + 226$ |
| 4 | 170 | 176 | 51 | $183x^2 + 134x + 51$ |
| 5 | 20 | 48 | 207 | $233x^2 + 170x + 207$ |

Each party also constructs and inverts the Vandermonde Matrix required by the polynomial degrees involved, in our case, $t = 2$ and $2t + 1 = 5$.

$$V = \begin{bmatrix} 1 & 1 & 1 & 1 & 1 \\ 1 & 2 & 4 & 8 & 16 \\ 1 & 3 & 9 & 27 & 81 \\ 1 & 4 & 16 & 64 & 256 \\ 1 & 5 & 25 & 125 & 625 \end{bmatrix}$$

$$V^{-1} = \begin{bmatrix} 5 & -10 & 10 & -5 & 1 \\ -77/12 & 107/6 & -39/2 & 61/6 & -25/12 \\ 71/24 & -59/6 & 49/4 & -41/6 & 35/24 \\ -7/12 & 13/6 & -3 & 11/6 & -5/12 \\ 1/24 & -1/6 & 1/4 & -1/6 & 1/24 \end{bmatrix}$$

Every party then evaluates their ploynomial at the indicies of all the parties and send to each other party the value of their polynomial evaluated at the index of that player. This results in the the values in Table A.4 being shared appropriately among the parties. In the table, each column indicates the values of the polynomial held by the party denoted in the subscript. Each row in the table indicates the values held by the party with the given index. This table represents the values denoted earlier by $R_{ij}$. To calculate the values given in the rightmost column each party computes the dot product of the first row of the inverted Vandermonde matrix and the vector of values received from the peers as well as the one value they calculated for themselves. For party with index 1, it would look like the following:

$$[5 \; -10 \; 10 \; -5 \; 1] \; [247 \; 27 \; 127 \; 117 \; 108]^T \mod 251 = 1$$

Table A.4: Values to complete the multiplication protocol

| Index | $h_1(x)$ | $h_2(x)$ | $h_3(x)$ | $h_4(x)$ | $h_5(x)$ | Shares of a · c |
|---|---|---|---|---|---|---|
| 1 | 247 | 27 | 127 | 117 | 108 | 1 |
| 2 | 240 | 52 | 140 | 47 | 224 | 61 |
| 3 | 11 | 54 | 14 | 92 | 53 | 1 |
| 4 | 62 | 33 | 0 | 1 | 97 | 72 |
| 5 | 142 | 240 | 98 | 25 | 105 | 23 |

Now, it can be verified that we do indeed possess shares of a second degree polynomial which has the desired product as its constant term, using the first three shares again:

$$L(0) = 1 \cdot \frac{-2}{1-2} \cdot \frac{-3}{1-3} + 61 \cdot \frac{-1}{2-1} \cdot \frac{-3}{2-3} + 1 \cdot \frac{-1}{3-1} \cdot \frac{-2}{3-2}$$

Which of course simplifies to:

$$L(0) = 1 \cdot 3 - 61 \cdot 3 + 1 \mod 251 = 72$$

and the product of 18 and 4 is in fact 72. This demonstrates that it is possible to rebuild the secret with three shares, which would only be true if the degree is less than or equal to two. This further indicates the protocol is correct since the appropriate value is the constant term of the polynomial reconstructed even though the order of the initially constructed polynomial has been halved. Also, note that we now have two representations of the secret value, 72, in our set of shares which are not in any way implicitly related to one another from a simple inspection of the values in the shares.

## A.2 Additive Secret Sharing Illustrated

**Constructing Shares** Using the same secret values that were employed previously, and generating 3 shares, since no more than this is required by any of the operations' protocols. the secret values 18, 54, and 4 can be computed to yield the shares in Table A.5. For the first secret value, 18, bolded below, the shares are chosen from a uniform random distribution on the domain of the secret, again defined by the field generated under the integer, $p = 251$.

$$117 = \mathbf{18} - 90 - 67 \mod p$$

The rest of the shares are generated in a similar manner.

**Rebuilding Secrets** This operation is as simple as the creation of the shares, one need simply add the shares modulus the prime defining the field. Here we shall reconstruct the secret value 54 using its shares.

$$29 + 154 + 122 \mod p = 54$$

Table A.5: Additive Secret Sharing Shares

|   | Shares of 18 | Shares of 54 | Shares of 4 |
|---|---|---|---|
| 1 | 85 | 29 | 121 |
| 2 | 67 | 154 | 63 |
| 3 | 117 | 122 | 71 |

Table A.6: Addition in Additive Secret Sharing

|   | Shares of 18 | Shares of 54 | Summed shares of 18 and 54 |
|---|---|---|---|
| 1 | 85 | 29 | 114 |
| 2 | 67 | 154 | 221 |
| 3 | 117 | 122 | 239 |

**Addition** As explained previously, addition in this scheme is also trivial requiring no communication and only one local operation, adding the individually held shares to obtain a share of the sum of the two secrets. The individually summed shares can be seen and verified correct in Table A.6.

This can easily be verified as correct by using the sums of the shares to rebuild the secret:

$$114 + 221 + 239 \mod p = 72$$

which is the expected sum of 18 and 54.

**Multiplication** Here the previously given protocol in Algorithm 1 is demonstrated. For each party to receive a share of the product of U, and V, in this example, 18 and 4, respectively. Next, we follow the necessary steps to complete the computation, again, all operations must occur according to the field used in the scheme:

Party 1 generates $r_{12} = 236, r_{13} = 233, s_{12} = 184, s_{13} = 85, t_{12} = 90$

Party 2 generates $r_{21} = 129, r_{23} = 108, s_{21} = 176, s_{23} = 96, t_{23} = 245$

Party 3 generates $r_{31} = 16, r_{32} = 56, s_{31} = 20, s_{32} = 71, t_{31} = 37$

All parties exchange appropriate values

Party 1 Computes:

$\hat{a}_{12} = u_1 + r_{31} = 101, \hat{b}_{12} = v_1 + s_{31} = 141, \hat{a}_{13} = u_1 + r_{21} = 214, \hat{b}_{13} = v_1 + s_{21} = 46$

Party 2 Computes:

$\hat{a}_{23} = u_2 + r_{12} = 52, \hat{b}_{23} = v_2 + s_{12} = 247, \hat{a}_{21} = u_2 + r_{32} = 123, \hat{b}_{21} = v_2 + s_{32} = 134$

Party 3 Computes:

$$\hat{a}_{31} = u_3 + r_{23} = 225, \hat{b}_{31} = v_3 + s_{23} = 167, \hat{a}_{32} = u_3 + r_{13} = 99, \hat{b}_{32} = v_3 + s_{13} = 156$$

All parties exchange appropriate values

Party 1 computes:

$$c_1 = u_1\hat{b}_{21} + u_1\hat{b}_{31} + v_1\hat{a}_{21} + v_1\hat{a}_{31} - \hat{a}_{12}\hat{b}_{21} - \hat{b}_{12}\hat{a}_{21} + r_{12}s_{13} + s_{12}r_{13} - t_{12} + t_{31} = 48$$

$$uv_1 = c_1 + u_1v_1 = 42$$

Party 2 computes:

$$c_2 = u_2\hat{b}_{32} + u_2\hat{b}_{12} + v_2\hat{a}_{32} + v_2\hat{a}_{12} - \hat{a}_{23}\hat{b}_{32} - \hat{b}_{23}\hat{a}_{32} + r_{23}s_{21} + s_{23}r_{21} - t_{23} + t_{12} = 47$$

$$uv_2 = c_2 + u_2v_2 = 1$$

Party 3 computes:

$$c_3 = u_3\hat{b}_{13} + u_3\hat{b}_{23} + v_3\hat{a}_{13} + v_3\hat{a}_{23} - \hat{a}_{31}\hat{b}_{13} - \hat{b}_{31}\hat{a}_{13} + r_{31}s_{32} + s_{31}r_{32} - t_{31} + t_{23} = 5$$

$$uv_3 = c_3 + u_3v_3 = 29$$

Once again, it is easy to reconstruct the desired secret from these shares to verify correctness.

$$42 + 1 + 29 \mod p = 72$$

This is the intended product of 18 and 4. Furthermore, we now again have two representations of the same secret value in the set of shares and there is no clear or discernible connection between them. All the beginning and constructed values are available in Table A.7. The basic notion of it is in unfolding and substituting the basic building blocks in the cooperatively constructed linear combination and seeing that, at the termination of the scheme, what is held by the three parties are shares of:

$$(u_1 + u_2 + u_3)(v_1 + v_2 + v_3)$$

which is the desired outcome.

Table A.7: Additive Secret Sharing Shares

|   | Shares of 18 | Shares of 54 | Shares of 4 | Shares of 72 | Shares of 72 |
|---|---|---|---|---|---|
| 1 | 85 | 29 | 121 | 114 | 24 |
| 2 | 67 | 154 | 63 | 221 | 216 |
| 3 | 117 | 122 | 71 | 239 | 117 |

# Appendix B

# Numerical Examples of Comparison Protocols

It is often the case for somewhat monolithic protocols such as ours to see step by step executions of the protocol to aid in understanding. What follows is two example executions of our protocol from Section 3.3, our main protocol. The other variations presented are closely related to this and it is our hope that demonstrating these two should be sufficient for aiding in the understanding of our other approaches. In this example we will operate on values to compare $a = 15$ and $b = 13$. Clearly in both cases our protocol should return $f = 1$ shared among the parties indicating that $a \geq b$. In the following explanation we focus on the main aspects of our approach to the comparison, and leave aside some of the details of the sharing and mapping between sharing groups which we include for security or efficiency. All values will be addressed by the steps in which they occur with the notation in accordance with the variables in the algorithm.

## B.1 Example for Algorithm 4 from Section 3.3

- *Input Transformation:* Steps 1-2 We either double or double and increment the input values to assure that there is at least one location with a differences

  - $a = 15 = 01111$
  - $b = 13 = 01101$
  - $2a + 1 = 31 = 011111$
  - $2b = 26 = 011010$
  - $s_a = 5$

- *Compute bitwise differences:* Steps 3-4 focuses on producing the bitwise exclusive or of the parties inputs and handling their mapping between sharing domains.

  - $e = 000101$

- *Locate the most significant difference* Steps 5 and 6 identify the location of the most significant difference in the bitwise `xor`.

  - $\gamma' = 0, 0, 0, 1, 1, 2$
  - $\gamma = 0, 0, 0, 1, 2, 3$
  - $\gamma = -1, -1, -1, 0, 1, 2$
  - $u = r, r, r, 0, r, r$ note: r simply denotes a uniform random value
  - $v = r, 0, r, r, r, r$
  - $h = 010000$

- *Deriving $s'_a$* Steps 7-9 use the constructed vector of bits $h$ with the private input $a$ to construct the vector of bits $h'$ which will be different from $a$ in exactly one bit index, in accordance with Claim 1. Then the protocol sums the bits of $h'$ calculates the difference between this result and the count of set bits in $a$ calculated earlier and maps this to the desired comparison result, again as justified in Claim 1.

- $h = 000100$
- $h' = 011011$
- $s'_a = 4$
- $f = 1$
- $f = f + 1 = 2$
- $f = f 2^{-1} = 1 \to a \geq b$

- *Share transformation:* Steps 10-11 The final step is to map the result of the transformed functionality back into the group in use for the secret sharing scheme in general, but no computations integral to the functionality are handled in this step.

## B.2  Example for Algorithm 7 from Section 3.4.3

- *Locate differences* Step 1 focuses on producing the bitwise exclusive or of the parties inputs and handling their mapping between sharing domains.

  - $a = 15 = 01111$
  - $b = 13 = 01101$
  - $2a + 1 = 31 = 011111$
  - $2b = 26 = 011010$
  - $s_a = 5$
  - $e = 000101$

- *Locate most significant difference* The focus of step 2 is the construction of the matrix $E$ which is a triangular matrix based on $e$, the `xor` of the values calculated in Step 1. Then in step 3 the product along the columns of elements in the triangular region of the matrix is calculated. This is denoted in the matrix and vector below by alternating gray columns in the matrix $E$ associated with the same locations in vector $v$. The sole exception is the case of the most significant location in $v$ which is copied directly from $e$. This location is denoted in salmon color.

$$E = \begin{bmatrix} & 0 & 0 & 1 & 0 & 1 \\ & & 0 & 1 & 0 & 1 \\ & & & 1 & 0 & 1 \\ & & & & 1 & 0 \\ & & & & & 1 \end{bmatrix} \quad \text{(B.1)}$$

$$v = \begin{array}{|c|c|c|c|c|c|} \hline 0 & 0 & 0 & 1 & 0 & 0 \\ \hline \end{array} \quad \text{(B.2)}$$

- *Mask input, sum, and map* Step 4 uses the constructed vector of bits $v$ with the private input $a$ to construct the vector of bits $h$ which will be different from $a$ in exactly one bit index, in accordance with Claim 1. This summation across $h$ yields $s'_a$ which is used to calculate the difference between this result and the count of set bits in $a$ calculated earlier. Finally the result is mapped into $\mathbb{Z}_2$, again as justified in Claim 1.

  – $h = 011011$

  – $s'_a = 4$

  – $f = 1$

  – $f = f + 1 = 2$

  – $f = f2^{-1} = 1 \rightarrow a \geq b$

## B.3 Dependencies and Details

### B.3.1 Bloom Filters

This probabilistic data structure is eminently beneficial in a setting for which tests of membership in a large set are both frequent, and can tolerate a small to negligible chance of false positives while guaranteeing no false negatives [96]. The mathematical underpinning of this data structure relies on hashing, the probability of hash collisions, and the probability of the entire vector of values eventually being changed to represent set membership. Constructing a Bloom filter involves generating a vector of bits of length $\beta$. The $\eta$ values to be represented in the filter are each hashed with $\kappa$ distinct hash functions. The result of the hash, modulo $\beta$, is used as an index to the bit array, and the resulting location is set. If the resulting index location was already 1, it is left as 1. Testing for set membership involves calculating the hashes for the object of the query, accessing each location, and verifying that every resulting location is 1. If any location is not 1, the queried object does not belong to the set [76].

A small and simple example is given in Figure B.1. Here the first vector denotes the initialization of a filter consisting of 8 locations indexed from left to right by values in $\{0\ldots7\}$. The second represents the insertion of the element 5. We have hashed the value 5 with three distinct hash functions. The result of each hash function, modulo 8, is 2, 4, and 5, respectively. As can be seen in the figure, each of these locations are set to 1. The filter only contains the element 5. If we test for set membership with a value 2, we calculate the hash values with the same hash functions, modulo 8, which are 5, 0, and 4 respectively. In the case of indices 5 and 4, the filter contains 1, but location 0 contains a 0. All tested locations are not equal to one; therefore, the filter correctly indicates that 2 is not a member of the set.

The probabilistic nature of this structure is related to the collision probability and the fact that, given a sufficiently large number of elements in the set of interest, and a sufficiently small filter, the rate of collision increases. This results in, eventually, an item which is not in the set testing positive for set membership under a query through a sufficient number of collisions with other elements. Given enough elements, the entire filter may fill with set bits, thus permitting every query to be evaluated as a set member. With appropriately chosen parameters, the event of a false positive test for set membership can be made insignificant. The relationship between these variables can be derived fairly intuitively. The probability that any individual bit location is still 0 after inserting $\eta$ elements, hashed by $\kappa$ hash functions, in a filter with $\beta$ locations is approximated as follows:

$$p = \left(1 - \frac{1}{\beta}\right)^{\kappa\eta} \approx e^{-\kappa\eta/\beta} \tag{B.3}$$

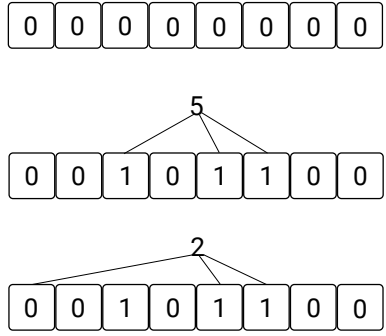

Figure B.1: Bloom filter creation, item insertion, and query

Consequently, the probability of a false positive is approximately $(1-p)^\kappa \approx .6185^{\beta/\eta}$. As discussed in [77], the optimal error rate occurs when $p = \frac{1}{2}$, so the parameters for which the following approximate equality holds, constrained by integer values for each parameter, are nearly optimal choices:

$$\kappa \approx \frac{\beta}{\eta} \ln 2 \tag{B.4}$$

## B.3.2 Secret Sharing

Secret Sharing schemes present the ability to break data up into multiple shares, distribute the shares, and perform computations based on these shares rather than the data. This can be used to construct a solution to the classic Yao's Millionaire problem [25, 30], or any other general MPC functionality desired. Our motivation is to present a solution to the problems we have described based on techniques and functionalities with a similar foundation.

**Shamir's Secret Sharing Scheme**

The scheme we focus on for this work is Shamir's secret sharing scheme [3]. In this setting, the data is set as the constant term of a polynomial of degree $t - 1$ for some threshold $t \geq 2$. Arbitrarily many parties $m$ may be involved in the scheme, but to keep complexities minimal, if there is no desire for multiplications, $m$ may be set equal to $t$. If multiplications are needed, $m$ may be set to $m = 2t - 1$. If resilience in either case is desired, it is possible through setting $m > t$ or $m > 2t - 1$ dependent on the need for multiplication [97].

**Generating Shares** Formally, a secret value s is placed within a $t - 1$ degree polynomial $f$ of the form:

$$f(x) = c_{t-1}x^{t-1} + c_{t-2}x^{t-2} + \cdots + c_2 x^2 + c_1 x + s \mod N$$

in which all constants $c_{t-1}$ through $c_1$ are chosen from a uniform random distribution on the domain denoted by $\mathbb{Z}_N$ where $N$ is a prime and sufficiently large to represent s. For all $m$ parties, each party $P_i$'s share of $s$ is constructed by:

$$[s]_N^{P_i} = f(i) \mod N$$

**Rebuilding Secrets** Lagrange Polynomial Interpolation is the method of choice to reconstruct $f$ and subsequently derive $s$:

$$s = f(0) = \sum_{j=1}^{m} [s]_N^{P_j} \prod_{k=1:k\neq j}^{m} \frac{-x_k}{x_j - x_k} \mod N$$

**Addition** In the Shamir Secret Sharing Scheme, addition is "free" in that no communication is required between parties, and no lengthy local computations are necessary. For two values, $a$ and $b$, which are shared among the parties, the shared sum is computed by each party $P_i$ locally computing their new share by adding their shares of the secrets $a$ and $b$:

$$[a+b]_N^{P_i} = [a]_N^{P_i} + [b]_N^{P_i} \mod N$$

**Multiplication** Multiplication of a public constant $c$ is affected simply through the multiplication of this constant into each share

$$[ac]_N^{P_i} = c[a]_N^{P_i} \mod N$$

Due to the fact that this scheme is inherently a linear scheme, multiplication of two shared values is a more complex process. We will adopt the method proposed by Gennaro et al [5].

**Additive Secret Sharing**

If the concern for the network is only passive adversaries, a large boost in efficiency may be realized via implementing our proposed solutions to follow under the additive secret sharing scheme with three parties or servers involved in the scheme.

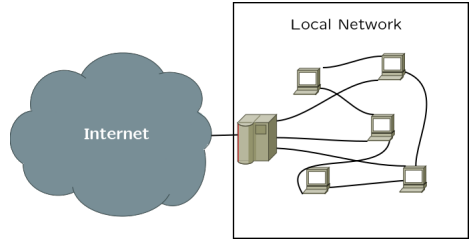

Figure B.2: Constrained network topology

**Generating Shares** To create any desired number of shares, $m$, in this scheme, the secret, $s$, to be shared is placed in an equation as follows in which all $[s]_N^{P_1}$ through $[s]_N^{P_{m-1}}$ are uniformly randomly selected integers and $[s]_N^{P_m}$ satisfies the equation, all in $\mathbb{Z}_N$:

$$[s]_N^{P_m} = s - [s]_N^{P_1} - [s]_N^{P_2} - \cdots - [s]_N^{P_{m-1}} \mod N$$

**Rebuilding Secrets** Rebuilding secrets in this case is also trivial since it can be seen from the preceding all one need do is sum all the $m$, shares held by the parties modulus the same integer $N$:

$$s = \sum_{i=1}^{m}[s]_N^{P_i} \mod N$$

This is the greatest area of difference for our efficiency concerns between the two schemes. Additive secret sharing requires only additions modulo $N$ to rebuild secrets while Shamir's scheme requires polynomial interpolations. Similar to Shamir scheme, addition between two secretly shared values, and multiplication with a constant can be performed locally at each party without communication. However, multiplication between two shared values is not as straight forward. When there are three parties, we can adopt the protocol given in the work of Bogdanov et al [98].

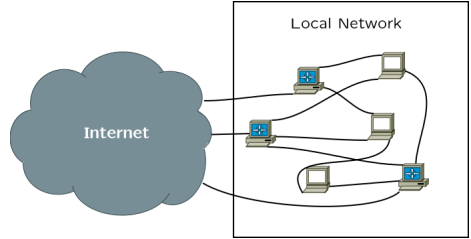

Figure B.3: Unconstrained network topology

### B.3.3 Network Topology

One situation for network structure si that in which the network topology is constrained such that there exists a single gateway server to the wider Internet, such as which is depicted in Figure B.2. The other situation is for less control than a pure centralized topology, but it is centralized up to an arbitrary number of links between the local network and the rest of the Internet. This is depicted in Figure B.3. For our following proposed approaches, unlike either of the previously discussed cases from Section 4.2, the structure of the firewall is distributed in a secret shared format. The control function is evaluated via secure multi-party computation techniques, and either topology is acceptable in our scheme.

**Centralized Topology**

In the centralized or constrained topology setting, the firewall exists on the main server acting as a gateway. With every packet arriving, the sole bridge between the local and external networks conducts the source checking against the rules of the firewall. The result of the check leads to the packet's forwarding or rejection. The rules in this case normally exist within the settings of this single machine and are thus potentially and easily able to be manipulated, particularly by a malicious insider. Once the malicious behaviors have been carried out, it may be difficult for parties within the network to identify the alteration which represents a serious security risk to the local network [56].

As mentioned previously, there may be additional motivations to keep the rules of the system itself secret from most parties after the system has been initialized. For our protocol to follow, this gateway would invoke the protocols among the servers who would evaluate the firewall function and send their shares of the result back to the gateway. The gateway can then reconstruct the result from combinations of the shares to identify malicious behavior by the servers and permit or reject network traffic in accordance with the firewall.

**Distributed Topology**

In a situation without the benefit of a constrained topology, the firewall must be distributed by some means. If many machines may have access to the external larger Internet, this mesh must have tools to consistently, efficiently and reliably maintain the functionality of the firewall in a distributed manner. This is achieved in many cases by associating the firewall rules with the forwarding tables for other network nodes [77]. However, these existing methods dramatically increase the risk of information leakage regarding firewall rules, alteration and inconsistency of the rules among each machine in the network.

For our proposed solution to follow, each machine on the fringe of the network, connected both internally and to the external network, can act similarly to the constrained case previously described. Each of these machines, serving as a gateway, can invoke the protocols among the servers sharing the filter who return shares of the result of the firewall evaluation according to our forthcoming protocols. Again, the invoking machine would use combinations of the shares to rebuild the result to identify malicious activity and control packet flow.

[40] Tomas Toft. Sub-linear, secure comparison with two non-colluding parties. In *International Workshop on Public Key Cryptography*, pages 174–191. Springer, 2011.

[41] Berry Schoenmakers and Pim Tuyls. Efficient binary conversion for paillier encrypted values. In Serge Vaudenay, editor, *Advances in Cryptology - EUROCRYPT 2006*, pages 522–537, Berlin, Heidelberg, 2006. Springer Berlin Heidelberg.

[42] Florian Kerschbaum, Debmalya Biswas, and Sebastiaan de Hoogh. Performance comparison of secure comparison protocols. In *Database and Expert Systems Application, 2009. DEXA'09. 20th International Workshop on*, pages 133–136. IEEE, 2009.

[43] David W Archer, Dan Bogdanov, Benny Pinkas, and Pille Pullonen. Maturity and performance of programmable secure computation. *IEEE security & privacy*, 14(5):48–56, 2016.

[44] Frederik Armknecht, Colin Boyd, Christopher Carr, Kristian Gjøsteen, Angela Jäschke, Christian A Reuter, and Martin Strand. A guide to fully homomorphic encryption. *IACR Cryptology ePrint Archive*, 2015:1192, 2015.

[45] Ivan Damgård, Matthias Fitzi, Eike Kiltz, Jesper Buus Nielsen, and Tomas Toft. Unconditionally secure constant-rounds multi-party computation for equality, comparison, bits and exponentiation. In *Theory of Cryptography Conference*, pages 285–304. Springer, 2006.

[46] Takashi Nishide and Kazuo Ohta. Multiparty computation for interval, equality, and comparison without bit-decomposition protocol. In *International Workshop on Public Key Cryptography*, pages 343–360. Springer, 2007.
138

# VITA

Ken Goss was born and raised in Lebanon, Missouri. He attended Truman State University where he obtained a BA in Music, an MA in Education, as well as a number of other skills of dubious utility. Following this debacle, he proceeded to study in computer science and engineering, at Missouri S&T where he obtained two BS degrees in Computer Science and Computer Engineering, as well as a minor in Mathematics. He is currently completing a PhD at the University of Missouri and is working in researching applied cryptography with Dr. Wei Jiang.

Ken is blissfully married to Heather and they have five children (so far), Miriam, Emet, Laila, Adi, and Ari. Following the completion of his present course of study, Ken plans to work in cryptography research and development at Sandia National Labs.